\documentclass[12pt]{article}
\usepackage{amsmath}
\usepackage{graphicx}
\usepackage[round]{natbib}
\usepackage{url} 
\usepackage[latin1]{inputenc} 
\newtheorem{theorem}{Theorem}

\newtheorem{lemma}{Lemma}

\newtheorem{remark}{Remark}
\newtheorem{proposition}{Proposition}

\usepackage{amssymb}
\usepackage{bm}
\usepackage{algorithmic}
\usepackage{booktabs}
\usepackage{amsmath}
\usepackage{algorithm}
\usepackage{algorithmic}
\renewcommand{\algorithmicrequire}{\textbf{Input:}}    \renewcommand{\algorithmicensure}{\textbf{Output:}}
\usepackage{color}
\usepackage{algorithm}
\usepackage{verbatim}
\usepackage{algorithmic}
\usepackage{makecell}
\usepackage{enumerate}
\usepackage{graphicx}
\usepackage{threeparttable}
\usepackage{multirow}
\usepackage{mathrsfs}
\usepackage{hyperref}
\usepackage{subfigure}
\usepackage{setspace}
 \hypersetup{
     colorlinks=true,
     linkcolor=magenta,
     filecolor=blue,
     citecolor = darkgreen,
     urlcolor=cyan,
     }

\newcommand*{\QEDA}{\hfill\ensuremath{\blacksquare}}

\newcommand{\blind}{0}

\definecolor{darkgreen}{rgb}{0,0.6,0.3}

\begin{document}

\def\spacingset#1{\renewcommand{\baselinestretch}%
{#1}\small\normalsize} \spacingset{1}

\if0\blind
{
  \title{\Large \textbf{On the efficacy of higher-order spectral clustering under weighted stochastic block models}}
  \author{ Xiao Guo$^{\dagger}$, Hai Zhang$^\dagger$, Xiangyu Chang$^\ddag$\thanks{Xiangyu Chang is the corresponding author.}\hspace{.2cm}  \\
    $^\dagger$ School of Mathematics, Northwest University, China\\
    $^\ddag$
     School of Management, Xi'an Jiaotong University, China}
 \maketitle
} \fi

\if1\blind
{
  \bigskip
  \bigskip
  \bigskip
  \begin{center}
    {\LARGE\bf }
\end{center}
  \medskip
} \fi

\bigskip

\begin{abstract}
Higher-order structures of networks, namely, small subgraphs of networks (also called network motifs), are widely known to be crucial and essential to the organization of networks. Several works have studied the community detection problem\textendash a fundamental problem in network analysis at the level of motifs. In particular, the higher-order spectral clustering has been developed, where the notion of \emph{motif adjacency matrix} is introduced as the algorithm's input. However, how the higher-order spectral clustering works and when it performs better than its edge-based counterpart remain largely unknown. To elucidate these problems, the higher-order spectral clustering is investigated from a statistical perspective. The clustering performance of the higher-order spectral clustering is theoretically studied under a {\it weighted stochastic block model}, and the resulting bounds are compared with the corresponding results of the edge-based spectral clustering. \textcolor{black}{The upper bounds and simulations show that when the network is dense with the weak signal of weights, higher-order spectral clustering can lead to a performance gain in clustering. Real data experiments also corroborate the merits of higher-order spectral clustering.
}

\end{abstract}

\vfill

\newpage
\spacingset{1.5} 
\section{Introduction}
\label{sec:intro}
The network is a standard representation of relationships among units of complex systems in many scientific domains, including biology, ecology, sociology, and information science, among many others \citep{newman2018networks,goldenberg2010survey,kolaczyk2009statistical}. One of the most common features of networks is that they have communities, clusters, or modules\textendash groups of nodes that are, in some sense, more similar to nodes within the same community than to other nodes. Community detection, namely, detecting such communities based on given network structures, has been one of the fundamental problems in network analysis which helps us gain knowledge about the behavior and functionality of networks.

Past decades have seen various community detection methods, including spectral clustering, modularity maximization, likelihood methods, and semidefinite programming. See \citet{abbe2017community} for a survey. Most of the current community detection procedures are edge-based, that is, those procedures essentially make use of the network adjacency matrix, which stores the similarity of each pair of nodes to detect communities. However, there has been increasing evidence that \emph{higher-order} connectivity patterns are crucial to the organization of real networks. They can help understand and gain more insights into the behavior of networks \citep{holland1977method,milo2002network,mangan2003coherent,yang2014overlapping,rosvall2014memory}. The higher-order structures of networks often refer to small network subgraphs, or the so-called network \emph{motifs}, in contrast to the \emph{lower-order} network edges. In particular, triangular motifs are common in many kinds of networks. For example, in social networks, individuals who share a common friend are more likely
to become friends, leading to a large proportion of triangles to the two-edge wedges~\citep{Rohe}.

Very recently, several works have studied the problem of network clustering at the level of motifs, which can provide new insights into network organization beyond the edge-based clustering if the network exhibits rich higher-order structures; see \citet{benson2016higher,tsourakakis2017scalable,laenen2020higher,underwood2020motif}, among others. Generally, to capture the higher-order connective patterns, a \emph{motif adjacency matrix} is often designed as a surrogate for the original edge adjacency matrix; see \citet{benson2016higher,paul2018higher} for example. To be specific, the $(i,j)$'s entry of the motif adjacency matrix is the number of certain motifs that nodes $i$ and $j$ participate in commonly in binary networks.
It is suggested that one could use the motif adjacency matrix in the subsequent clustering analysis to obtain a good empirical result. In particular, the \emph{higher-order spectral clustering}, namely, the spectral clustering with the motif adjacency matrix as its input, is well-suited to the problem and is the focus of this work. However, it remains largely unknown how higher-order spectral clustering actually works. Moreover, when it is better than its edge-based counterpart remains to be seen. From a statistical perspective, the aim should be to show the efficacy of higher-order spectral clustering on a particular network dataset and to understand its merits under some underlying mechanisms.

In this paper, we attempt to illustrate the efficacy of higher-order spectral clustering in the context of weighted networks. To this end, we develop the \emph{weighted stochastic block models} (WSBMs) to mimic the weighted networks with communities, which is a generalization of the stochastic block models (SBMs) \citep{holland1983stochastic} to weighted networks. In particular, we assume the edges are first generated according to the SBM mechanism, where
nodes are partitioned into several distinct communities and conditioned on the underlying community assignments. The edges are generated independently according to the community membership of their end nodes. After that, positive weights independently generated from a common distribution are assigned to all the existing edges. We specify the expectation of weights without requiring their distributions. As a generalization of the aforementioned motif adjacency matrix, we define the \emph{weighted motif adjacency matrix} to capture the network weights, whose $(i,j)$'s entry is the weights summation of a certain motif that nodes $i$ and $j$ both involve. In particular, we focus on the undirected triangular motif. With these at hand, we study the approximation error and misclustering error of the higher-order spectral clustering under the WSBM mechanism and compare the \textcolor{black}{upper bounds} with the corresponding results of edge-based spectral clustering  (i.e., spectral clustering with the weighted edge adjacency matrix as the input). \textcolor{black}{It turns out that when the network is dense {\emph{but}} the conditional expectation of edge weight is small, and the higher-order spectral clustering leads to a lower misclustering error than the edge-based one does. The rationality is as follows. The signal of communities could be weak under small edge weight, which may lead to unsatisfactory clustering performance of the edge-based method. However, when the network is dense, the triangles are rich, which enables each node pair to borrow the strength and information of incident edges. Thus, the clustering quality could be improved by utilizing the triangular motif.} Note that the unequal weights are crucial to ensure the advantage of the higher-order spectral clustering over the edge-based one. In addition, it is worth noting that, theoretically, the weighted motif adjacency matrix leads to a larger eigen-gap of the corresponding population matrix than the edge adjacency matrix does. In this sense, the signals of the weighted networks are thus enhanced. \textcolor{black}{We conduct simulations to verify the theoretical findings. In particular, the higher-order spectral clustering has performance gain over the edge-based counterpart in various regimes where the requirements for theory are not necessarily met. The real data experiments on two statisticians networks and one ${\rm PM}_{2.5}$ network provide insightful and interesting results, showing the efficacy of higher-order spectral clustering on weighted networks.}

The reminder of this paper is organized as follows. Subsection \ref{relatedwork} reviews the related works. Section \ref{model} provides the general framework including the notation, weighted stochastic block models and higher-order spectral clustering. Section \ref{theory} presents the theoretical results. Section \ref{discussion} discusses the theoretical and methodological aspects of our work and also poses possible extensions. Sections \ref{simulation} and \ref{realdata} include the simulations and real data analysis, respectively. Section \ref{conclusion} concludes the paper. The proofs can be found in the Supplementary Material.

\subsection{Related works}
\label{relatedwork}
Analyzing networks at the level of motifs has received increasing attention in recent years. In particular, \citet{benson2016higher} proposed a conductance-based method, which generalized the original edge-based conductance to the motif-based one. The motif adjacency matrix was then introduced to simplify the minimization of the motif-based conductance; more specifically, minimizing the motif-based conductance can be reduced to minimizing the edge-based conductance but using the motif adjacency matrix. \citet{serrour2011detecting} studied the higher-order network clustering in the modularity maximization framework, where they generalized the notion of modularity by transforming its building block from edges to motifs. They employed the tensor to capture the triangular patterns within the network but then transformed it to the aforementioned motif adjacency matrix to accelerate the algorithm. There are also works focusing on directed networks \citep{laenen2020higher,cucuringu2020hermitian,underwood2020motif} and local spectral clustering \citep{yin2017local}.

The statistical aspects of the motif-based methods are also studied by several authors. \citet{Rohe} show the blessings of transitivity in sparse networks. They developed an algorithm that exploits the triangles built by network transitivity and shows its clustering ability statistically under the newly developed local stochastic block models. \citet{tsourakakis2017scalable} also studied the motif-based conductance and provided a random walk explanation of it. They showed that the random walk on the network corresponding to the motif adjacency matrix is more likely to stay in the same true community of the stochastic block model \citep{holland1983stochastic} than the random walk on the original network. \citet{paul2018higher} proposed a superimposed stochastic block model, which is a superimposition of a classical dyadic (edge-based) random graph and a triadic (triangle-based) random graph. They rigorously analyzed the misclustering error bound of the higher-order spectral clustering under such models. However, the analysis fails to disclose the advantage of higher-order spectral clustering over the edge-based counterpart. In addition, the model is somewhat restrictive since the networks generated by the superimposed stochastic block models have edges lying in $\{0,1,2\}$, which infrequently happens in real networks. \citet{cucuringu2020hermitian} studies the spectral clustering on complex-valued Hermitian matrix representations, which implicitly use the higher-order structure of directed networks. They studied the algorithm theoretically under stochastic block models.

There exists another line of works focusing on the hyper-graph clustering problem; see  \citet{ghoshdastidar2017consistency,ghoshdastidar2017uniform,ghoshdastidar2014consistency}, among many others. In hyper-graphs, the hyper-edges are directly known as a prior rather than constructed using the motifs. Theoretically,
the motif adjacency matrix brings extra dependence between entries. Nevertheless, in the context of hyper-graph clustering, the entries of the adjacency tensor are independent.

There are also some authors making use of the motifs to study the testing problem of whether it has only one community or multiple communities under SBMs or its variants; notably by \cite{jin2018network,gao2017achieving,jin2021optimal,cammarata2022power}.

\textcolor{black}{On the other hand, there exist a few works on community detection for weighted networks; see \citet{aicher2015learning,xu2020optimal,cerqueira2023pseudo} for example and
references therein. Compared with these works, we are motivated to enhance the signal strength in weighted networks by using the weighted motif adjacency matrix and studying its effect theoretically. Generally, using the weighted motif adjacency matrix
can be regarded as a network preprocessing step, and the weighted motif adjacency matrix can essentially be the input of other community detection algorithms.}


\section{Framework}
\label{model}
In this section, we introduce the general framework of analysis. In particular, the WSBMs and the higher-order spectral clustering are introduced.

We first introduce some notes and notation. The readers could also refer to Table \ref{notation} for a brief summary. $\mathbb M_{n,K}$ denotes the set of all $n\times K$ matrices which have exactly one 1 and $K-1$ 0's in each row. Any $\Theta\in \mathbb M_{n,K} $ is called a \emph{membership matrix}, where each row corresponds to the community membership of the corresponding node. For example, node $i$ belongs to community $g_i\in\{1,...,K\}$ if and only if $\Theta_{ig_i}=1$. For $1\leq k\leq K$, denote $G_k=G_k(\Theta)=\{1\leq i\leq n:g_i=k\}$ and $n_k=|G_k|$, where $G_k$ consists of nodes with their community membership being $k$. For any matrix $A_{n\times n}$ and $I,J\subseteq \{1,...,n\}$, $A_{I\ast}$ and $A_{\ast J}$ denote the submatrix of $A$ consisting of the corresponding rows and columns of $A$, respectively. $\|A\|_{\tiny {\rm F}}$, $\|A\|_2$, and $\|A\|_\infty$ denote the Frobenius norm, spectral norm, and the element-wise maximum absolute value of $A$, respectively. ${\rm diag}(A)$ denotes the diagonal matrix with its diagonal elements being the same as those of $A$. We also use $\|a\|_2$ to denote the Euclidean norm of any vector $a$.

Now we introduce the WSBMs which generalizes the SBMs \citep{holland1983stochastic} such that the potential network can have weighted edges rather than the binary edges in the SBMs. For an underlying network with $n$ nodes and $K$ communities, the two main parameter matrices of the WSBMs are the membership matrix $\Theta\in\mathbb M_{n,K}$, and the connectivity matrix $B\in[0,1]^{K\times K}$ where $B$ is of full rank, symmetric, and the entry $B_{kl}$ of $B$ represents the edge probability between any node in community $l$ and any node in community $k$. For simplicity, throughout the paper we assume the underlying network has balanced community size $n/K$ with the within-community probability being $p_n$ and the between-community probability being $p_n(1-\lambda)$ for any pair of nodes, where $\lambda$ is a constant in $(0,1)$. \textcolor{black}{Whereas note that the theoretical results, say the approximation error bound to be established in Theorem \ref{spbound}, also holds for general $B$'s; see Remark \ref{remarkforapp}.} Under the simplified structural assumption, the connectivity matrix $B$ takes the simple form
\begin{equation}
\label{2.1}
B=p_n\lambda I_K+p_n(1-\lambda)1_K1_K^\intercal,
\tag{2.1}
\end{equation}
where $1_K$ represents a $K$-dimensional vector of 1's. Further, we introduce a probability distribution $f$ supported on the positive line which is actually used to generate the edge weights. We do not specify the distribution of $f$. We assume that the expectation of its corresponding random variable is $\alpha_n$, and the variance is larger than 0. \textcolor{black}{Although we assume an community-independent weight here, we note that the results can be extended to community-independent case; see Remarks \ref{eigen-gap-B} and \ref{remarkforapp}.}

Given $\Theta$, $B$, and $f$, the weighted network adjacency matrix $W=(W_{ij})_{1\leq i,j\leq n}$ is generated as
\[W_{ij}=\begin{cases}
\label{2.2}
{\rm Bernoulli } (B_{g_ig_j})\cdot V_{ij} & \mbox{if }\; i<j, \\
0,& \mbox{if } \;i=j,\\
W_{ji}, & \mbox{if } \;i>j.\tag{2.2}
\end{cases}\]
where ${\rm Bernoulli }(p)$ denotes the bernoulli random variable which is 1 with probability $p$ and is 0 with probability $1-p$, $V_{ij}$'s denote the random variables generated from distribution $f$, and $V_{ij}$'s are mutually independent and they are also independent of $B_{g_ig_j}$'s. Define $\mathcal W=\alpha_n\Theta B\Theta^{\intercal}$, and it is then easy to see that $\mathcal W$ is the population of $W$ in the sense that $\mathbb E(W)=\mathcal W-{\rm diag}(\mathcal W)$. In order to capture the higher-order connectivity patterns of the weighted network, we define the following weighted motif (triangular) adjacency matrix $W^M=(W^M_{ij})$ to be
\begin{equation}
\label{2.3}
W^{M}_{ij}=\sum_{k=1}^n\mathbf 1(W_{ij}\cdot W_{ik}\cdot W_{jk}>0)(W_{ij}+W_{ik}+W_{jk}),
\tag{2.3}
\end{equation}
where $\mathbf 1(\cdot)$ stands for the indicator function. Recall that in \cite{benson2016higher,tsourakakis2017scalable}, the motif (triangular) adjacency matrix was defined to be a matrix with its $(i,j)$'s entry being the number of triangles that nodes $i$ and $j$ participate in commonly. Thus (\ref{2.3}) is a simple generalization of the motif adjacency matrix proposed by \cite{benson2016higher,tsourakakis2017scalable} in order to handle the weighted networks. Now, to facilitate further analysis, let us have a closer look at the expectation of $W^M$. Recall (\ref{2.1}) and (\ref{2.2}), and then we can easily obtain the following observations. When $g_i=g_j$,
\begin{align}
\label{2.4}
\mathbb E(W_{ij}^M)&=3\alpha_n\cdot p_n\{(\frac{n}{K}-2)p_n^2+(K-1)\frac{n}{K}p_n^2(1-\lambda)^2\}
:={\equiv h_1},
\tag{2.4}
\end{align}
and when $g_i\neq g_j$,
\begin{align}
\label{2.5}
\mathbb E(W_{ij}^M)=3\alpha_n\cdot p_n(1-\lambda)&\{2(\frac{n}{K}-1)p_n^2(1-\lambda)
+(K-2)\frac{n}{K}p_n^2(1-\lambda)^2\}{:= h_2}.
\tag{2.5}
\end{align}
Define
\begin{equation}
\label{2.6}
\mathcal W^M=\Theta((h_1-h_2)I_K+h_21_K1_K^\intercal)\Theta^\intercal,
\tag{2.6}
\end{equation}
then $\mathcal W^M$ is the population of $W^M$ by noting
$$\mathbb E(W^M)=\mathcal W^M-\mbox{diag}(\mathcal W^M).$$

\begin{remark}
\label{eigen-gap-B}
\textcolor{black}{(\ref{2.4}) and (\ref{2.5}) can be similarly computed under more general $B$'s and community-dependent weights. For example, assume that the within- and between-community probabilities are $a_n$ and $b_n$, respectively, and the within- and between-community weights are $\alpha_n$ and $\beta_n$, respectively, then within- and between-community expectations of $W^M$ are
$$h_1:=a_n(\frac{n}{K}-2)a_n^2 \cdot 3\alpha_n +a_n (K-1)\frac{n}{K} b_n^2 (\alpha_n+2\beta_n),$$
$$h_2:=2b_n (\frac{n}{K}-1)a_nb_n (\alpha_n+2\beta_n)+b_n(K-2)\frac{n}{K}b_n^2 \cdot (3\beta_n),$$
which are crucial for bounding the eigen-gap and thus the misclustering rate; see (\ref{3.4}).
}
\end{remark}

With these formulations at hand, we now introduce the algorithm for community detection where the goal is to recover the membership matrix $\Theta$ up to some column permutations. In this paper, we study the spectral clustering \citep{von2007tutorial} which generally consists of two steps. The first step is to perform the eigenvalue decomposition of a suitable matrix representing the network. The next step is to run $k$-means on the resulting $K$ leading eigenvectors.  We consider two kinds of spectral clustering. We will mainly deal with the so-called {\it higher-order spectral clustering or the motif-based spectral clustering}; see {Algorithm \ref{spectral} for details.} {And we would compare it with the {edge-based spectral clustering}, namely, Algorithm \ref{spectral} without \emph{step 1} and with \emph{step 2} replaced by the eigenvalue decomposition of the weighted adjacency matrix $W$.} The goal is to study how the higher-order spectral clustering performs under the WSBMs and to understand how it can enhance the clustering performance compared with that of the edge-based spectral clustering.

\textcolor{black}{\begin{remark}
\label{intuition}
Similar to Algorithm \ref{spectral}, \citet{benson2016higher} and \citet{serrour2011detecting} also proposed spectral-based algorithms by utilizing higher-order motifs. The sweep-cut algorithm in \citet{benson2016higher} essentially minimizes the motif-based conductance, which is the normalized number of motifs that are cut of two detected communities. The spectral-based algorithm in \citet{serrour2011detecting} approximately maximizes the motif-based modularity, which further reduces to the modularity with respect to motif adjacency matrix. In this sense, we expect that Algorithm \ref{spectral} would lead to larger modularity (w.r.t. weighted motif adjacency matrix) than the edge-based counterpart does. This actually will be verified by simulated experiments in Section \ref{simulation}.
\end{remark}}

The next lemma shows that $\mathcal W^M$, namely, the population version of $W^M$, has eigenvectors that reveal the true communities.
\begin{lemma}
\label{lemma}
For a WSBM with $K$ communities and $B$ structured as in (\ref{2.1}), suppose $\mathcal W^M=U\Sigma U^{\intercal}$ is the eigenvalue decomposition of $\mathcal W^M$ (see \ref{2.6}), then $U=\Theta X$. Specifically, for $\Theta_{i\ast}=\Theta_{j\ast}$, we have ${U}_{i\ast}={U}_{j\ast}$; while for $\Theta_{i\ast}\neq\Theta_{j\ast}$, we have $\|{U}_{i\ast}-{U}_{j\ast}\|_2=\sqrt{(n_{g_i})^{-1}+(n_{g_j})^{-1}}$.
\end{lemma}

It is easy to see from Lemma \ref{lemma} that the population eigenvector has $K$ distinct rows ($n$ rows in total) and two rows are identical if and only if the corresponding nodes are in the same underlying community. Therefore, the higher-order spectral clustering would cluster well if the sample version eigenvectors are concentrated around its expectation. We will discuss its theoretical properties in the next section.

\begin{algorithm}[htb] 

\renewcommand{\algorithmicrequire}{\textbf{Input:}}

\renewcommand\algorithmicensure {\textbf{Output:} }

\caption{{Higher-order spectral clustering for $K$ communities}} 

\label{spectral} 

\begin{algorithmic}[1] 

\REQUIRE ~\\ 

Community number $K$, a weighted adjacency matrix $W\in \mathbb{R}^{n\times n}$ representing the network;\\

\ENSURE ~\\ 
Community membership;\\
~\\
\STATE \textbf{Motif matrix construction:} Construct the weighted motif adjacency matrix $W^M$ according to (\ref{2.3}).\\
\STATE \textbf{Eigen-decomposition:} Find the $K$ leading eigenvectors of $W^M$ and form the $n\times K$ eigen-matrix. \\
\STATE \textbf{$k$-means:} Treat each row of the eigen-matrix as a point in $\mathbb{R}^K$ and run the Lloyd's algorithm on these points to solve $k$-means.
 \\
\end{algorithmic}
\end{algorithm}

\begin{table}[H]
\centering
\caption{A summary of the notes and notations
}\vspace{0.1cm}
\def\arraystretch{1.3}
\setlength{\tabcolsep}{0.04cm}
\small
\begin{tabular}{cl}
\hline
\textbf{Notation}&\textbf{Definition}\\
\hline
$n$&Number of nodes\\
$K$&Number of communities\\
$\Theta$&Membership matrix\\
$B$&Connectivity matrix\\
$\mathcal W$&Population edge-based weighted adjacency matrix\\
$\mathcal W^M$& Population motif-based weighted adjacency matrix\\
$W$&Edge-based weighted adjacency matrix\\
$W^M$& Motif-based weighted adjacency matrix\\
$g_i$&The community which node $i$ belongs to\\
$G_k$&Set of nodes from community $k$\\
$n_k$&The cardinality of $G_k$\\
$p_n$&The maximum link probability in $\mathcal W$\\
$\alpha_n$& The expectation of edge weights when edge exists\\
\hline
\end{tabular}
\label{notation}
\end{table}

\section{Theoretical analysis}
\label{theory}
In this section, we theoretically justify the clustering performance of the higher-order spectral clustering under the model set-up of the WSBMs and then we compare the theoretical bounds with those of the edge-based spectral clustering.
\subsection{Higher-order spectral clustering}
Recall that $W$ is the weighted adjacency matrix of a weighted graph generated from the WSBMs with $n$ nodes and $K$ communities (see (\ref{2.2})), and $\mathcal W$ denotes the population of $W$. $p_n$ is the maximum linking probability between two nodes in the WSBMs (see (\ref{2.1})) and $\alpha_n$ is the conditional expectation of an edge given that the edge exists. \textcolor{black}{For simplicity, we assume that $W_{ij}\leq 1$ for any $1\leq i,j\leq n$. This is reasonable since we will see in (\ref{3.8}) that we require a vanishing $\alpha_n$. On the other hand, one could specify the distribution of weights and using concentration inequalities to derive a ${\rm log}n$-type bound.}
$W^M$ is the corresponding weighted motif adjacency matrix (see (\ref{2.3})), where each entry is defined to be the weighted sum of the triangles that the corresponding nodes both join, and $\mathcal W^M$ denotes the population of $W^M$ except the diagonal elements. The next theorem provides the concentration bound of the weighted motif adjacency matrix around its expectation.
\begin{theorem}\label{spbound}
Assume {$p_{n}\geq c' \sqrt{\frac{{\rm log} n }{n}}$} for some constant $c'>0$, and {$\alpha_np_{n}\geq c''\frac{({\rm log}n)^\beta}{n} $} for some constants $c''>0$ and any $0<\beta<1$. Let
${\tau_{\rm max}=np_n^2}$ and $${D=\tau_{\rm max}^2n\alpha_np_{n}= n^{3}\alpha_np_{n}^5}.$$ Then for some constant $r>0$, there exists a constant $c$ such that with probability at least $1-n^{-r}-{\rm exp}(-r({\rm log}n)^\beta)$,
\begin{equation}
\label{3.1}{\|W^M-\mathbb E(W^M)\|_{2}\leq c\sqrt{D}},\tag{3.1}
\end{equation}
where $c$ depends on $r$ and $\beta$.
\end{theorem}

\textcolor{black}{
\begin{remark}
\label{remarkforapp}
Note that we make use of neither the structure of the connectivity matrix $B$ nor the weights distribution to derive the results. Indeed, (\ref{3.1}) holds for more general WSBMs, as long as the weights' expectations and connectivity matrix's entries are upper bounded.
\end{remark}
}
\textcolor{black}{Theorem \ref{spbound} requires that the network is not too sparse in the sense that $p_{n}\geq c' \sqrt{\frac{{\rm log} n }{n}}$ for some constant $c'>0$, where actually ensures that the network motifs are not rare. We also require the overall signal strength is lower bounded, namely, $\alpha_np_{n}\geq c''\frac{({\rm log}n)^\beta}{n}$ for some constants $c''>0$ and $0<\beta<1$}. Theorem \ref{spbound} indicates that the weighted motif adjacency matrix concentrates around its population version at the rate of $\sqrt{n^{3}\alpha_np_{n}^5}$.  The resulting bound is similar to the Theorem 2 of \citet{paul2018higher} except that $\alpha_n$ arises in our result. Recall that for the edge adjacency matrix, its spectral bound is the square root of the maximum expected degree \citep{lei2015consistency}. Applying this law to the weighted motif adjacency matrix, its spectral bound would read as the square root of the maximum expected ``motif degree'', which is actually $\sqrt{n^2\alpha_np_n^3}$ as can be seen in the proof. Comparing with $\sqrt{n^2\alpha_np_n^3}$, the bound $\sqrt{n^{3}\alpha_np_{n}^5}$ in (\ref{3.1}) is loose in this sense if we note that $p_{n}\geq c' \sqrt{\frac{{\rm log} n }{n}}$ for some constant $c'>0$.

Now we are ready to study the clustering performance of the higher-order spectral clustering. Specifically, we use the following metric to evaluate the quality of clustering,
\begin{align}
\label{3.2}
L(\hat{{\Theta}}, \Theta)=\underset{J\in E_K}{\rm min}\,\underset{1\leq k\leq K}{\sum}\;(2n_k)^{-1}\|(\hat{{\Theta}}J)_{G_{k\ast}}-\Theta_{G_{k\ast}}\|_0,
\tag{3.2}
\end{align}
where $\hat{{\Theta}}$ denotes the estimated membership matrix by the higher-order spectral clustering, and $E_K$ denotes the set of all $K\times K$ permutation matrices. Obviously, $L$ measures the sum of the fractions of the misclustered nodes within each community. The following theorem provides an upper bound on $L$.

\begin{theorem}\label{mis}
Let $\hat{{\Theta}}$ be the estimated membership matrix by the higher-order spectral clustering, and denote the minimum non-zero eigenvalue of $\mathcal W^M$ as $\lambda_K(\mathcal W^M)$. Suppose the assumptions in Theorem \ref{spbound} hold, and there exists a constant $c>0$ such that, if
\begin{equation}
\label{3.3}\frac{K^3}{n\alpha_np_n}\leq 1/c,\tag{3.3}
\end{equation}
then with probability at least $1-n^{-r}-{\rm exp}(-r({\rm log}n)^\beta)$ for some constant $r>0$ and any $0<\beta<1$, there exist subsets $S_k\subseteq  G_k$ for $k=1,...,K$ such that
\begin{equation}
\label{3.4}L(\hat{{\Theta}}, \Theta)\leq\sum_{k=1}^K \frac{|S_k|}{n_k}\leq c\frac{K\|W^M-\mathcal W^M\|_2^2}{\lambda_K^2(\mathcal W^M)}\leq c\frac{K^3}{n\alpha_np_n}.\tag{3.4}
\end{equation}
Moreover, for $G=\cup _{k=1}^K(G_k\backslash S_k)$, there exists a $K\times K$ permutation matrix $J$ such that
\begin{equation}
\label{3.5}\hat{\Theta}_{G_\ast}J=\Theta_{G_\ast},\tag{3.5}
\end{equation}
{where we recall that $\hat{\Theta}_{G_\ast}$ and ${\Theta}_{G_\ast}$ denote the submatrix of $\hat{\Theta}$ and $\Theta$ consisting of the rows indexed by $G$.}
\end{theorem}

{Theorem \ref{mis} indicates that the misclustering rate of the motif-based spectral clustering is bounded by $O(\frac{K^3}{n\alpha_np_n})$ with high probability. (\ref{3.3}) is a technical condition that provides the range of parameters $(K,n,\alpha_n,p_n)$ under which the conclusions hold. $S_k$ in (\ref{3.4}) is actually the set of misclustered nodes in the true community $G_k$. The result in (\ref{3.4}) is high-dimensional in that each parameter can vary with the number of nodes $n$. The bound vanishes under several parameter settings. For example, when $\alpha_n\asymp {\rm log}n/{n} $ and $p_n\asymp 1/({\rm log}n)^{{1-\beta}}$ where $\beta\in (0,1)$ and $f(n)\asymp g(n)$ if $c g(n)\leq f(n)\leq C g(n)$ for some constants $0<c<C<\infty$, then $K=o(({\rm log} n)^{\tiny \beta/3})$ ensures a vanishing misclustering bound.}

\subsection{Comparison with edge-based spectral clustering}
To understand how the higher-order spectral clustering enhances the misclustering performance, we compare its misclustering bounds with those from edge-based spectral clustering. Note that edge-based spectral clustering has been studied by \cite{lei2015consistency,rohe2011spectral} under the \emph{unweighted} SBMs. One can easily modify the proofs in \cite{lei2015consistency} to obtain the corresponding results under the WSBMs, hence we omit the details. It is easy to learn from Theorem \ref{mis} that the spectral bound ($\|A-\mathcal A\|_2$) and the minimum non-zero eigenvalue of the population input matrix ($\lambda_K(\mathcal A)$) have the following relationship with the misclustering rate $L$ defined in (\ref{3.2}),
\begin{equation}
\label{3.6}L\leq  c\frac{K\|A-\mathcal A\|_2^2}{\lambda_K(\mathcal A)^2},\tag{3.6}
\end{equation}
where $c$ is some constant, $A$ can be $W$ or $W^M$, and $\mathcal A$ can be $\mathcal W$ or $\mathcal W^M$ correspondingly. Hence, we list these three metrics of the two spectral clustering methods in Table \ref{theotable}.
The bounds in Table \ref{theotable} show that when
\begin{equation}\label{3.7}
\frac{cK^3}{n\alpha_n p_n}\leq  \frac{c''K^3 \mbox{log}n}{n^2\alpha_n^2p_n^2},
\tag{3.7}
\end{equation}
the higher-order spectral clustering has a lower misclustering rate than does the edge-based counterpart. Combining (\ref{3.7}) with the parameter assumption in Theorem \ref{spbound} leads to the following parameter ranges
\[\begin{cases}
\label{3.8}
\alpha_n\leq c\frac{{\rm log}n}{n}/p_n, \\
p_n\geq c'\sqrt{\frac{{\rm log}n}{n}},\\
\alpha_np_n\geq c''\frac{({\rm log}n)^\beta}{n},
\tag{3.8}
\end{cases}\]
which can be all met when, for example, $\alpha_n\asymp {\rm log}n/{n} $ and $p_n\asymp 1/({\rm log}n)^{{1-\beta}}$, where $\beta\in (0,1)$ is any constant that arises in Theorem \ref{spbound}. {The first inequality in (\ref{3.8}) requires that the signal strength is weak if we notice the second inequality. The second inequality says that the network density should be large at a certain level. And the third inequality could be thought of as a requirement of the overall signal strength of weight and network density}. As a result, when the network is dense with a weak signal of weights, the higher-order spectral clustering can be better than its edge-based counterpart in terms of the clustering error. In addition, although we only specify the expectation of conditional weights, we implicitly assume that the variance of weight is larger than 0. In fact, the unweighted SBMs can not lead to the advantage of the higher-order spectral clustering over the edge-based counterpart, {which is consistent with the results in \cite{paul2018higher} (see Section 3.5 in \cite{paul2018higher}).
}

\begin{remark}
It is worth noting that there exists statistical minimax misclustering error rate under SBMs and a general weighted SBMs \citep{gao2017achieving,xu2020optimal}. However, we here focus on enhancing the clustering performance of the spectral clustering by using the weighted motif adjacency matrix. We show that the higher-order spectral clustering leads to a lower misclustering upper bound if the network is dense with a weak expected signal of weights, {which is partially because the motif-based adjacency matrix enlarges the eigen-gap between the smallest non-zero eigenvalue and 0 of the population matrix (see Table \ref{theotable}). Similar observations are obtained in \citet{wang2018network}}, and our empirical results in Section \ref{simulation} also validate the theoretical results. It is left as future work to compare these two spectral clustering methods more delicately, for example, deriving tighter bounds or studying the strong consistency of both methods.
\end{remark}

\begin{table}[H]
\centering
\footnotesize
\caption{A summary of the main results of two spectral clustering methods under the WSBMs.
}\vspace{0.1cm}
\def\arraystretch{2.3}
\setlength{\tabcolsep}{0.04cm}
\begin{tabular}{lcc}
\hline
\textbf{Bounds}&\textbf{Edge-based }&\textbf{Motif-based}\\
\hline
Spectral bound&$O_p(\sqrt{\mbox{max}\{n\alpha_n p_n,c\mbox{log}n\}})$&$O_p(\sqrt{n^{3}\alpha_np_{n}^5})$\\
$\lambda_K$&$O(\frac{n\alpha_np_n}{K})$&$O(\frac{n^2\alpha_np_n^3}{K})$\\
misclustering rate&$O_p(\mbox{max}\{\frac{K^3}{n\alpha_n p_n},\frac{K^3 {\rm log}n}{n^2\alpha_n^2p_n^2}\})$&$O_p(\frac{K^3}{n\alpha_np_n})$\\
\hline
\end{tabular}
\label{theotable}
\end{table}

\section{{Discussions}}
\label{discussion}
In this section, we discuss our work from methodology to theory. Possible extensions are also posed.

\paragraph{{\textbf{Methodology.}}} Conceptually, although we focused on the spectral clustering, the notion of motif adjacency matrix could be used as the input of many community detection methods. From this point of view, it can be regarded as a data preprocessing technique. \cite{qin2013regularized} showed that the regularized graph Laplacian in which the degree matrix is regularized with a small constant can lead to better clustering results than the original spectral clustering. {Similarly, using the motif adjacency matrix instead of the adjacency matrix can be thought of as one network denoising technique}.

On the other hand, driven by the evidence that network edges often contain sensitive information, there is a growing body of works on the privacy-preserving analysis of networks; see \citet{karwa2016inference,karwa2017sharing}, among others. From this perspective, the entries of the motif adjacency matrix can be regarded as the summary statistics, and thus it can help protect the original edges to some extent.

Computationally, there is no doubt that counting the number of triangles for each pair of nodes is costly on large-scale networks. One could use sampling techniques
to improve the computational performance \citep{seshadhri2014wedge,benson2016higher,chen2018efficient,guo2020randomized,zhang2022randomized}.

\paragraph{\textbf{Theory.}}
Theorem \ref{spbound} and Theorem \ref{mis} are our main results. Theorem \ref{spbound} investigates the deviation of the weighted motif adjacency matrix from its population version in the sense of the spectral norm. We generally use the $\epsilon$-net technique to make a discretization, and then use a combinatorial method to bound the approximation error. This technique was first developed by \cite{feige2005spectral}. Since that time it has been widely used by statistics and machine learning communities; see \cite{lei2015consistency,paul2018higher,gao2017achieving,chin2015stochastic}, among many others. Note that the entries of $W^M$ are dependent which creates difficulty in deriving the bound. \textcolor{black}{To tackle this issue, we use similar arguments developed in \cite{paul2018higher}.} In particular, we make use of the \emph{typical bounded difference inequality} established in \cite{warnke2016method} to handle the dependency. Note that as the network edges in our model are not Bernoulli distributed and the resulting weighted motif adjacency matrix is not identical to theirs, the details of our proof are slightly different from \cite{paul2018higher}. It remains unclear whether the bound in (\ref{3.1}) could be improved by using other techniques. Theorem \ref{mis} bounds the misclustering rate of the motif-based spectral clustering.
The general idea is to use the Davis-Kahan theorem \citep{davis1970rotation} to bound the perturbation of eigenvectors from the approximation error bound (\ref{3.1}). Such framework is widely used in \cite{lei2015consistency,rohe2011spectral}, among others.

\paragraph{\textbf{Extensions.}} {We mainly study the full rank WSBMs, that is, the rank of $\mathcal W$ equals the target community number $K$. Actually, all the results could be generalized to rank-deficient WSBMs (i.e., ${\rm rank}(\mathcal W)\leq K$) by investigating the population eigen-structure of such WSBMs and adding extra conditions on $B$, just as the argument in \cite{zhang2022randomized}. In addition, though we mainly deal with the clustering performance, we can estimate the connectivity matrix $B=(B_{ql})_{q,l=1,...,K}$ via the following simple plug-in estimator,
\begin{equation*}
\hat{B}_{ql}:=\frac{\sum_{1\leq i,j\leq n}W_{ij}^M\hat{\Theta}_{iq}\hat{\Theta}_{jl}}{\sum_{1\leq i,j\leq n}\hat{\Theta}_{iq}\hat{\Theta}_{jl}},\quad 1\leq q,l\leq K.
\end{equation*}
Moreover, we could evaluate its theoretical performance by using Theorem \ref{spbound} and \ref{mis}. Following the advantage of motif-based spectral clustering in terms of clustering, we can imagine that the motif-based method would lead to better estimates of $B$ than the edge-based method does. For the sake of space, we leave all these extensions as our future work.}

\section{Simulation studies}
\label{simulation}
In this section, we empirically compare the finite sample performance of the higher-order spectral clustering, namely, the motif-based spectral clustering (denoted by \emph{motif-based}) with that of the edge-based spectral clustering (denoted by \emph{edge-based}) under the WSBMs. Note that we use higher-order spectral clustering and motif-based spectral clustering interchangeably in this section. \textcolor{black}{To that end, we first carry out a sensitivity analysis to evaluate our theoretical findings. Then, we provide several extended experiments to show that high-order spectral clustering has an advantage under a broad regime even when some theoretical assumptions are violated.}

First, we evaluate the performance of two methods and validate the theoretical results. To be consistent with Section \ref{theory}, we use the following three metrics to assess the performance of each method. The first is the spectral deviation of the weighted motif adjacency matrix $W^M$ from its population version $\mathcal W^M$, denoted by \emph{spectral bound}. The second is the minimum non-zero eigenvalue of the population $\mathcal W^M$, denoted by \emph{eigen gap}. The third is the summation of the ratio of misclustered nodes within each true community (see (\ref{3.2})), denoted by \emph{misclustering rate}. We study the effect of sample size $n$, the effect of maximum link probability $p$, the effect of out-in-ratio (the ratio of the between-community probability $q$ over the within-community probability $p$), and the effect of the number of communities $K$ respectively with the following experimental set-ups,
\begin{itemize}
  \item \emph{{Effect of $n$:}} $K=2,p=0.5,q=p(1-0.4)$, and $n$ varies;
  \item \emph{{Effect of $p$:}} $K=2,n=60,q=p(1-0.4)$, and $p$ varies;
  \item \emph{{Effect of out-in-ratio:}} $K=2,n=60,p=0.5$, and $\frac{q}{p}$ varies;
  \item \emph{{Effect of $K$:}} $n=120,p=0.5,q=p(1-0.4)$, and $K$ varies;
\end{itemize}
where the weights are all \emph{i.i.d.} generated from uniform distribution ${\rm Uniform}(0.01,1)$ provided that there is an edge.

\begin{figure*}[!htbp]{}
\centering
\subfigure[Spectral bound]{\includegraphics[height=4.8cm,width=5cm,angle=0]{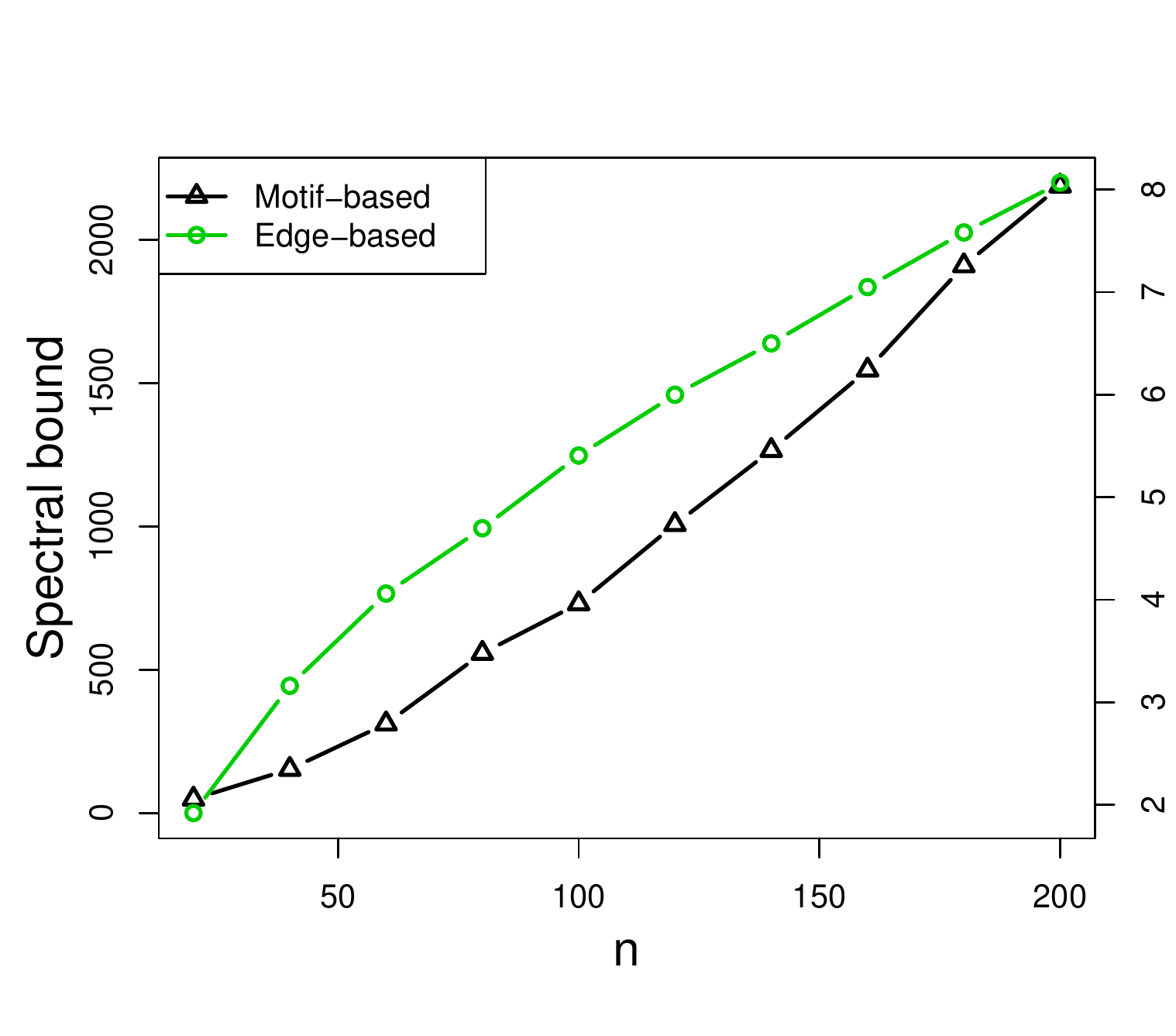}}
\subfigure[Eigen gap]{\includegraphics[height=4.8cm,width=5cm,angle=0]{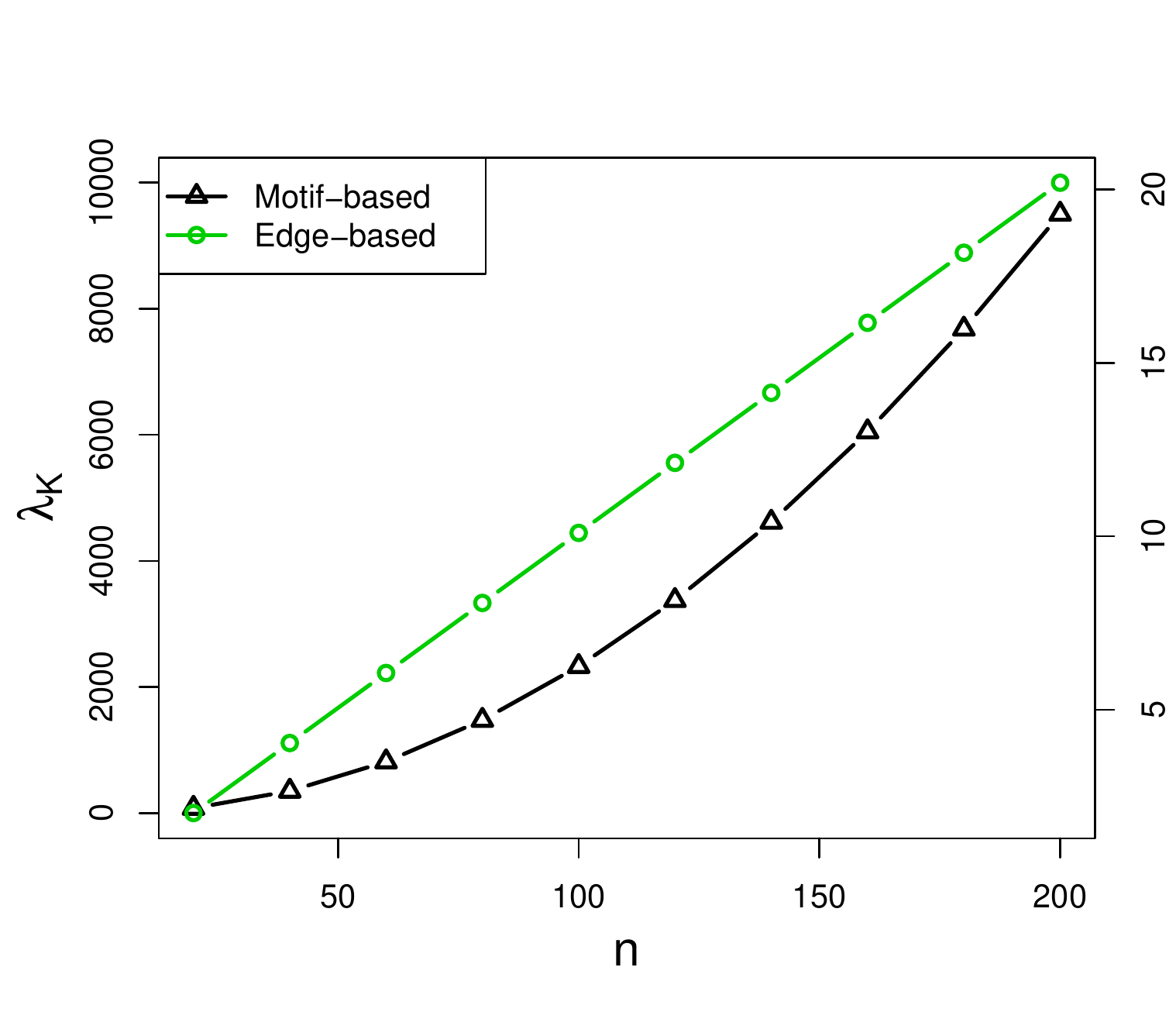}}
\subfigure[Misclustering rate]{\includegraphics[height=4.8cm,width=5cm,angle=0]{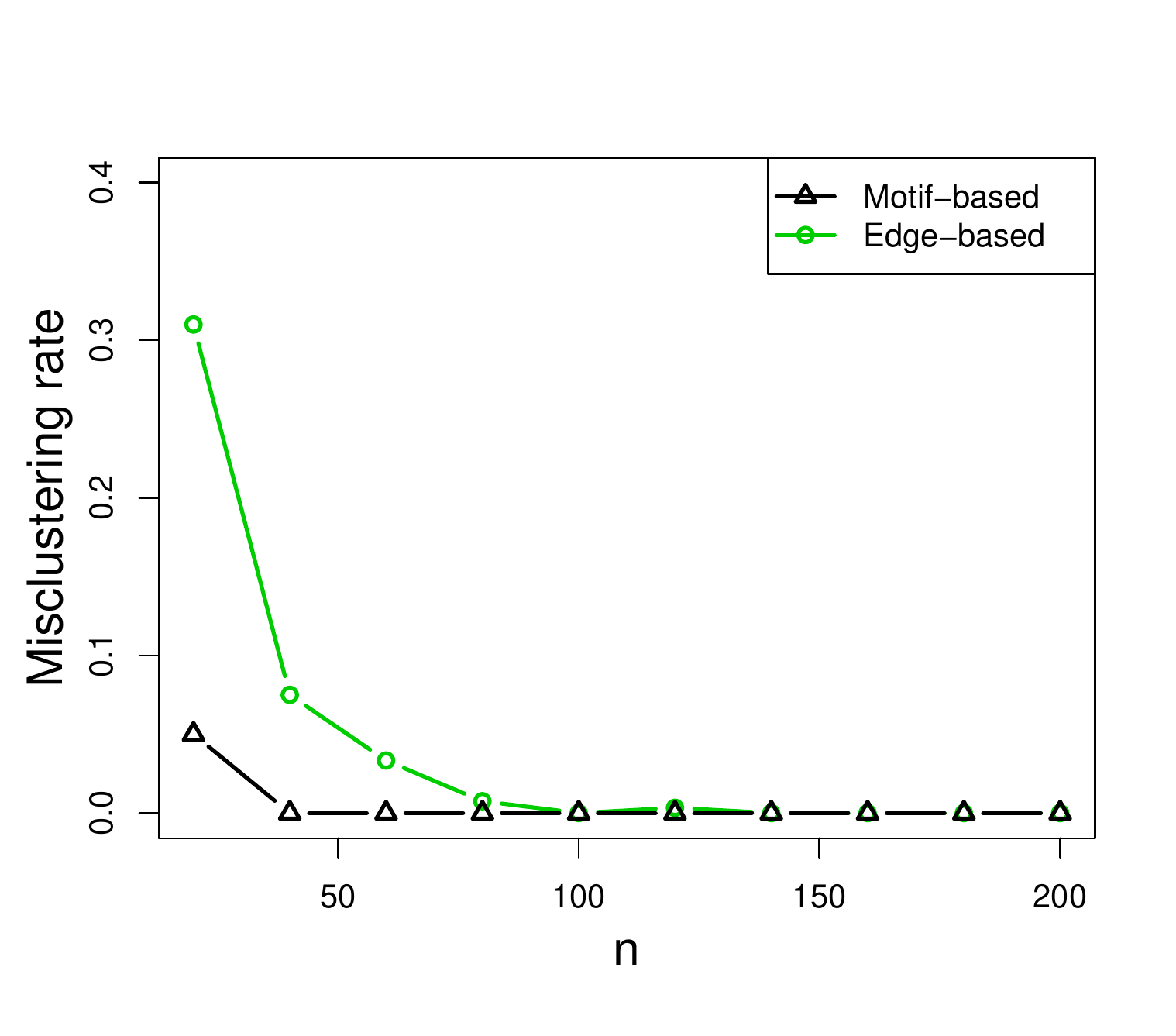}}
\caption{The average effect of $n$ on the three metrics over 100 replications. Other parameters $K=2,p=0.5,q=p(1-0.4)$, and the edge weights are \emph{i.i.d.} generated from ${\rm Uniform}(0.01,1)$ provided that the edges exist. In (a) and (b), the left and right axes correspond to motif-based and edge-based spectral clustering, respectively.}\label{effectofn}
\end{figure*}

\begin{figure*}[!htbp]{}
\centering
\subfigure[Spectral bound]{\includegraphics[height=4.8cm,width=5cm,angle=0]{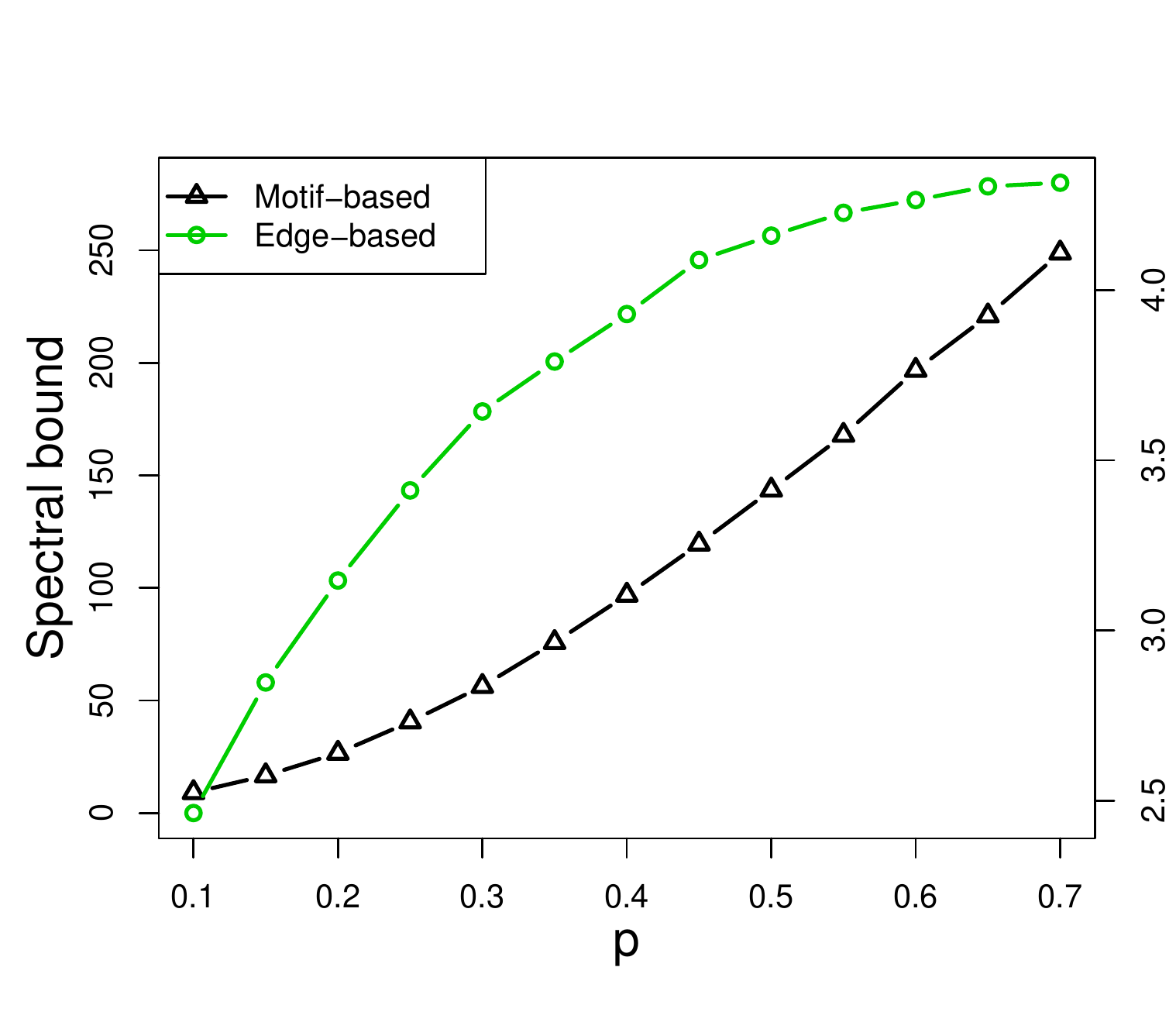}}
\subfigure[Eigen gap]{\includegraphics[height=4.8cm,width=5cm,angle=0]{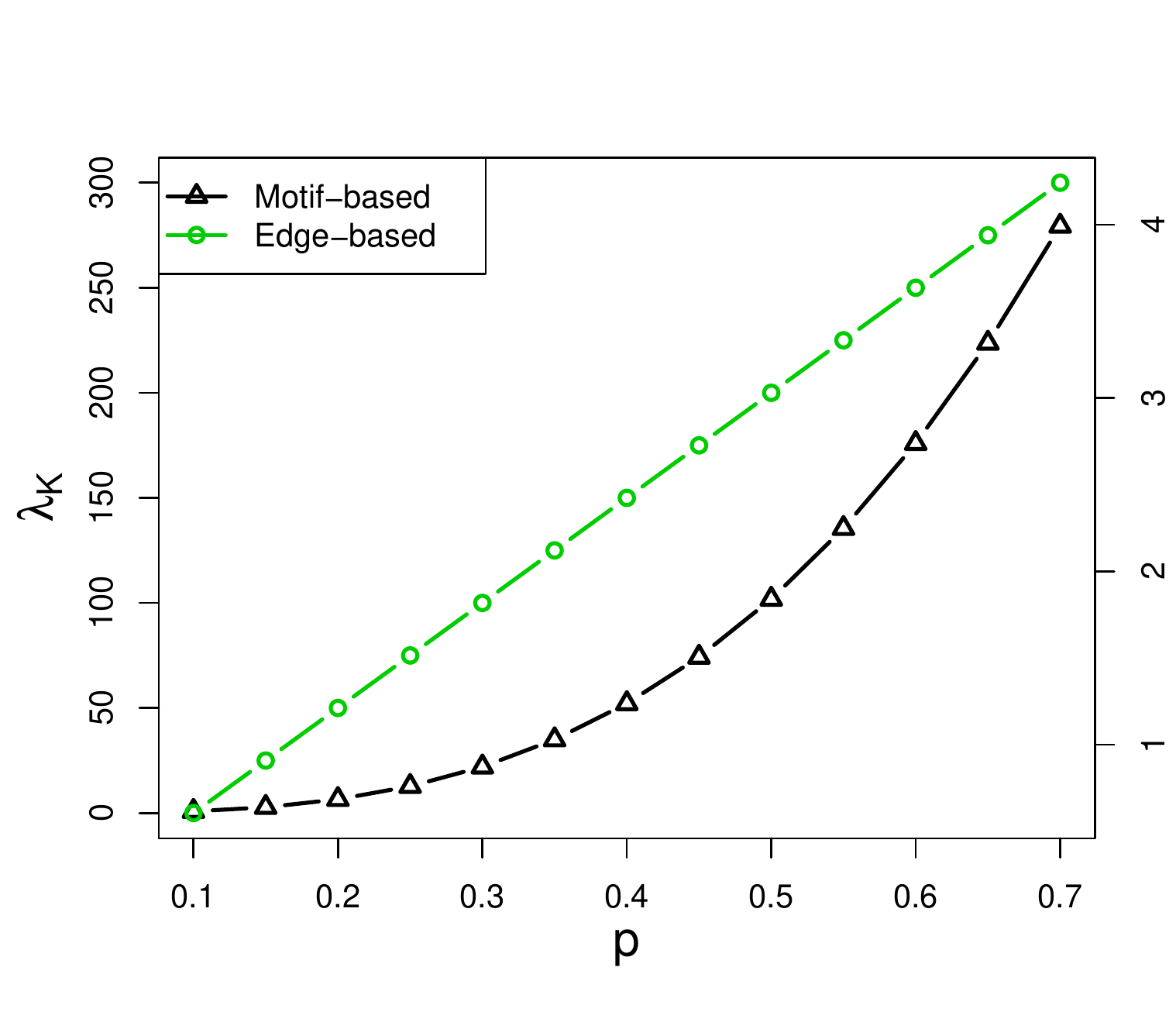}}
\subfigure[Misclustering rate]{\includegraphics[height=4.8cm,width=5cm,angle=0]{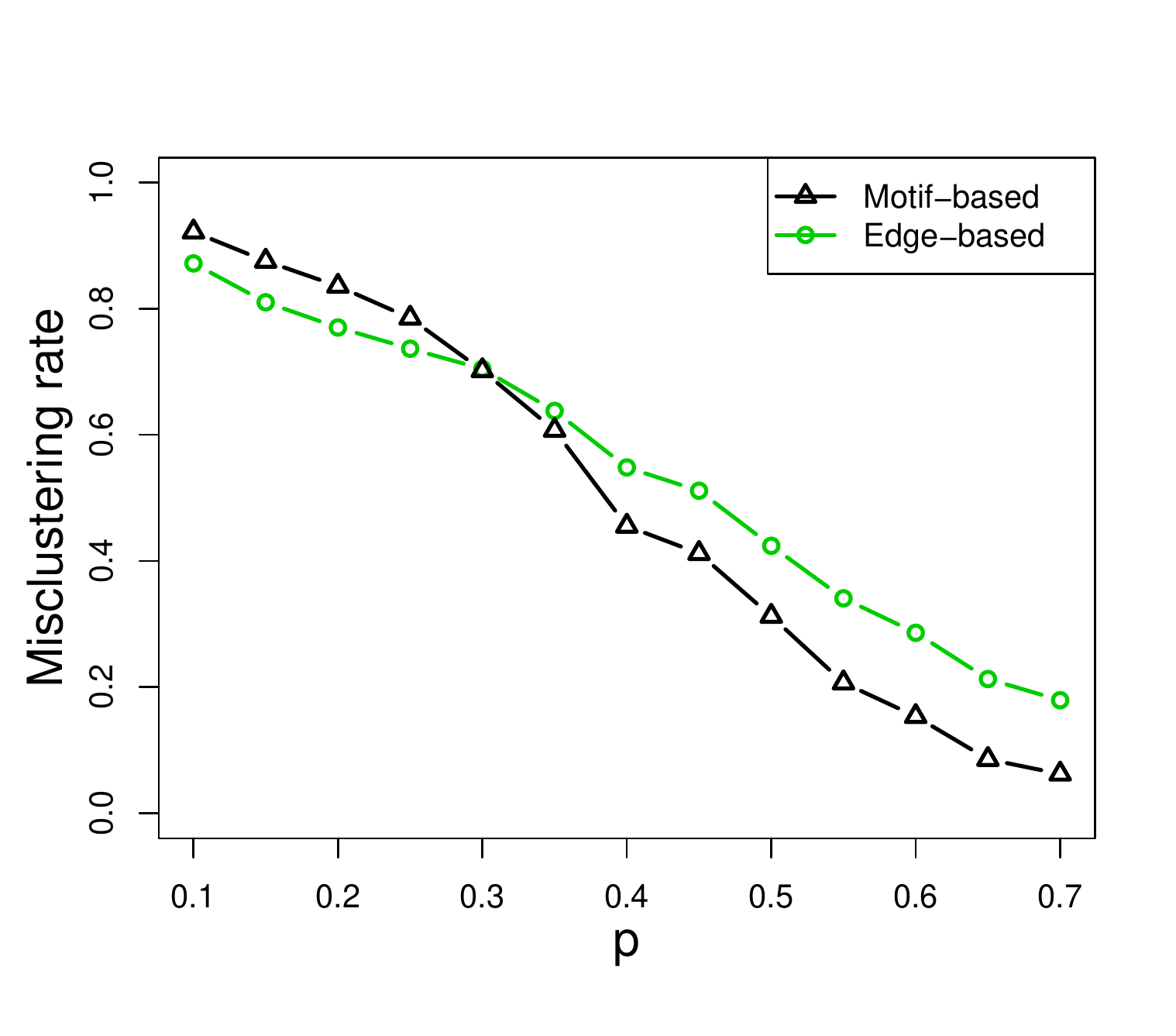}}
\caption{The average effect of $p$ on the three metrics over 100 replications. Other parameters $K=2,n=60,q=p(1-0.4)$, and the edge weights are \emph{i.i.d.} generated from ${\rm Uniform}(0.01,1)$ provided that the edges exist. In (a) and (b), the left and right axes correspond to motif-based and edge-based spectral clustering, respectively. }\label{effectofp}
\end{figure*}

\begin{figure*}[!htbp]{}
\centering
\subfigure[Spectral bound]{\includegraphics[height=4.8cm,width=5cm,angle=0]{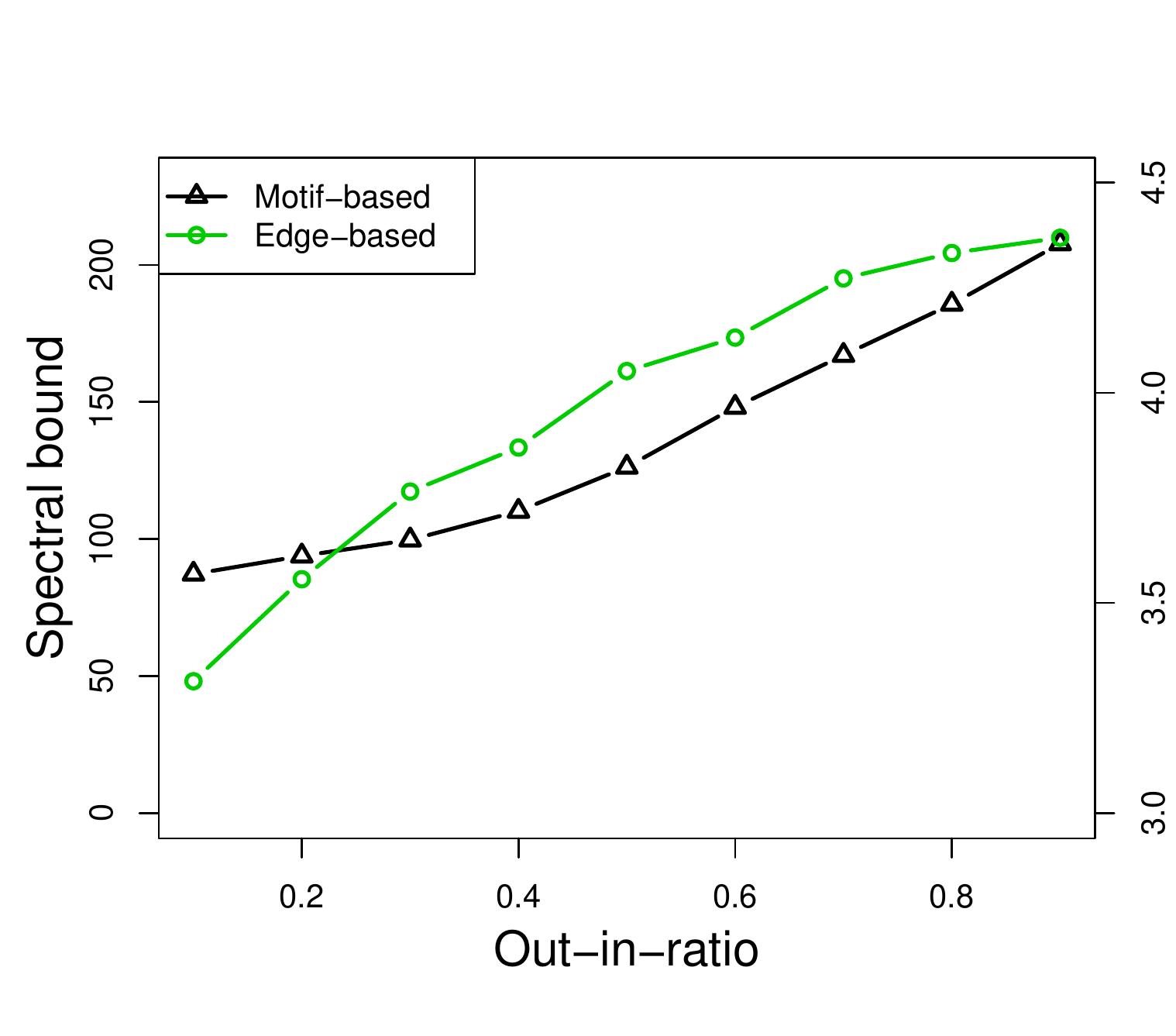}}
\subfigure[Eigen gap]{\includegraphics[height=4.8cm,width=5cm,angle=0]{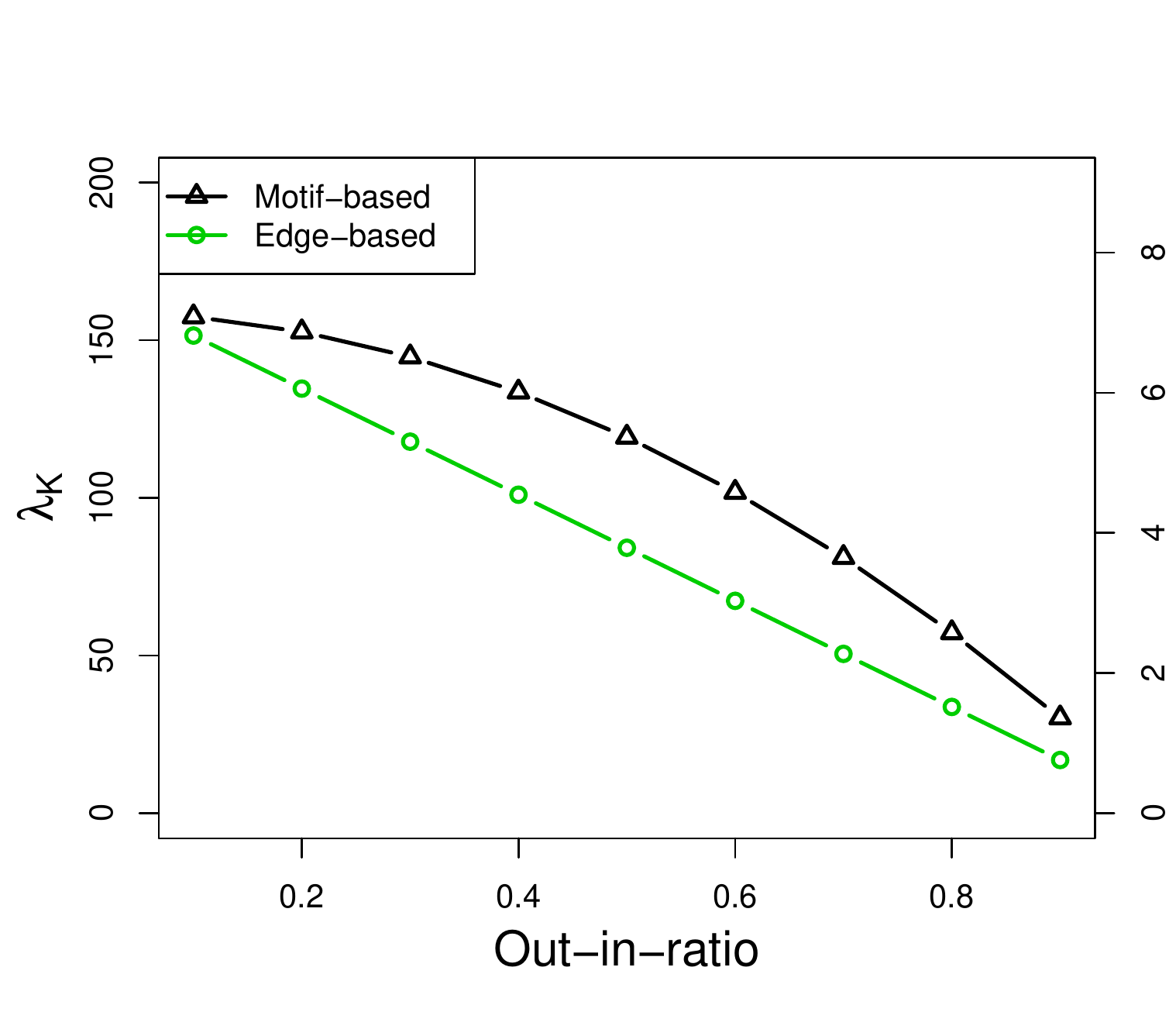}}
\subfigure[Misclustering rate]{\includegraphics[height=4.8cm,width=5cm,angle=0]{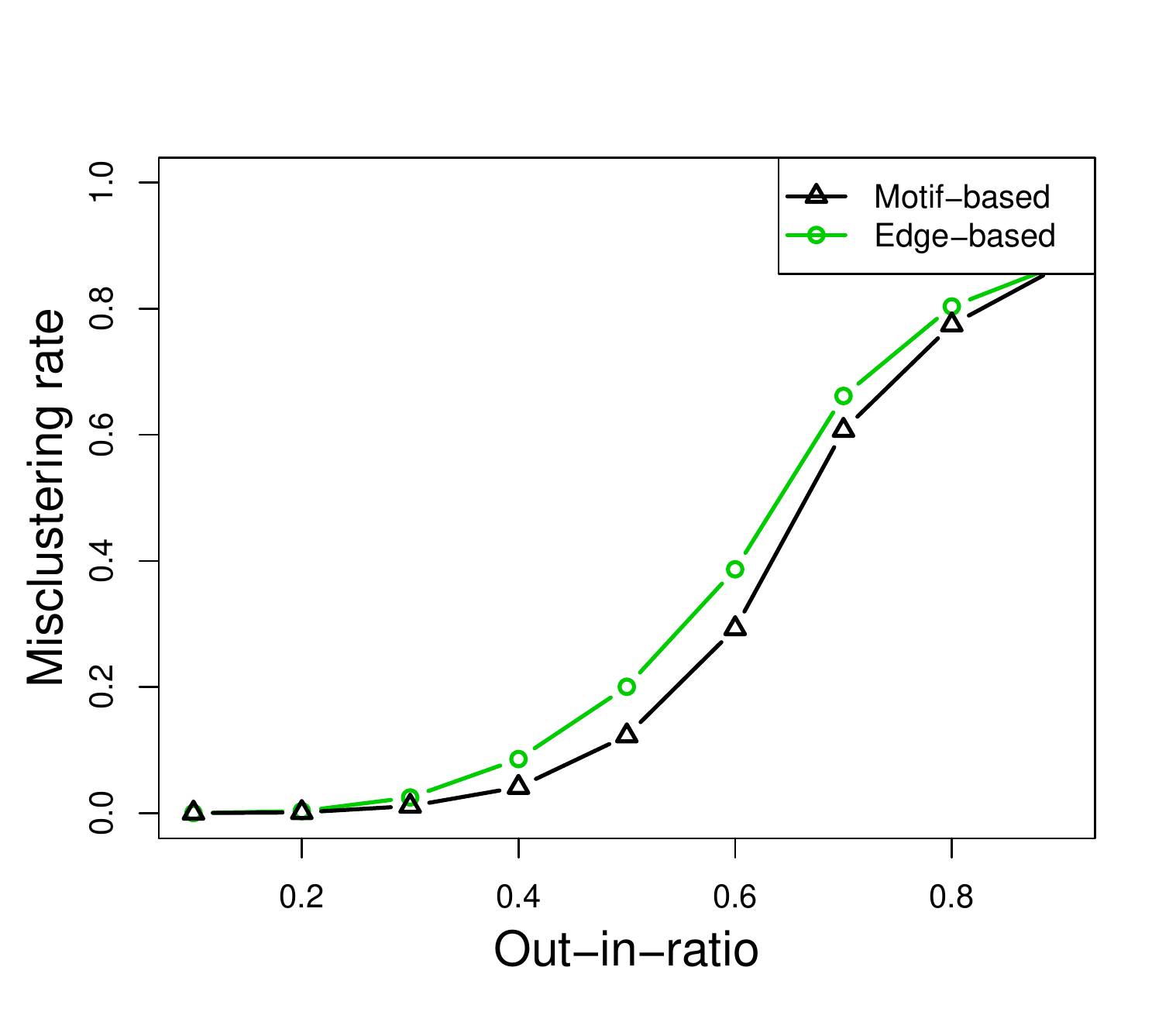}}
\caption{The average effect of out-in-ratio on the three metrics over 100 replications. Other parameters $K=2,n=60,p=0.5$, and the edge weights are \emph{i.i.d.} generated from ${\rm Uniform}(0.01,1)$ provided that the edges exist. In (a) and (b), the left and right axes correspond to motif-based and edge-based spectral clustering, respectively. }\label{effectofout}
\end{figure*}

\begin{figure*}[!htbp]{}
\centering
\subfigure[Spectral bound]{\includegraphics[height=4.8cm,width=5cm,angle=0]{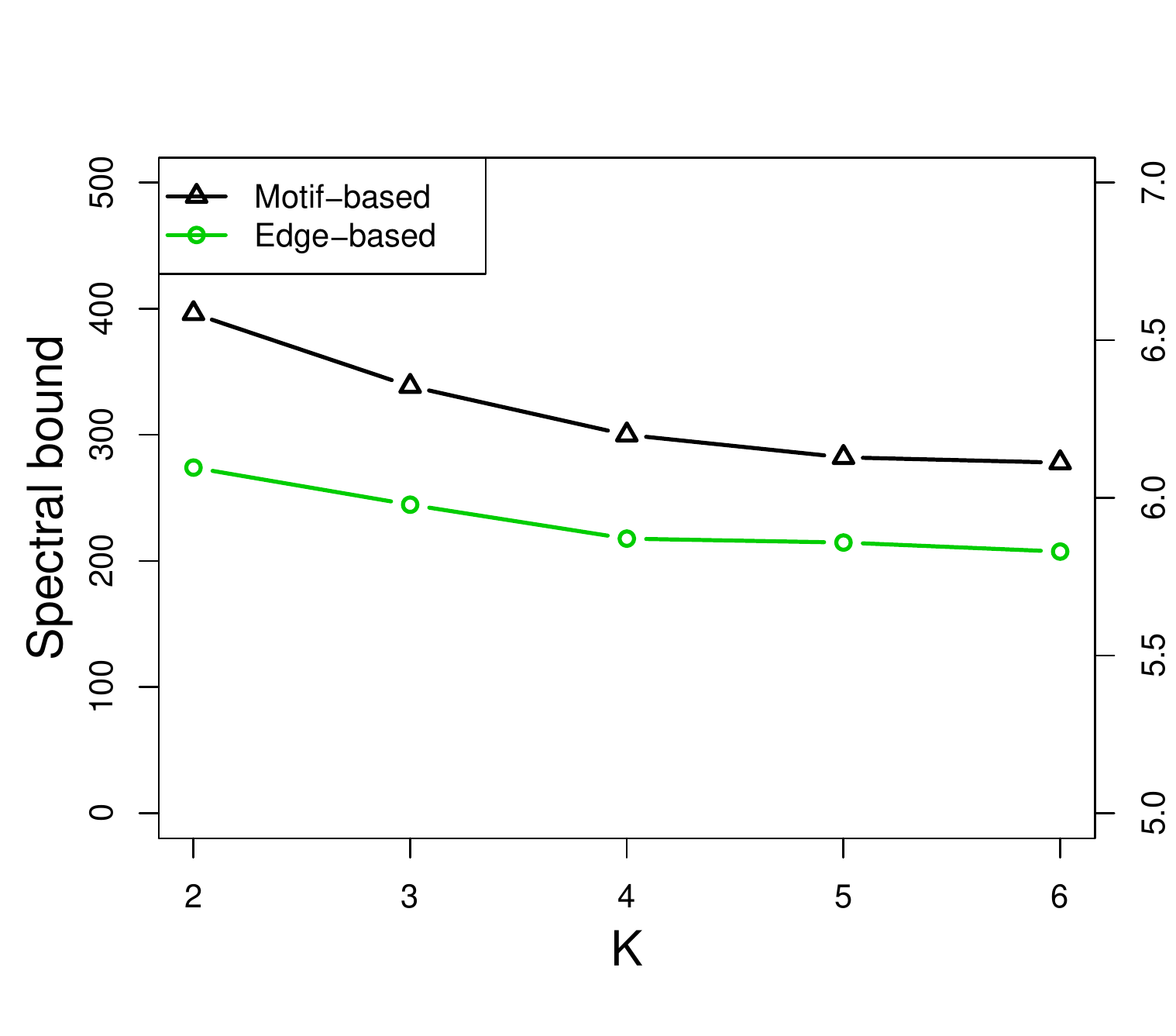}}
\subfigure[Eigen gap]{\includegraphics[height=4.8cm,width=5cm,angle=0]{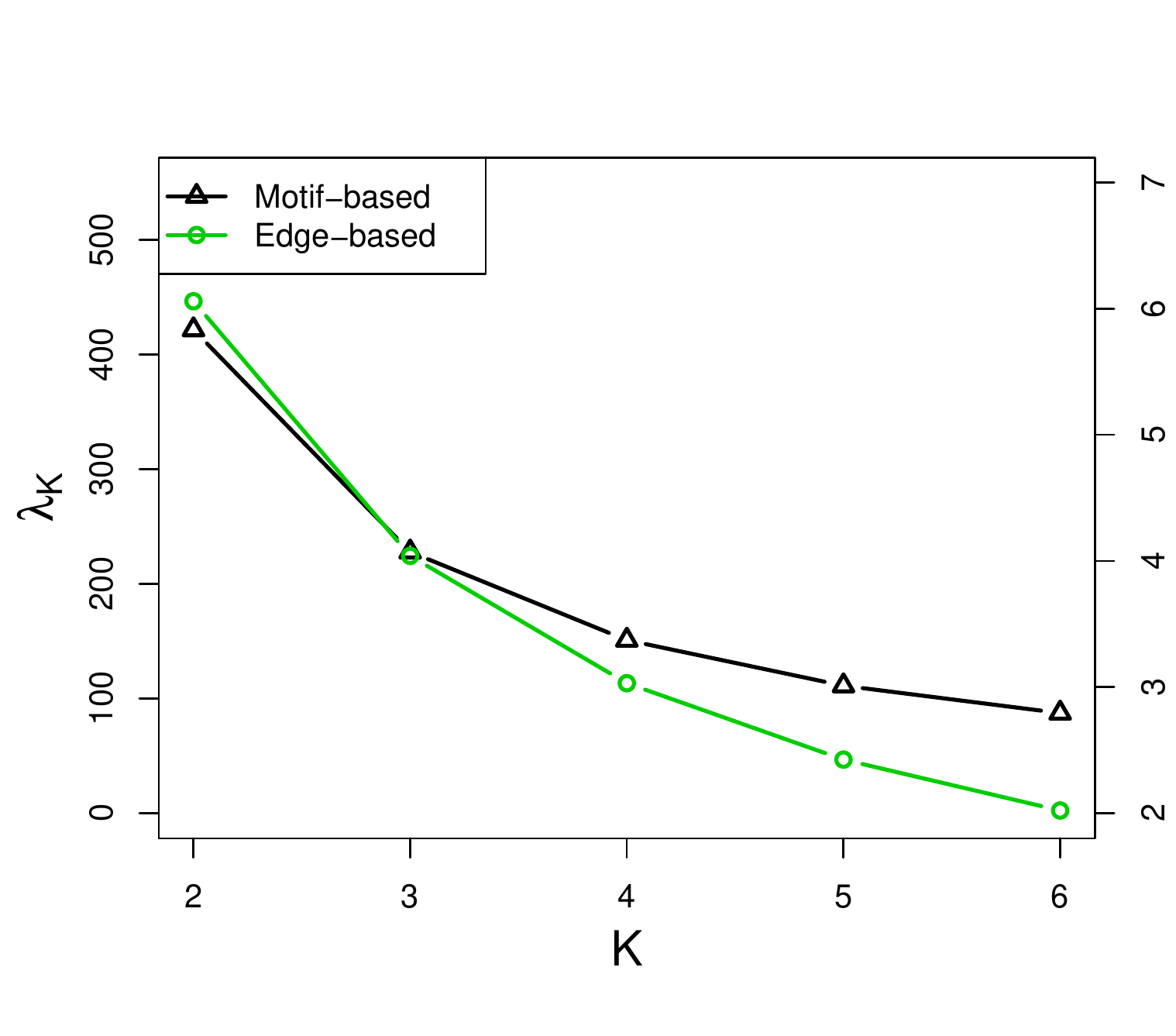}}
\subfigure[Misclustering rate]{\includegraphics[height=4.8cm,width=5cm,angle=0]{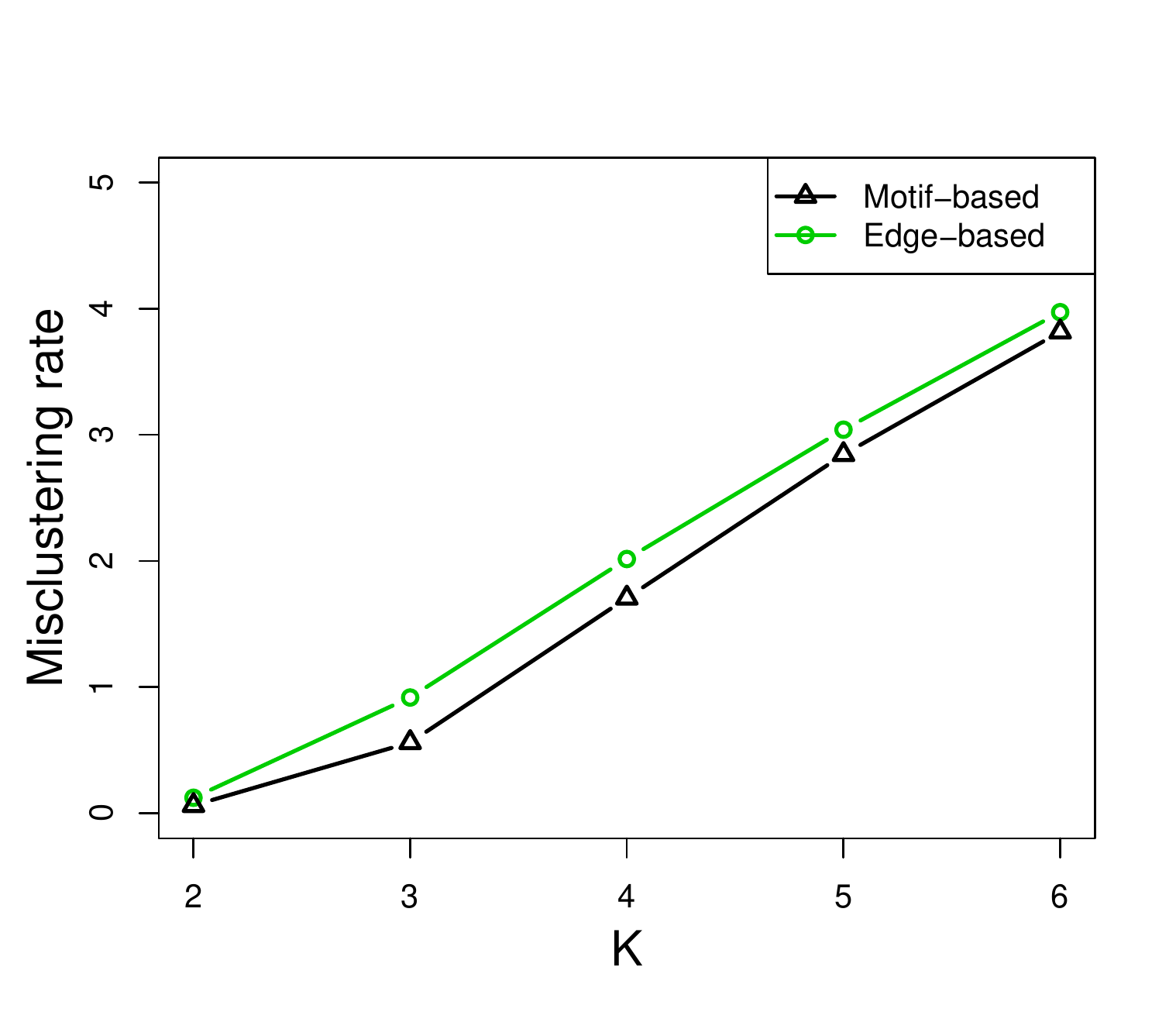}}
\caption{The average effect of $K$ on the three metrics over 100 replications. Other parameters $n=120,p=0.5,q=p(1-0.4)$, and the edge weights are \emph{i.i.d.} generated from ${\rm Uniform}(0.01,1)$ provided that the edges exist. In (a) and (b), the left and right axes correspond to motif-based and edge-based spectral clustering, respectively. }\label{effectofk}
\end{figure*}

Figures \ref{effectofn}-\ref{effectofk} show the results for the above four experimental set-ups. As indicated in Section \ref{theory}, the spectral bound and the eigen-gap with respect to the two spectral clustering methods have different scales. Hence we use two axes to show their tendency. Specifically, the left and right axes correspond to motif-based and edge-based spectral clustering, respectively. The numerical results show satisfactory consistency with the theoretical results. For the spectral bound, the motif-based spectral clustering grows super-linearly with $n$ and $p$, while the edge-based spectral clustering grows sub-linearly with $n$ and $p$ (see Figures \ref{effectofn}(a) and \ref{effectofp}(a)). Note that we have not taken the community structure into consideration when bounding the spectral error. While as indicated in Figures \ref{effectofout}(a) and \ref{effectofk}(a), the spectral bounds grow with the increasing of out-in-ratio and drop with the increasing number of communities. Hence, it would be beneficial to incorporate this information when bounding the spectral error. We leave this for our future work. For the eigen-gap, as $n$ and $p$ increase, the motif-based spectral clustering grows faster than the edge-based spectral clustering does, where the latter has linear growth with $n$ and $p$ (see Figures \ref{effectofn}(b) and \ref{effectofp}(b)). In addition, the eigen-gap decreases with the growth of $K$ for both methods. As for the misclustering rate, we can see from Figures \ref{effectofn}(c), \ref{effectofout}(c), and \ref{effectofk}(c) that when $n$, out-in-ratio, and $K$ are intermediate, the motif-based spectral clustering has great advantage over the edge-based method in clustering. When these terms are too small or too large, the signal for the communities is weak such that neither method can recover the communities, or the signal is strong such that the edge-based clustering can do well. In particular, we see from Figure \ref{effectofp}(c) that when the network is sparse, the edge-based spectral clustering is better than the motif-based method. As the maximum linking probability $p$ increases, the motif-based spectral clustering performs gradually better than the edge-based method does. This is consistent with our theoretical findings.

\textcolor{black}{Next, we provide additional experiments to show that motif-based spectral clustering has the advantage in wide scenarios where the theoretical requirements are not necessarily met. We use Adjusted Rand Index (ARI) \citep{manning2010introduction}, Normalized Mutual Information (NMI) \citep{manning2010introduction}, and Modularity \citep{newman2006modularity} to justify the clustering performance of the two methods, where larger scores indicate better clustering performance. In particular, ARI and NMI are computed by comparing the estimated communities of each of the methods with the true underlying communities in SBMs, respectively. Modularity is computed based on the \emph{weighted motif adjacency matrix}, which measures the difference between the strength of edges (i.e., the weighted edges in the weighted motif adjacency matrix) between any two nodes in the same community and the expected strength of edges between them. As explained in Remark \ref{intuition}, we expect that motif-based spectral clustering would lead to larger modularity. In particular, we conduct the following experiments.}

\begin{figure*}[!htbp]{}
\centering
\subfigure[ARI]{\includegraphics[height=4.8cm,width=5cm,angle=0]{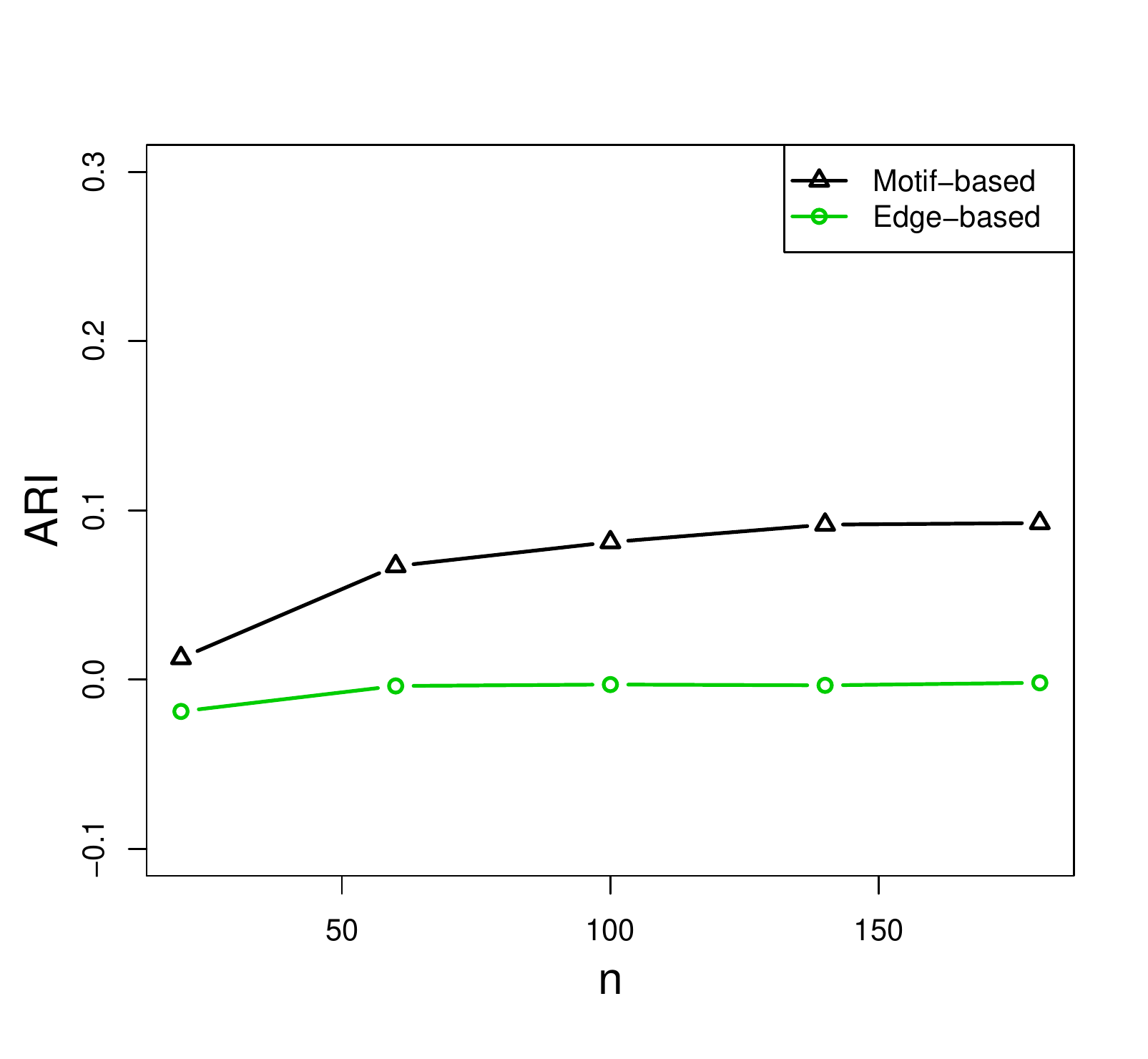}}
\subfigure[NMI]{\includegraphics[height=4.8cm,width=5cm,angle=0]{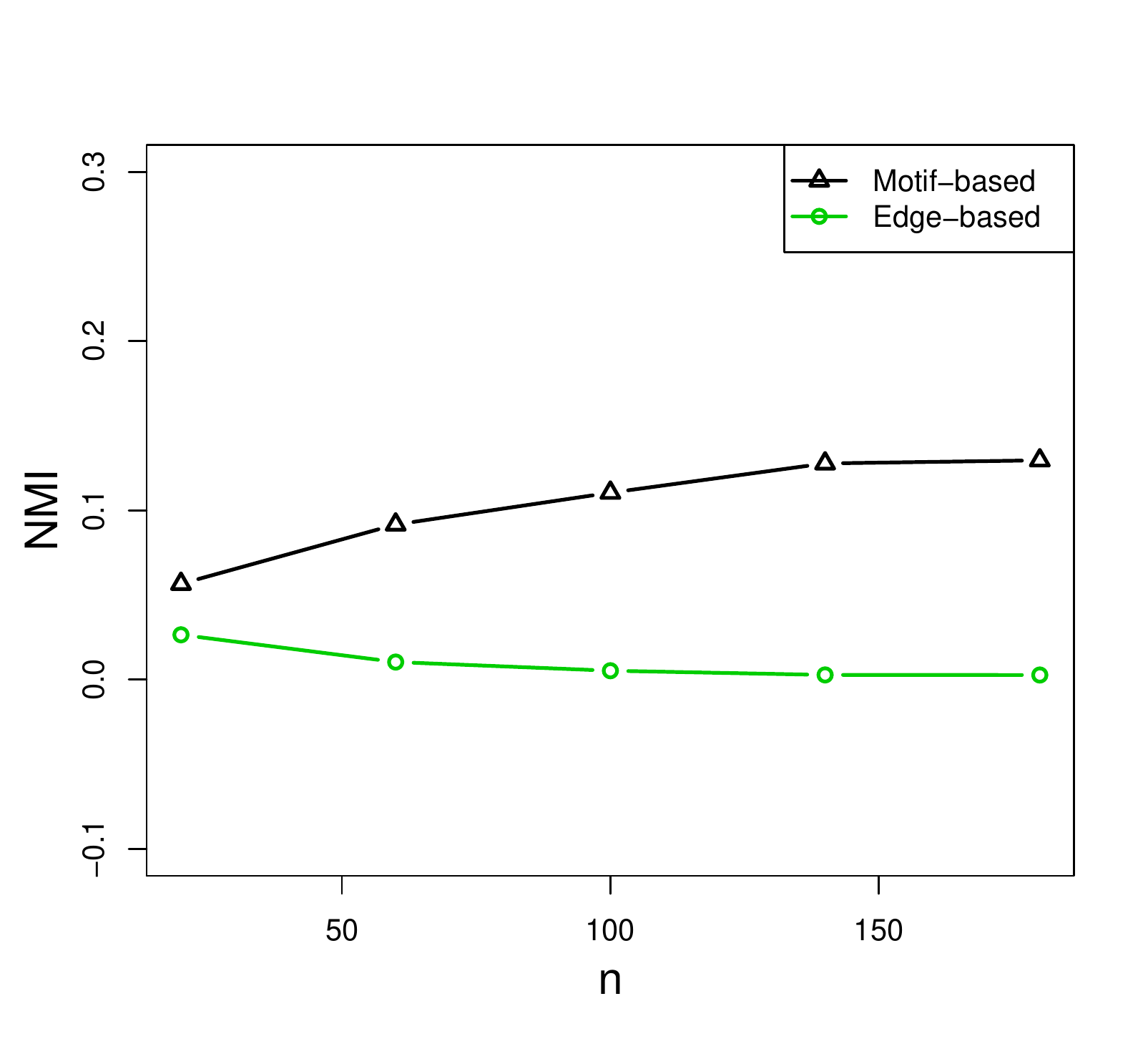}}
\subfigure[Modularity]{\includegraphics[height=4.8cm,width=5cm,angle=0]{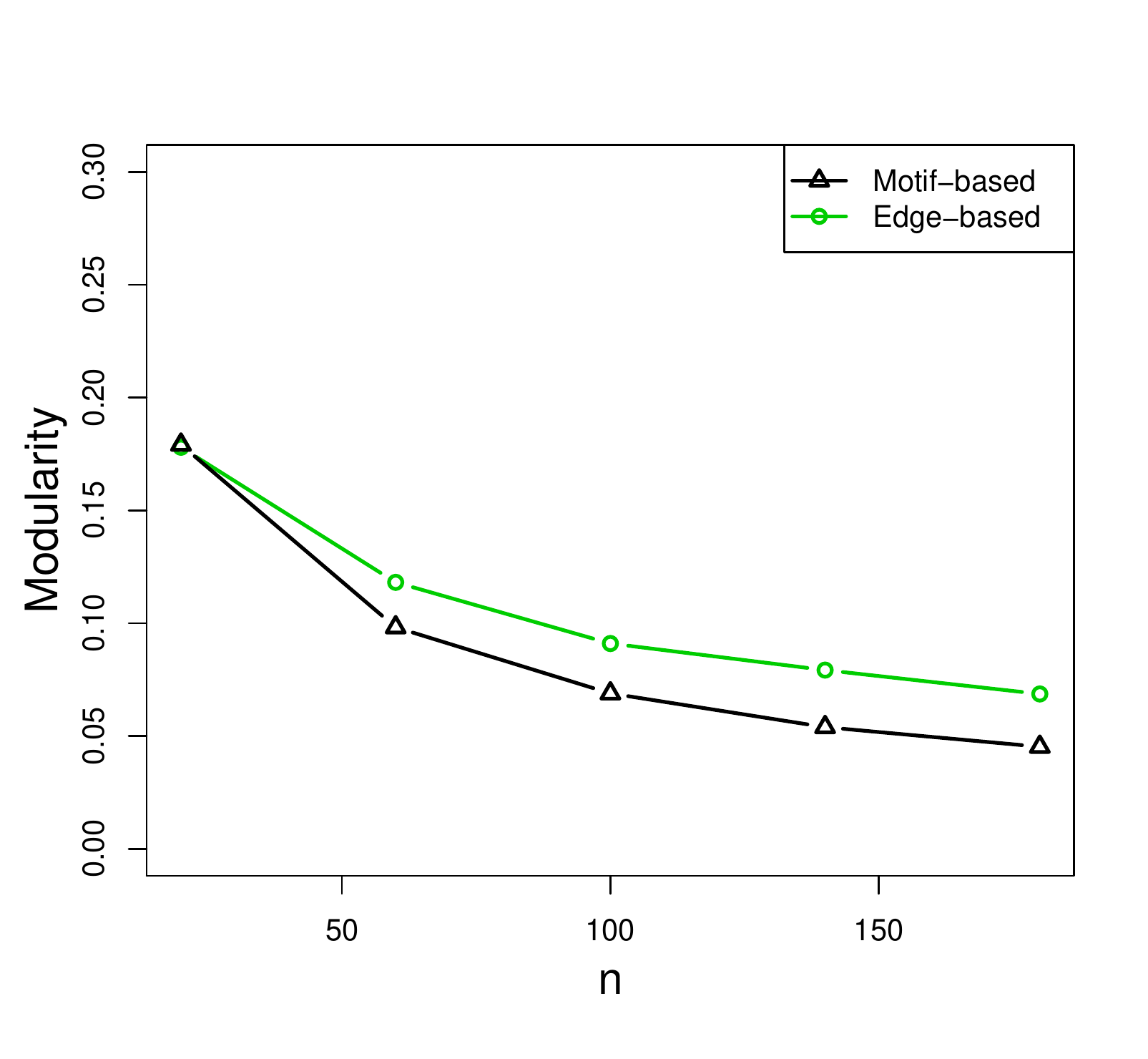}}\\
\caption{The average performance of two methods on networks with core-periphery structure over 100 replications. The parameters are $B:=[0.5,0.3;0.3; 0.05]$, $K=2, n=60$. The weights are \emph{i.i.d.} generated from ${\rm Uniform}(0.5,0.6)$ provided that the edges exist.}\label{effectofstructure}
\end{figure*}

\textcolor{black}{\paragraph{Core-periphery structure.} The previous experiments focused on recovering the \emph{affinity} structure in networks. That is, the two communities have relatively high within-community link probability compared to the between-community link probability. We here assume that the network exhibit \emph{core-periphery} structure, that is, one of the two communities has a relatively high within-community link probability compared to both the other cluster's within-community link probability and the between-community link probability \citep{priebe2019two}. Specifically, we assume the connectivity matrix $B:=[0.5,0.3;0.3; 0.05]$ and let the sample size $n$ vary. Figure \ref{effectofstructure} shows that in our set-up, the motif-based method has the advantage over the edge-based counterpart in terms of all three metrics, which corroborates the ability of the motif-based method for finding core-periphery structure \citep{benson2016higher}.}

\begin{figure*}[!htbp]{}
\centering
\subfigure[ARI]{\includegraphics[height=4.8cm,width=5cm,angle=0]{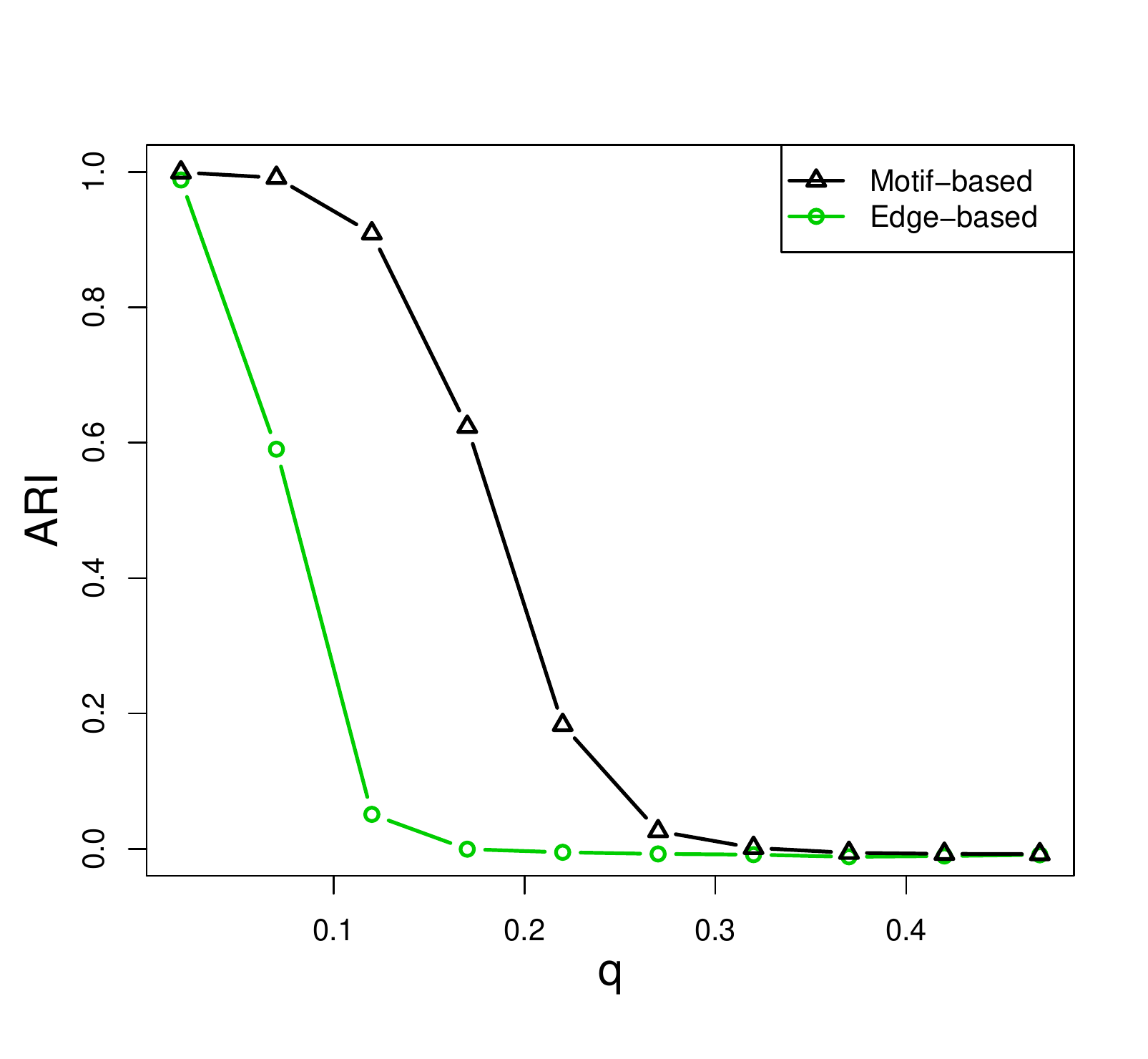}}
\subfigure[NMI]{\includegraphics[height=4.8cm,width=5cm,angle=0]{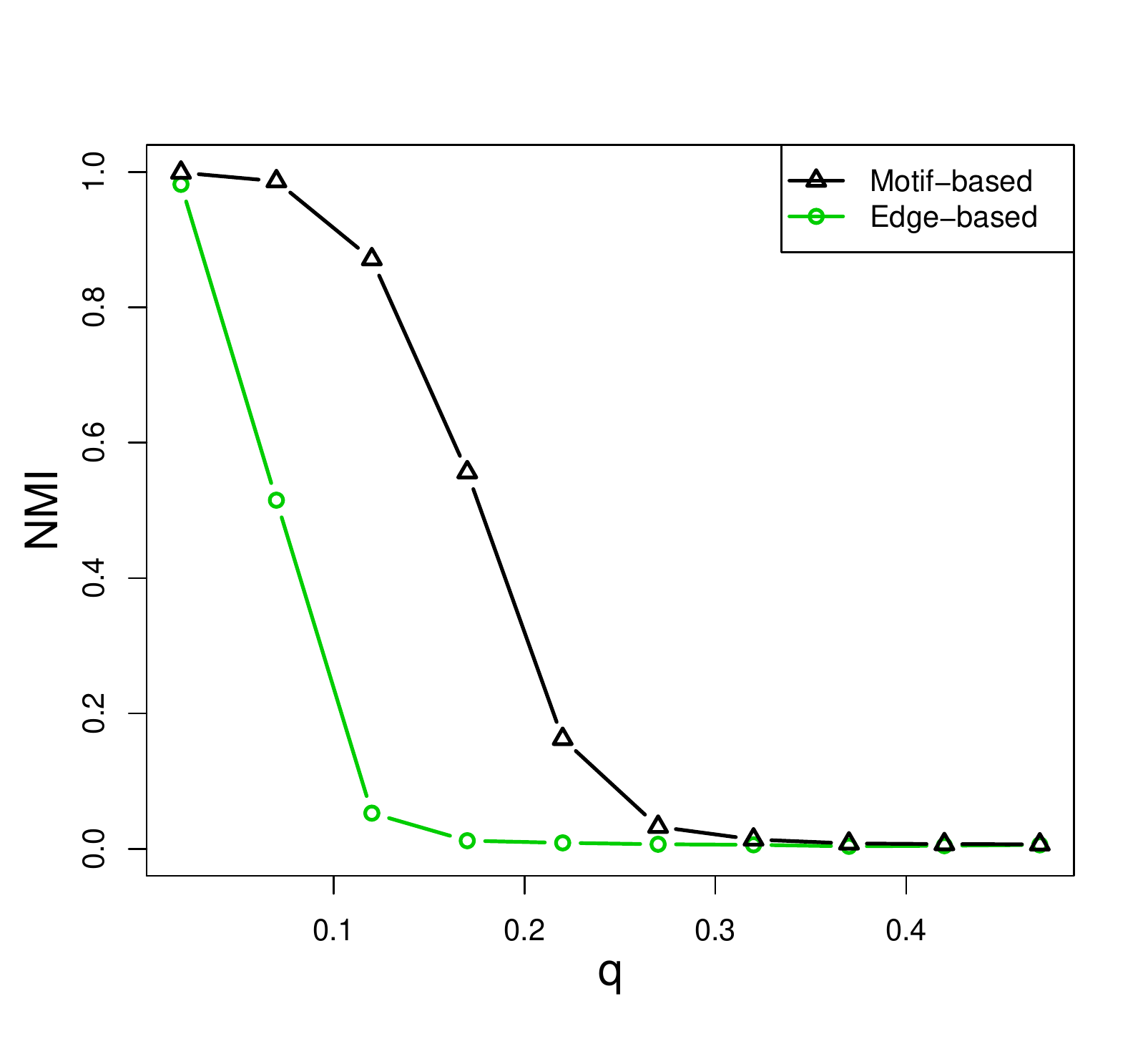}}
\subfigure[Modularity]{\includegraphics[height=4.8cm,width=5cm,angle=0]{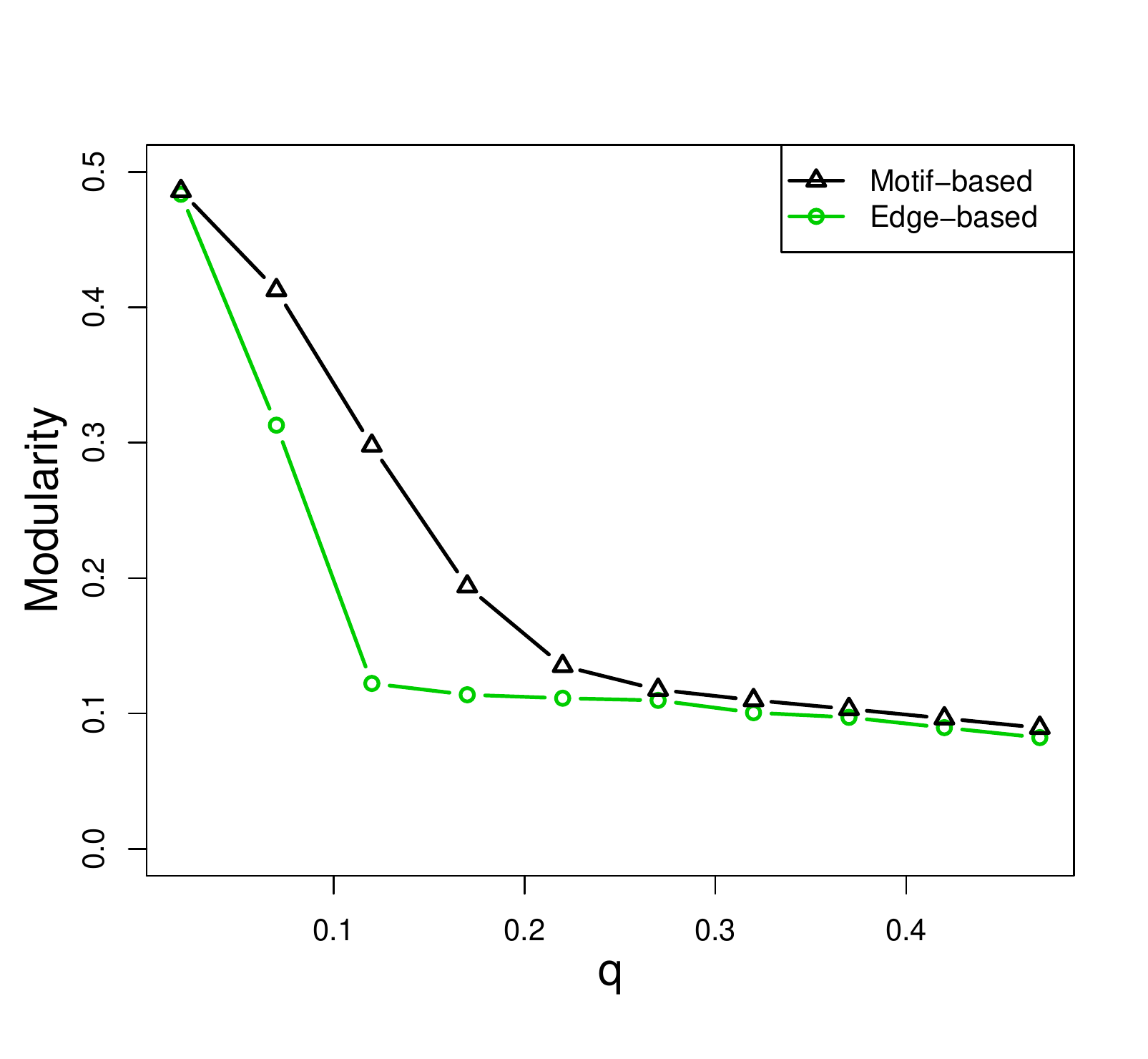}}\\
\caption{The average performance of two methods on networks with disassortative weights over 100 replications. The parameters are $K=2, n=60, p=0.5$. The between-community weights are \emph{i.i.d.} ${\rm Uniform}(0.5,1)$, and the within-community weights are \emph{i.i.d.} ${\rm Uniform}(0.01,0.5)$.}\label{effectofweights}
\end{figure*}

\textcolor{black}{\paragraph{Disassortative weights.} In theoretical analysis, we assumed that all the weights are \emph{i.i.d.} regardless of the communities. We assume that the weights are \emph{disassortative} in that the within-community edges have less weight than the between-community edges. In particular, the between-community weights are \emph{i.i.d.} ${\rm Uniform}(0.5,1)$, and the within-community weights are \emph{i.i.d.} ${\rm Uniform}(0.01,0.5)$. While the communities are \emph{assortative} in that the within-community edges have a larger link probability than between-community edges.
It is obvious that this structure brings difficulty to the estimation of communities. Figure \ref{effectofweights} shows that the motif-based method remains to be better than the edge-based method in this regime. It is worth mentioning that when the weights and communities are both assortative, the edge-based method is good enough to recover the communities since the signal is strong enough then.
}

\begin{figure*}[!htbp]{}
\centering
\scriptsize (I) Weights from $\chi^2(1)$ \\
\subfigure[ARI]{\includegraphics[height=4.8cm,width=5cm,angle=0]{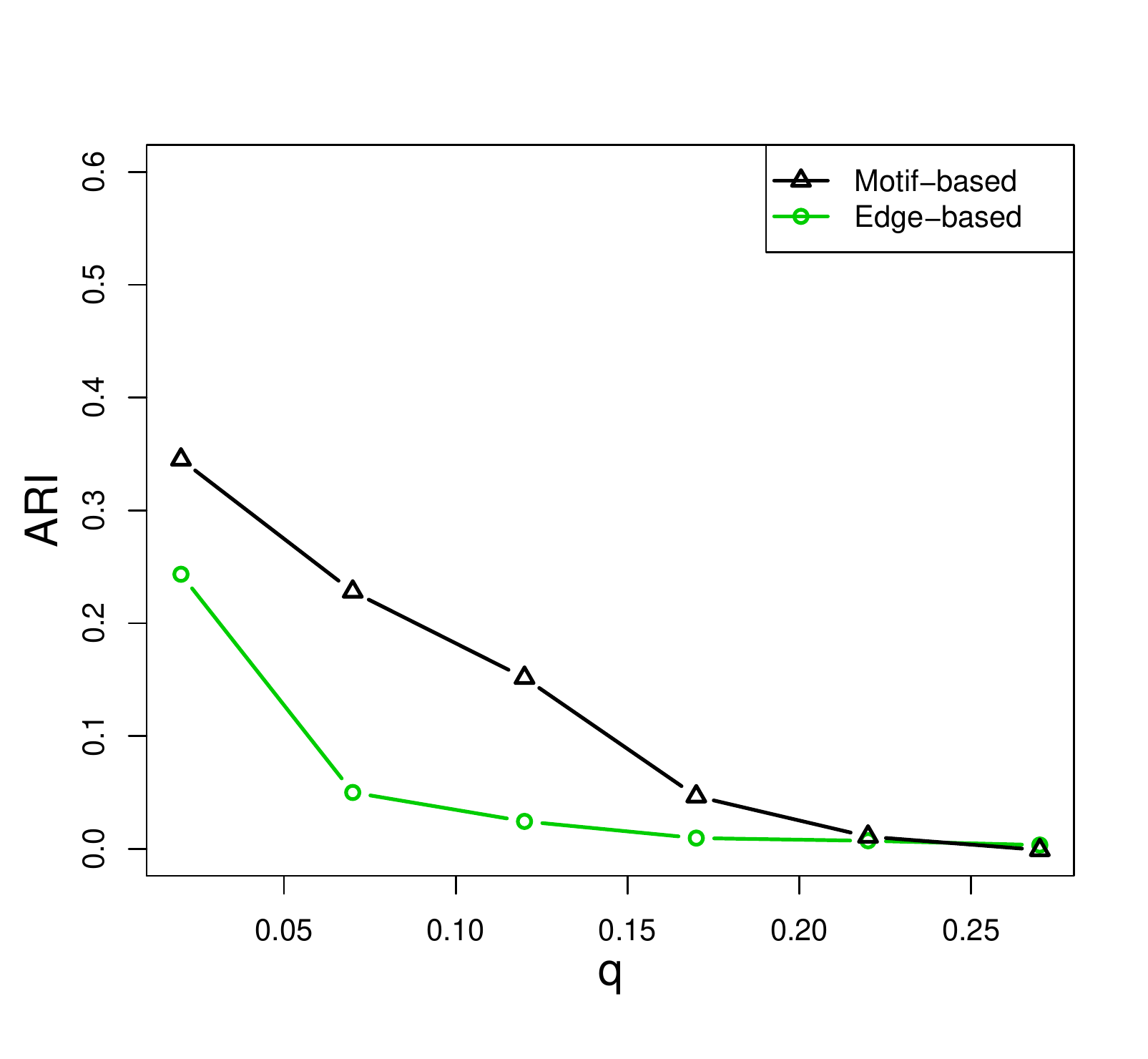}}
\subfigure[NMI]{\includegraphics[height=4.8cm,width=5cm,angle=0]{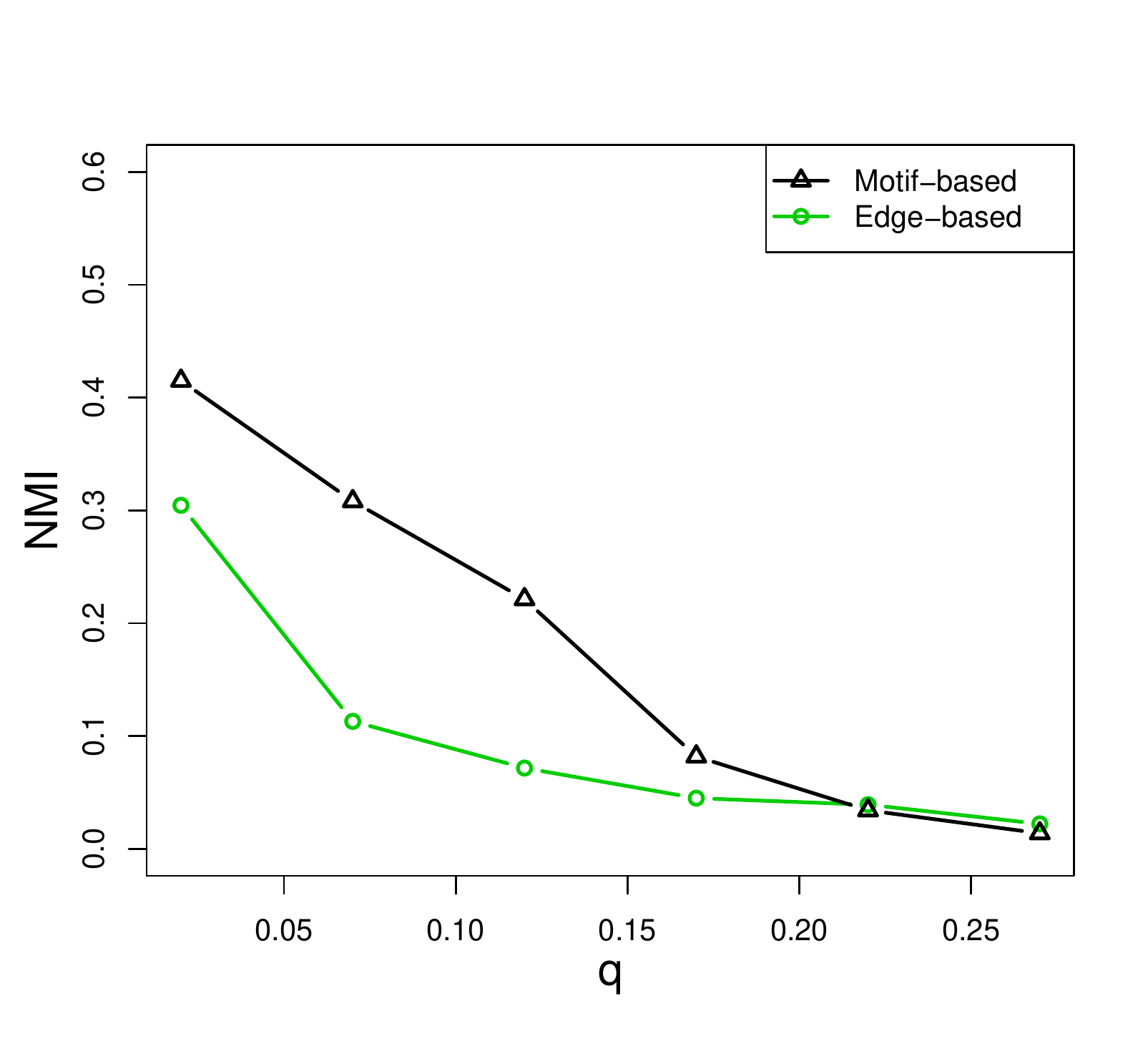}}
\subfigure[Modularity]{\includegraphics[height=4.8cm,width=5cm,angle=0]{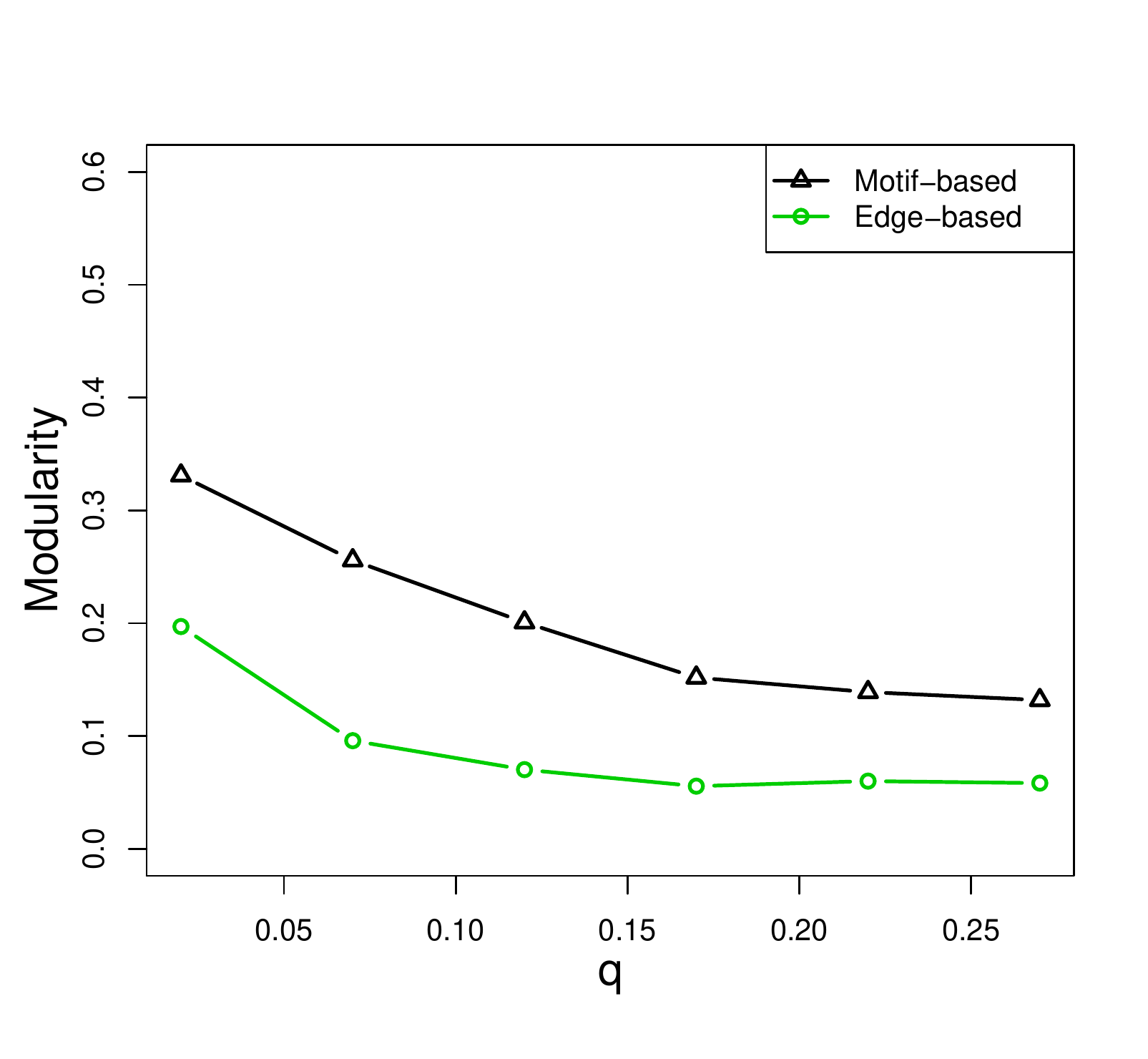}}\vspace{0.3cm}\\
\scriptsize (II) Weights from ${\rm exp}(1)$ \\
\subfigure[ARI]{\includegraphics[height=4.8cm,width=5cm,angle=0]{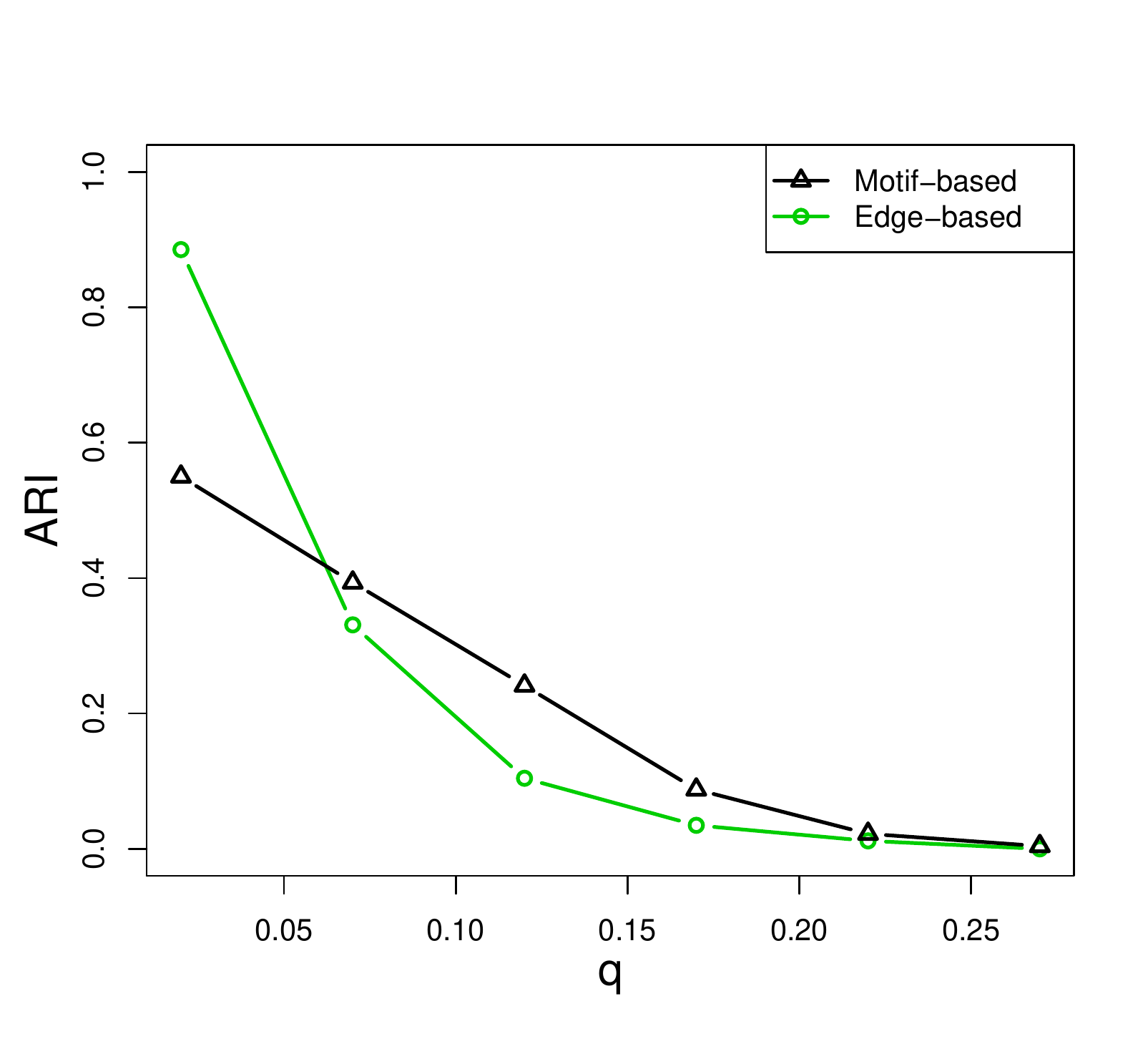}}
\subfigure[NMI]{\includegraphics[height=4.8cm,width=5cm,angle=0]{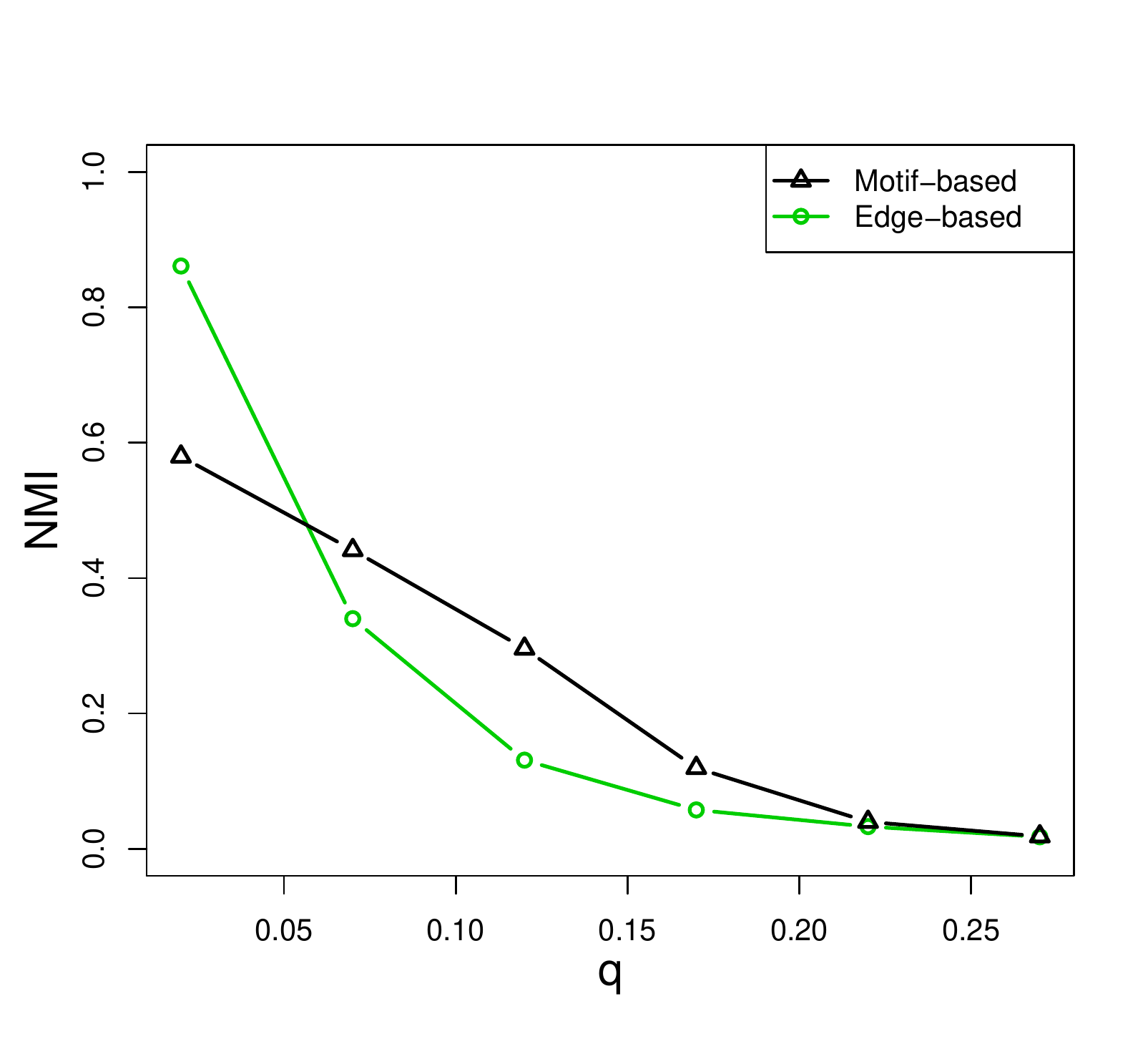}}
\subfigure[Modularity]{\includegraphics[height=4.8cm,width=5cm,angle=0]{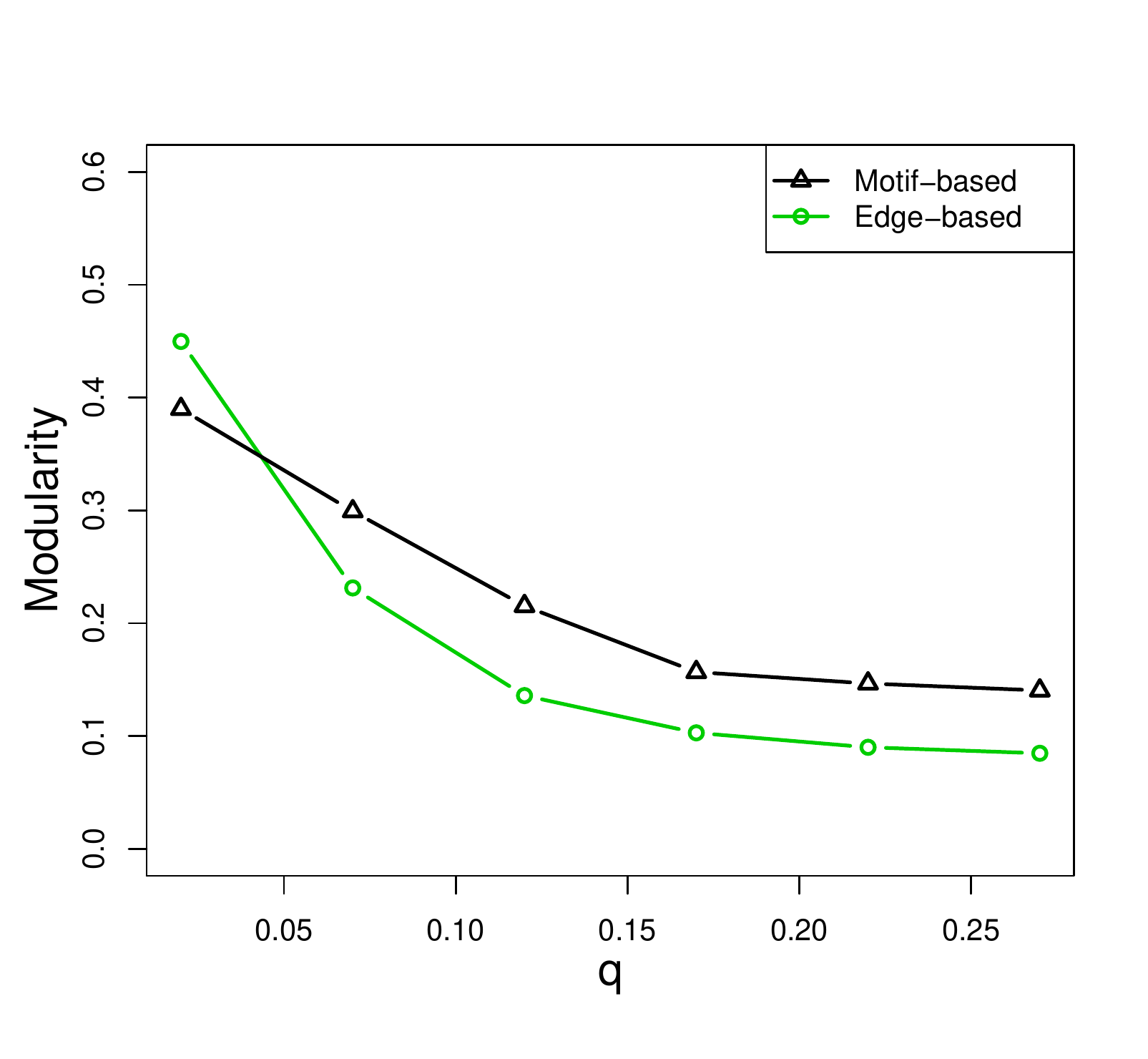}}\\
\caption{The average performance of two methods on networks with two kinds of weight distributions. (I) and (II) correspond to the results with the network weights \emph{i.i.d.} generated from $\chi^2(1)$ and ${\rm exp}(1)$, respectively. Other parameters are $n=60, p=0.3, K=2$.}\label{effectofdis}
\end{figure*}

\textcolor{black}{\paragraph{Other weight distributions.} The aforementioned experiments involved uniformly distributed network weights. Here, we test the efficacy of the motif-based method on two other distributions. The first is the chi-squared distribution with degree of freedom 1, denoted by $\chi^2(1)$. The second is the exponential distribution with mean 1, denoted by ${\rm exp}(1)$. The corresponding results are shown in Figure \ref{effectofdis}. It can be seen that the motif-based spectral clustering performs better than the edge-based counterpart, which shows that the efficacy of the motif-based method is not restrictive in terms of the distribution of weights.}

\begin{figure*}[!htbp]{}
\centering
\scriptsize (I) Wedge motif\\
\subfigure[ARI]{\includegraphics[height=4.8cm,width=5cm,angle=0]{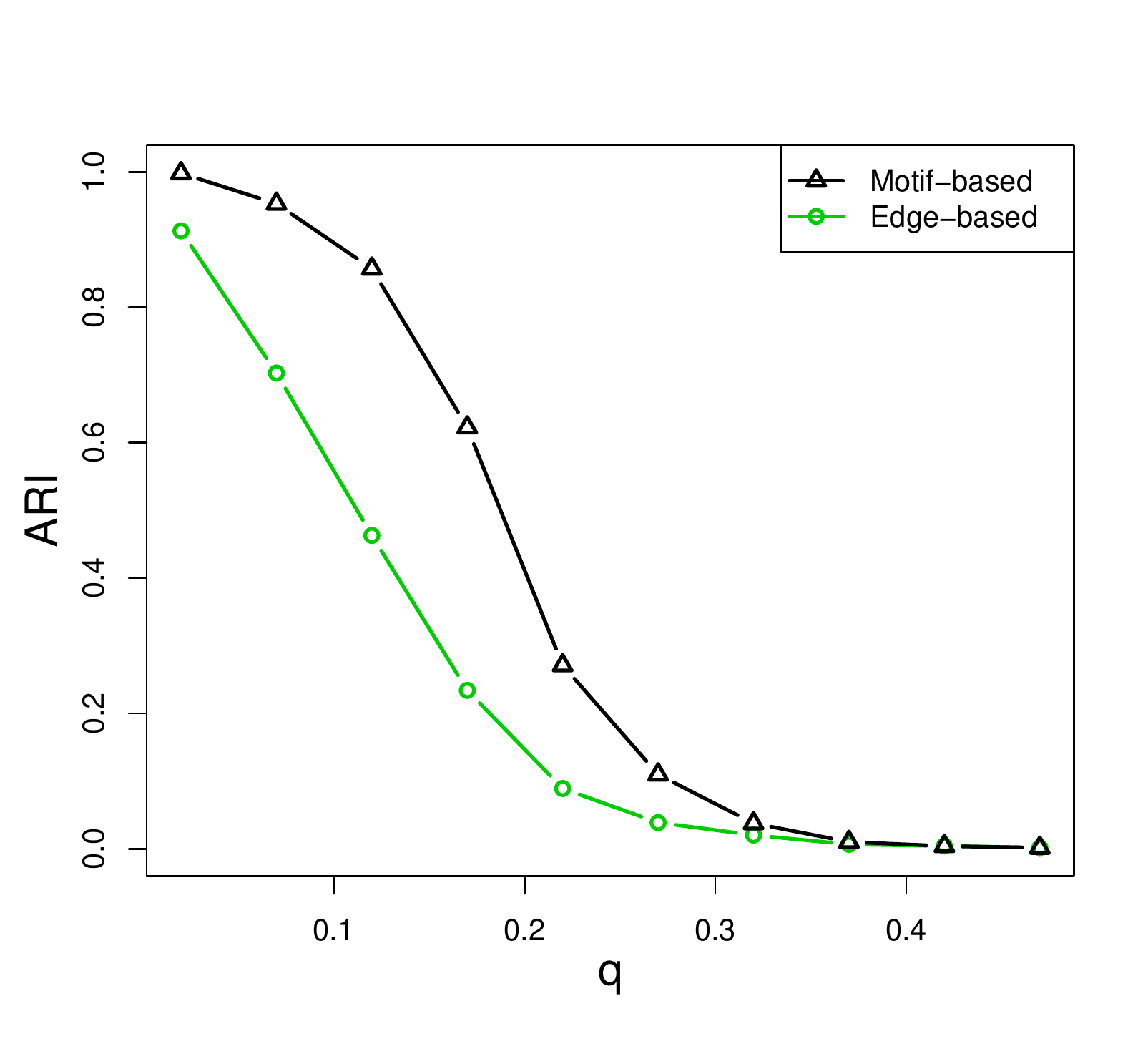}}
\subfigure[NMI]{\includegraphics[height=4.8cm,width=5cm,angle=0]{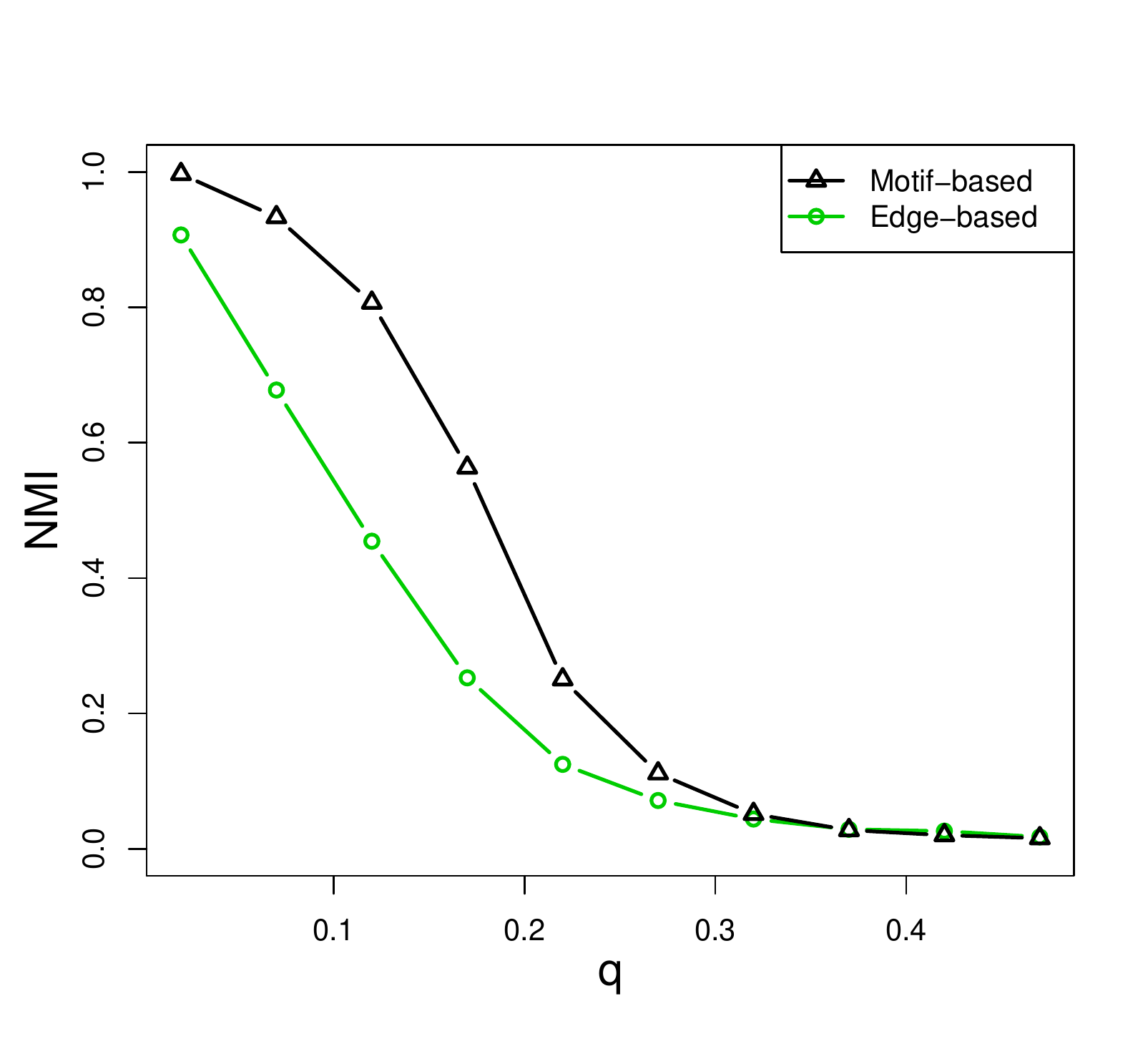}}
\subfigure[Modularity]{\includegraphics[height=4.8cm,width=5cm,angle=0]{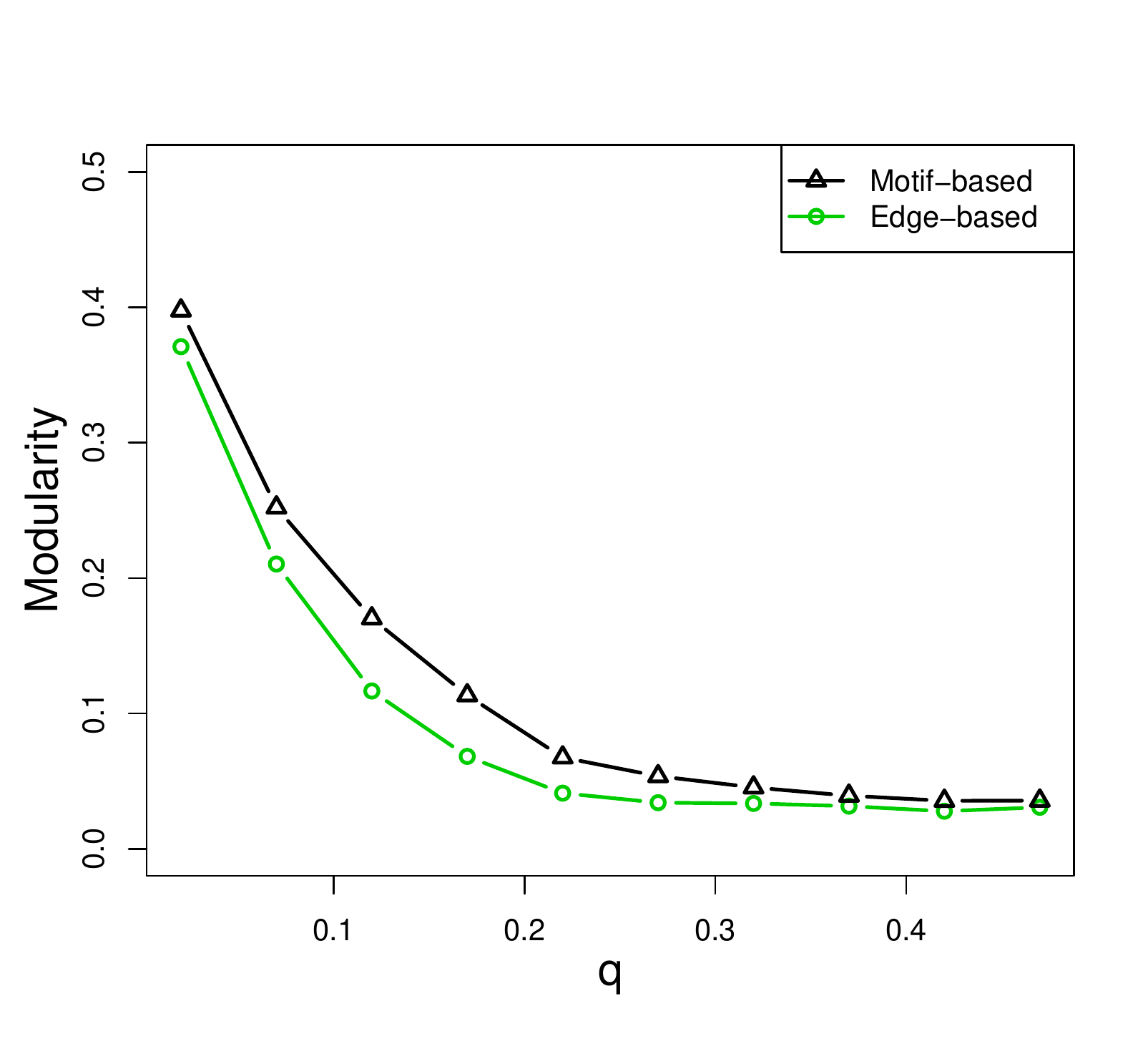}}\vspace{0.3cm}\\
(II) Four-nodes clique\\
\subfigure[ARI]{\includegraphics[height=4.8cm,width=5cm,angle=0]{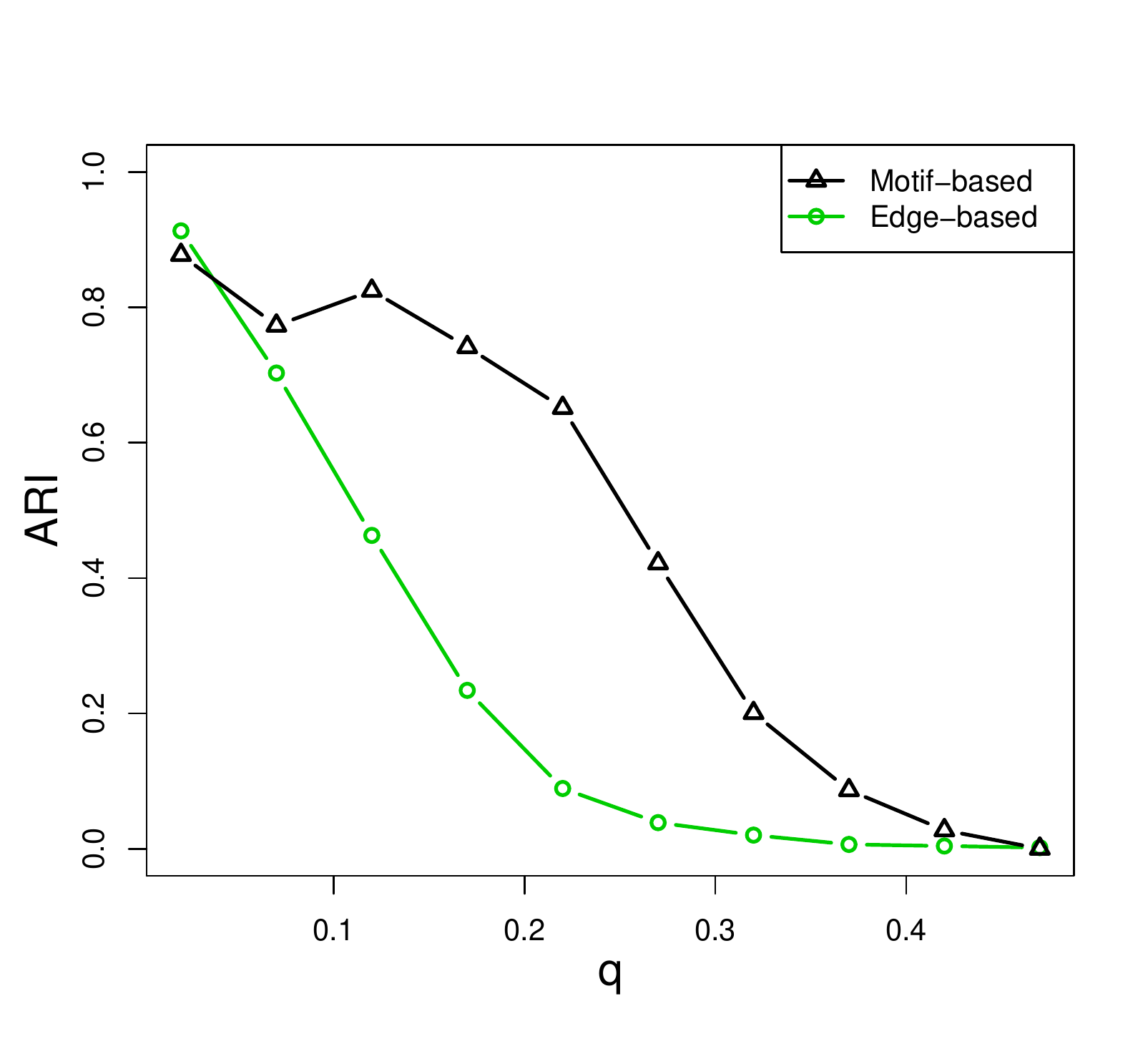}}
\subfigure[NMI]{\includegraphics[height=4.8cm,width=5cm,angle=0]{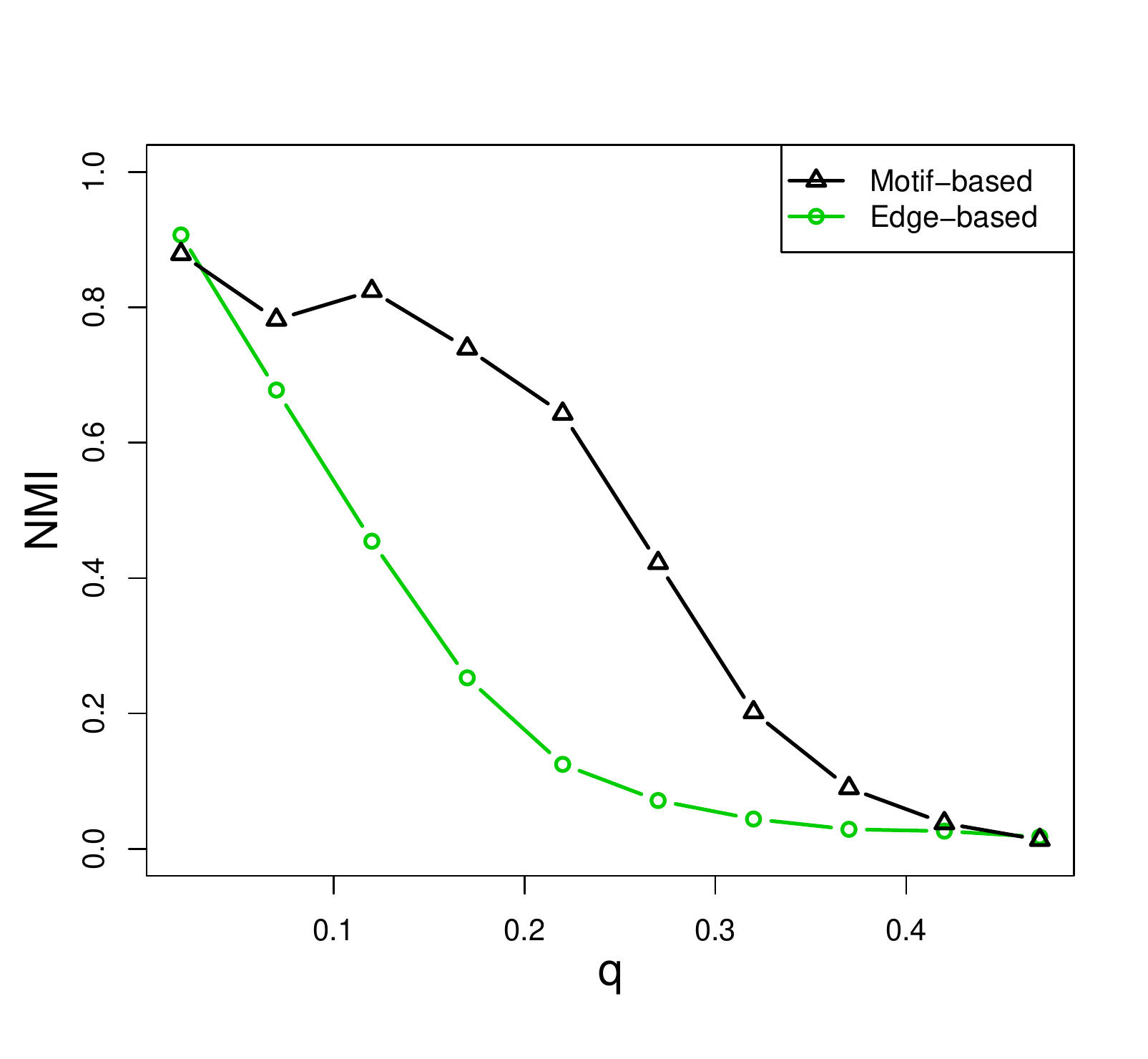}}
\subfigure[Modularity]{\includegraphics[height=4.8cm,width=5cm,angle=0]{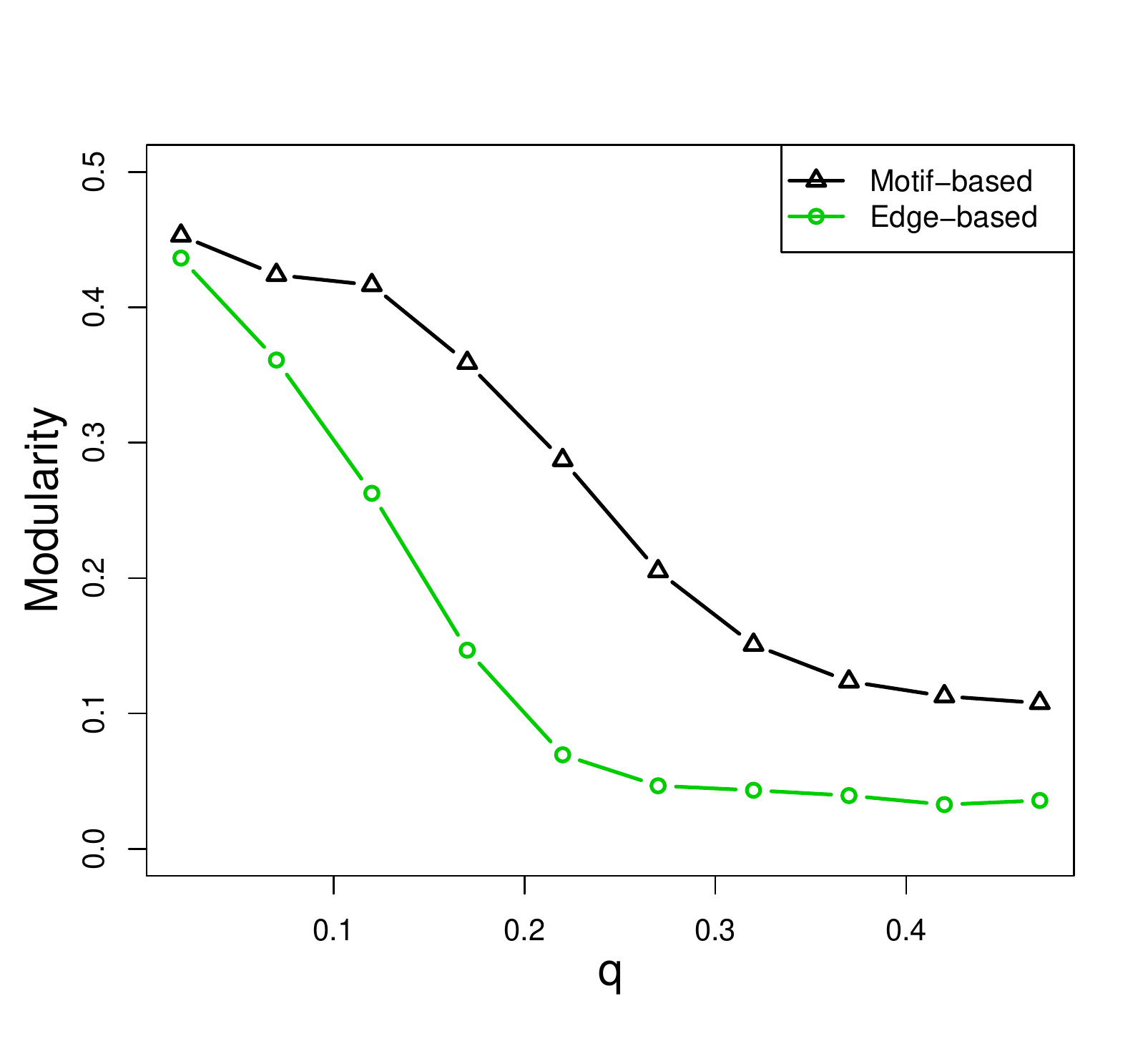}}\\
\caption{The performance of two methods, where the motif-based spectral clustering is based on (I) the wedge motif and (II) the four-nodes clique, respectively. The parameters are $K=2, n=60, p=0.5$, and the edge weights are \emph{i.i.d.} $\chi^2 (1)$.}\label{effectofmotif}
\end{figure*}

\textcolor{black}{\paragraph{Other motifs.} From both the theoretical and algorithmic sides, we have mainly focused on the triangular motif in motif-based spectral clustering. It is of interest to see whether motif-based spectral clustering remains powerful on other motifs. We here test the algorithm on two other motifs. The first is the wedge motif, which is the subgraph with three nodes and two edges. The second is the four-nodes clique motif, which is the totally connected subgraph with four nodes and six edges. The weighted motif adjacency matrix is then formulated by adding the weights in the motif that the node pairs participate in. Figure \ref{effectofmotif} shows the results. It turns out that motif-based spectral clustering is also effective on these two motifs. Hence, it would be interesting to theoretically study the consistency.}

\textcolor{black}{To sum up, the above experiments show that the motif-based spectral clustering has the advantage over the edge-based counterpart when the clustering problem is relatively hard, namely, the signal of communities is weak. And the applicable range of motif-based spectral clustering is far beyond the limits of theory. It is also worth noting that by our simulated examples, the modularity is a good measure to judge between the two methods when the true clustering labels are not known.}

\section{Real data analysis}
\label{realdata}
In this section, we show the merits of higher-order spectral clustering using three real datasets, including the statisticians' citation network, statisticians' coauthor network, and the ${\rm PM}_{2.5}$ data set. In the sequel, we introduce the datasets and their corresponding clustering results, respectively. \textcolor{black}{To save space, all the clustering results are included in the Supplementary Material.}

\begin{figure*}[!htbp]{}
\centering
\subfigure[Detected communities]{\includegraphics[height=10cm,width=10cm,angle=0]{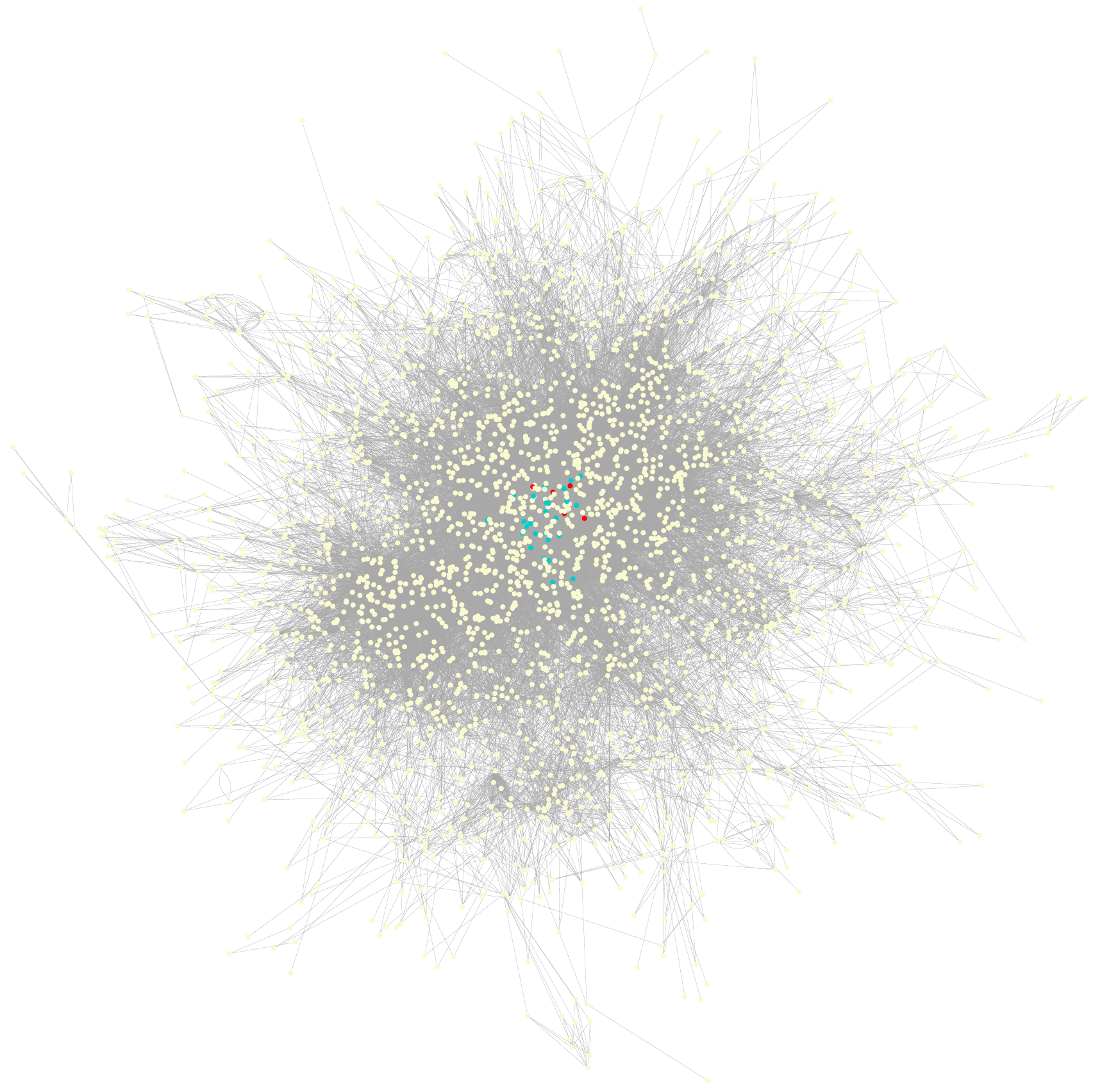}}\\
\subfigure[High-dimensional statistics community]{\includegraphics[height=6cm,width=6cm,angle=0]{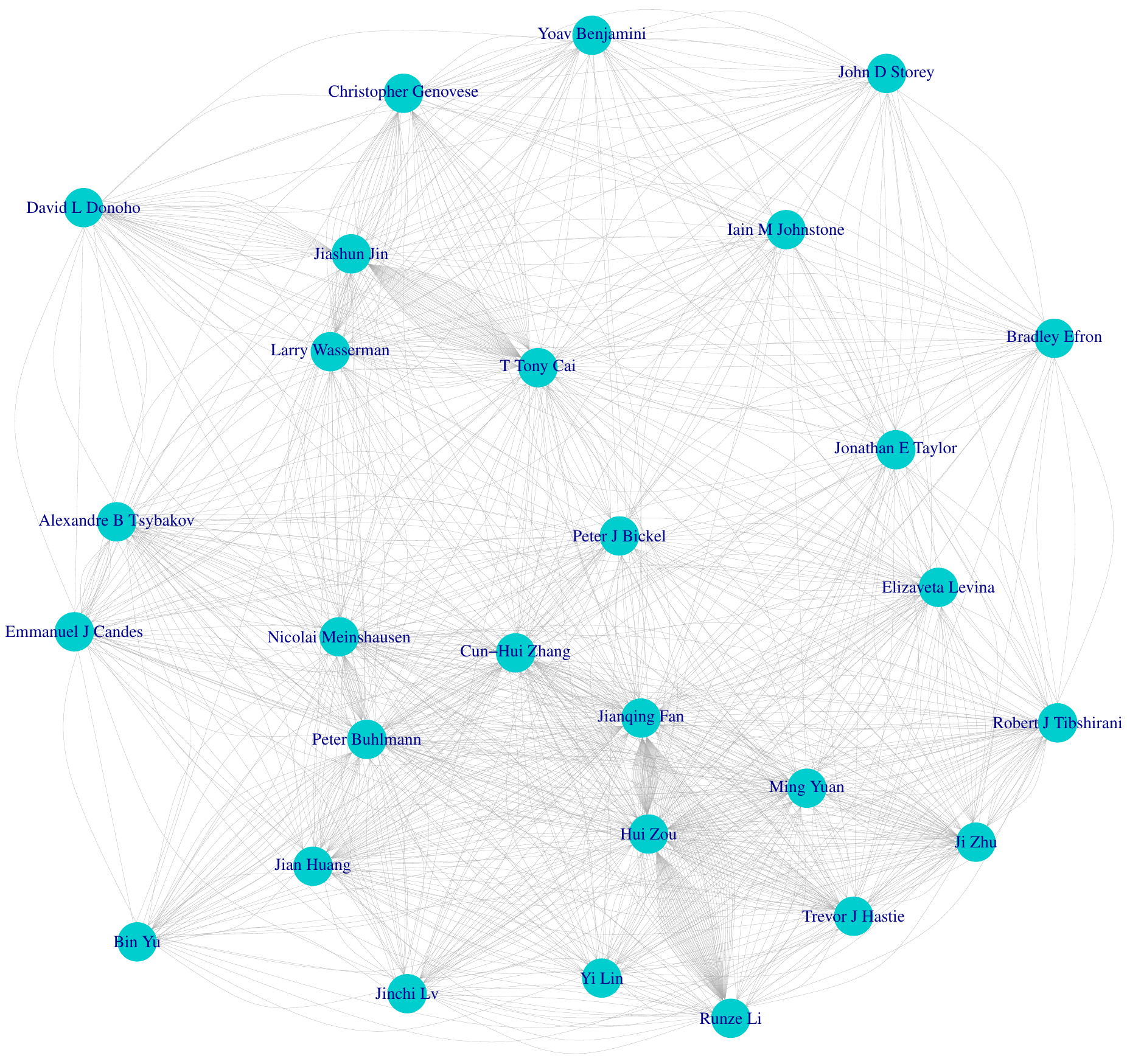}}
\hspace{0.1\textwidth}
\subfigure[Non-parametric statistics and Functional analysis community]{\includegraphics[height=6cm,width=6cm,angle=0]{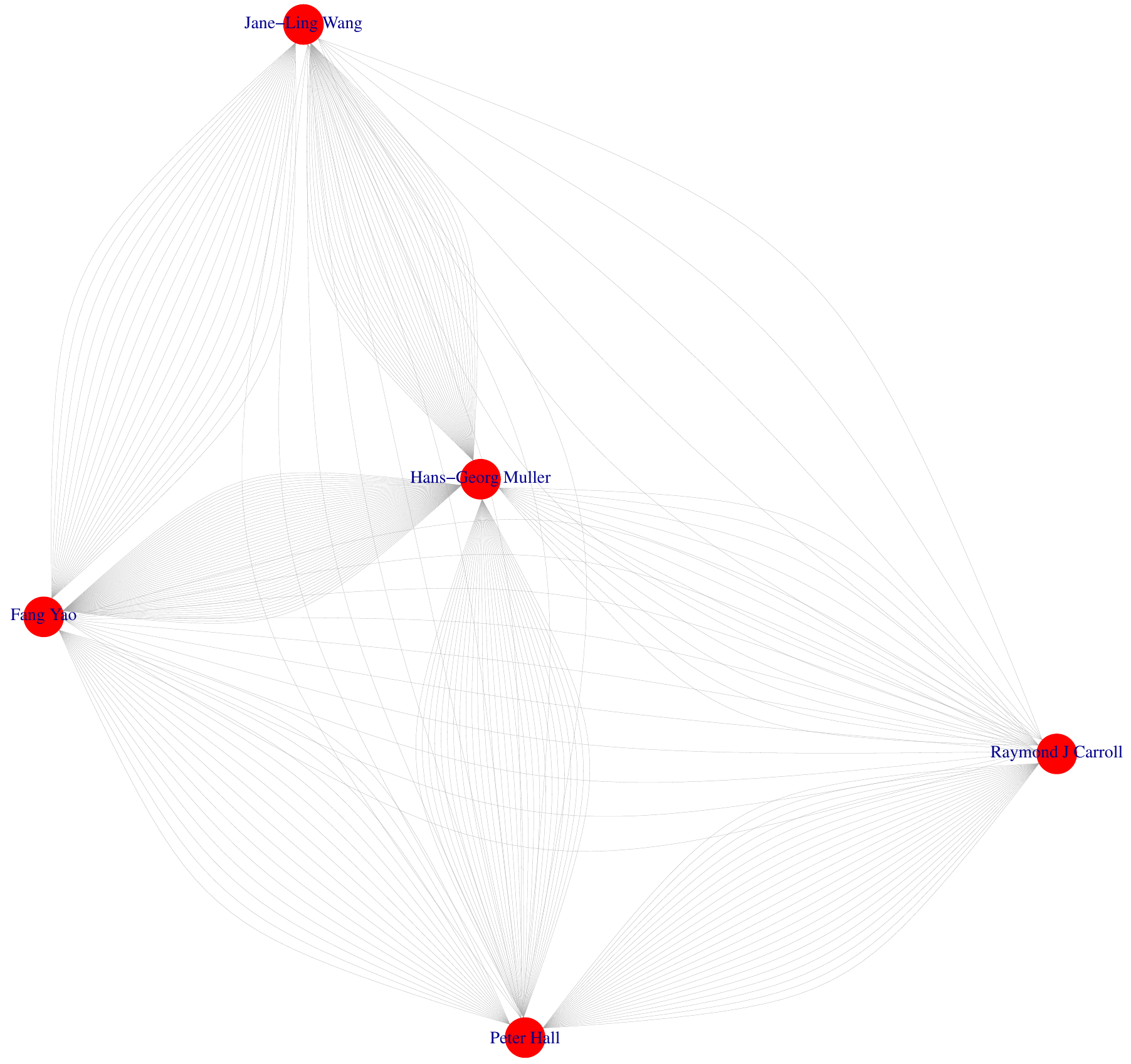}}
\caption{The clustering results for the statisticians' citation network by the motif-based spectral clustering. (a) is the general clustering results, and (b) and (c) are specific communities. }\label{citationmotif}
\end{figure*}

\begin{figure*}[!htbp]{}
\centering
\subfigure[Detected communities]{\includegraphics[height=10cm,width=10cm,angle=0]{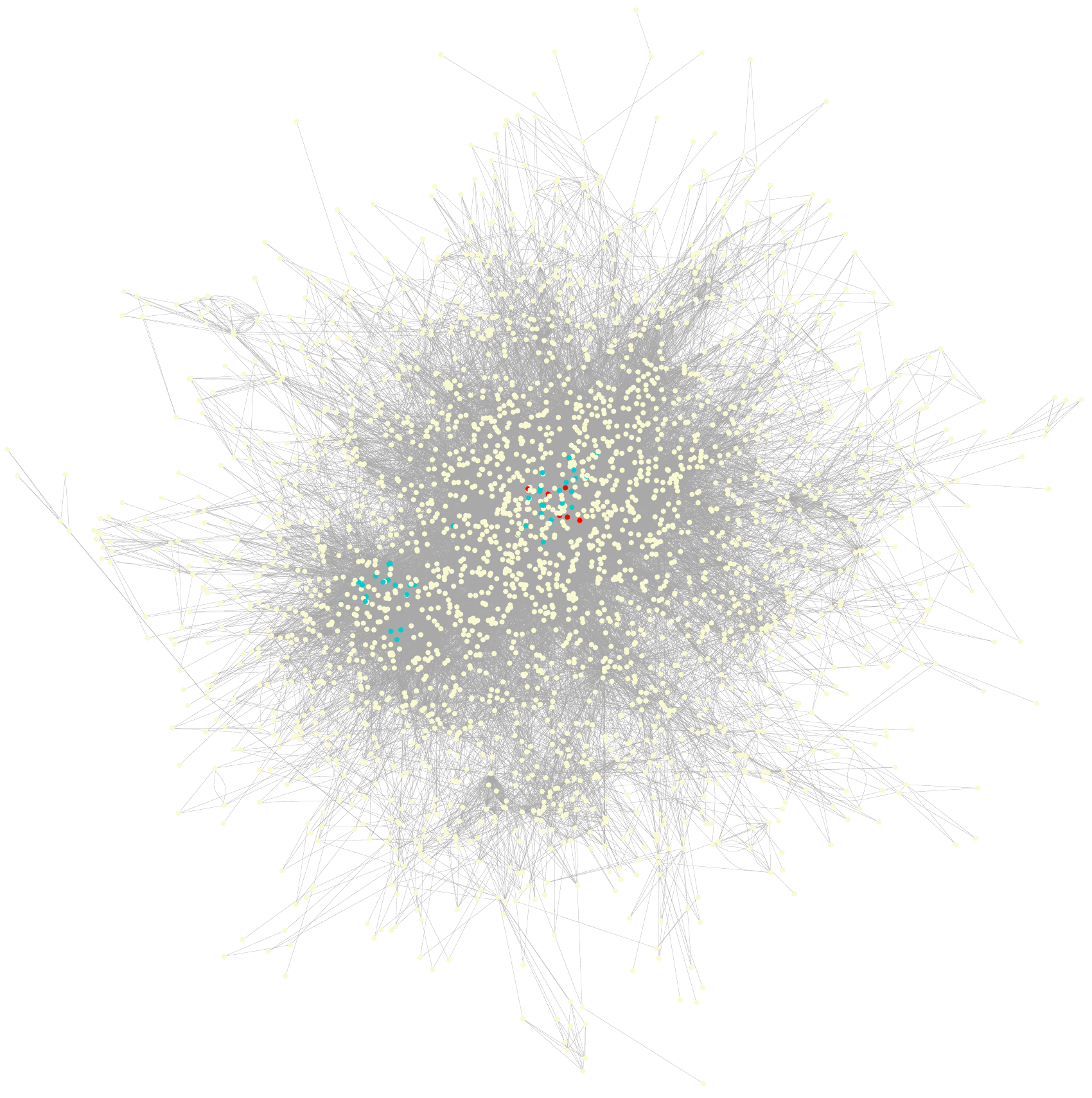}}\\
\subfigure[High-dimensional and Bayesian statistics \emph{mixed} community]{\includegraphics[height=6cm,width=6cm,angle=0]{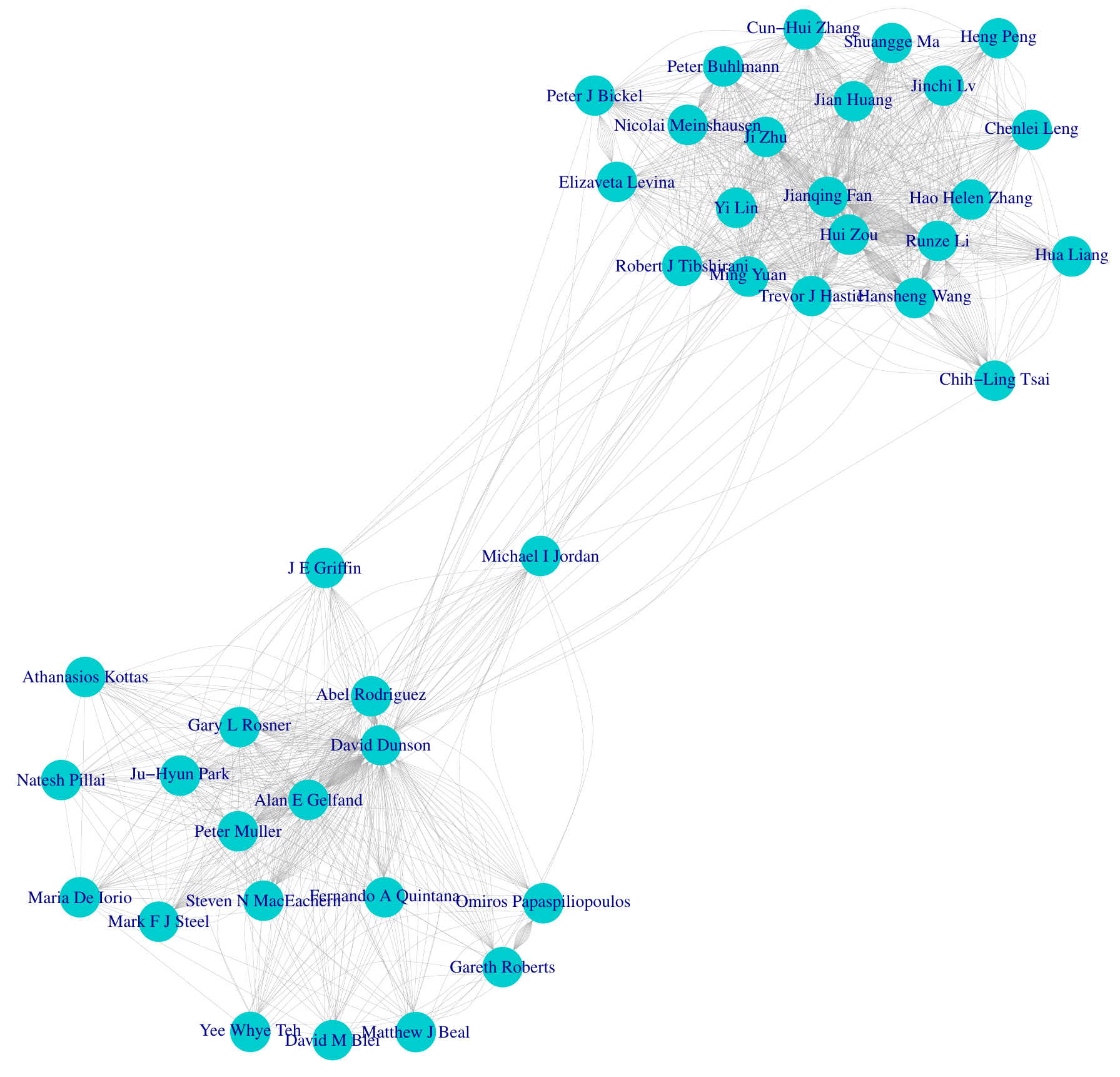}}
\hspace{0.1\textwidth}
\subfigure[Non-parametric statistics and Functional analysis community]{\includegraphics[height=6cm,width=6cm,angle=0]{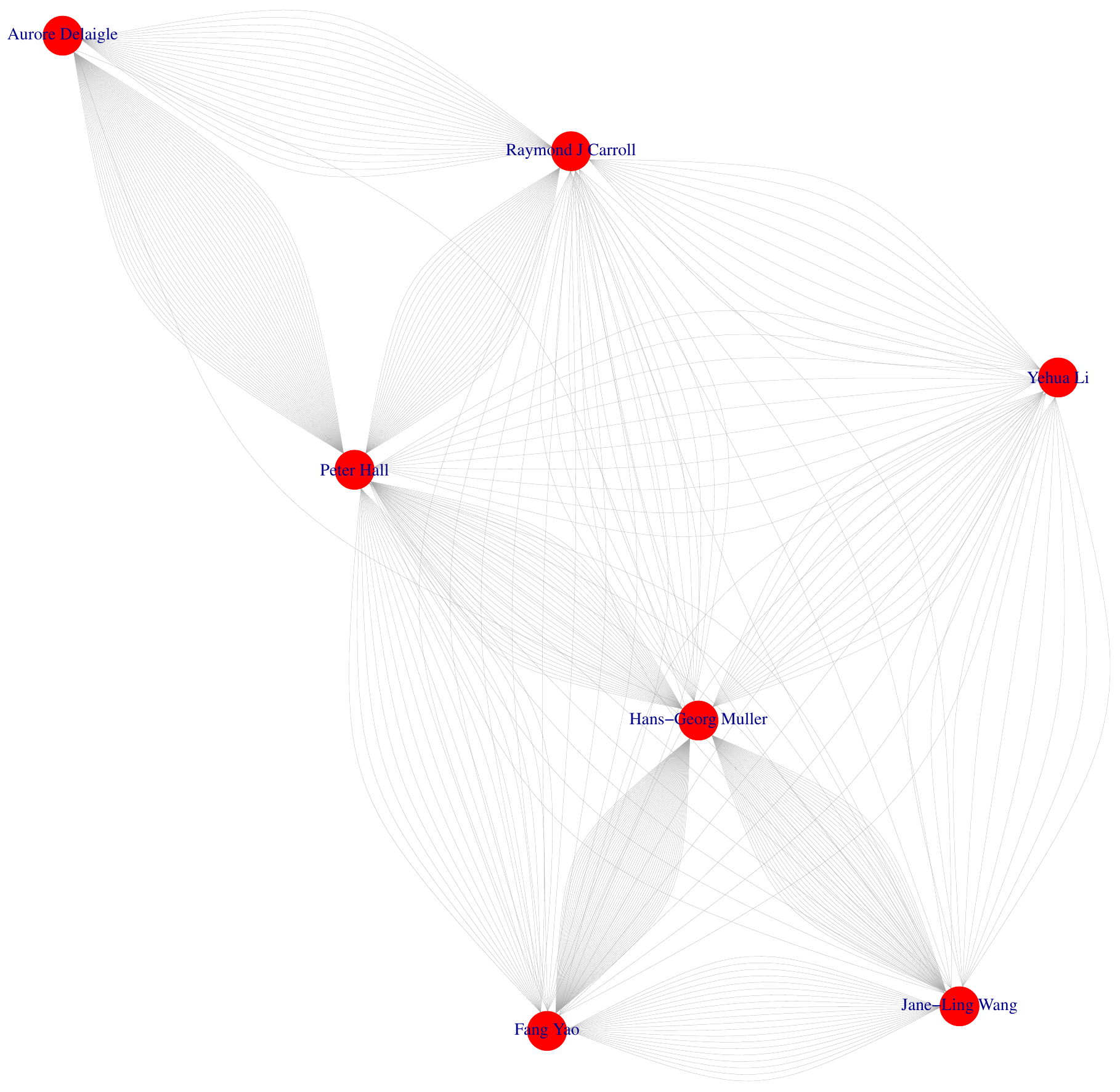}}
\caption{The clustering results for the statisticians' citation network by the edge-based spectral clustering. (a) is the general clustering results, and (b) and (c) are specific communities.
}\label{citationedge}
\end{figure*}

\textbf{Statisticians citation network.} This dataset was initially collected by \citet{ji2016coauthorship} based on all published papers from 2003 to the first half of 2012 in four of the top statistical journal: Annals of Statistics, Biometrika, Journal of American Statistical Association and Journal of Royal Statistical Society (Series B), which results in 3248 papers and 3607 authors in total. We here study a citation network, where each node represents an author and the number of edges between any pair of nodes is equal to the number of one-way citations between the two authors. We consider the largest connected components of this network which includes 2654 nodes. We set $K=3$ as in \citet{ji2016coauthorship}. Figure \ref{citationmotif} and \ref{citationedge} shows the clustering results for the higher-order and the edge-based spectral clustering, respectively. We can see from Figure \ref{citationmotif}(a) and \ref{citationedge}(a) that both of the two methods find two small communities and one large community, which can be regarded as the background community. Hence, we take a closer look at the two small communities. As shown in Figure \ref{citationmotif}(b) and \ref{citationmotif}(c), the motif-based spectral clustering detects two communities with many triangles. One community (see Figure \ref{citationmotif}(b)) consists of statisticians in high-dimensional statistics, including but without limiting to the authors of the lasso, group lasso, adaptive lasso, SCAD, graphical models, which are pioneering works of high-dimensional statistics in the past 20 years. The other community (\ref{citationmotif}(c)) consists of 5 statisticians engaged in a functional analysis or non-parametric statistics. Turning to the edge-based spectral clustering, we find that one community (see Figure \ref{citationedge}(c)) includes 7 statisticians engaged in functional analysis and non-parametric statistics, which is very similar to the non-parametric statistics community found by the motif-based clustering because only two more statisticians are included. After examining the other community (see Figure \ref{citationedge}(b)) carefully, we find that this community is comprised of two statistician communities, namely, high-dimensional statisticians and Bayesian statisticians, and statistician Michael Jordan bridges these two communities. The results are quite interesting, but from the clustering point of view, motif-based spectral clustering leads to better results since the resulting communities are purer.

\begin{figure*}[!htbp]{}
\centering
\subfigure[Detected communities]{\includegraphics[height=10cm,width=10cm,angle=0]{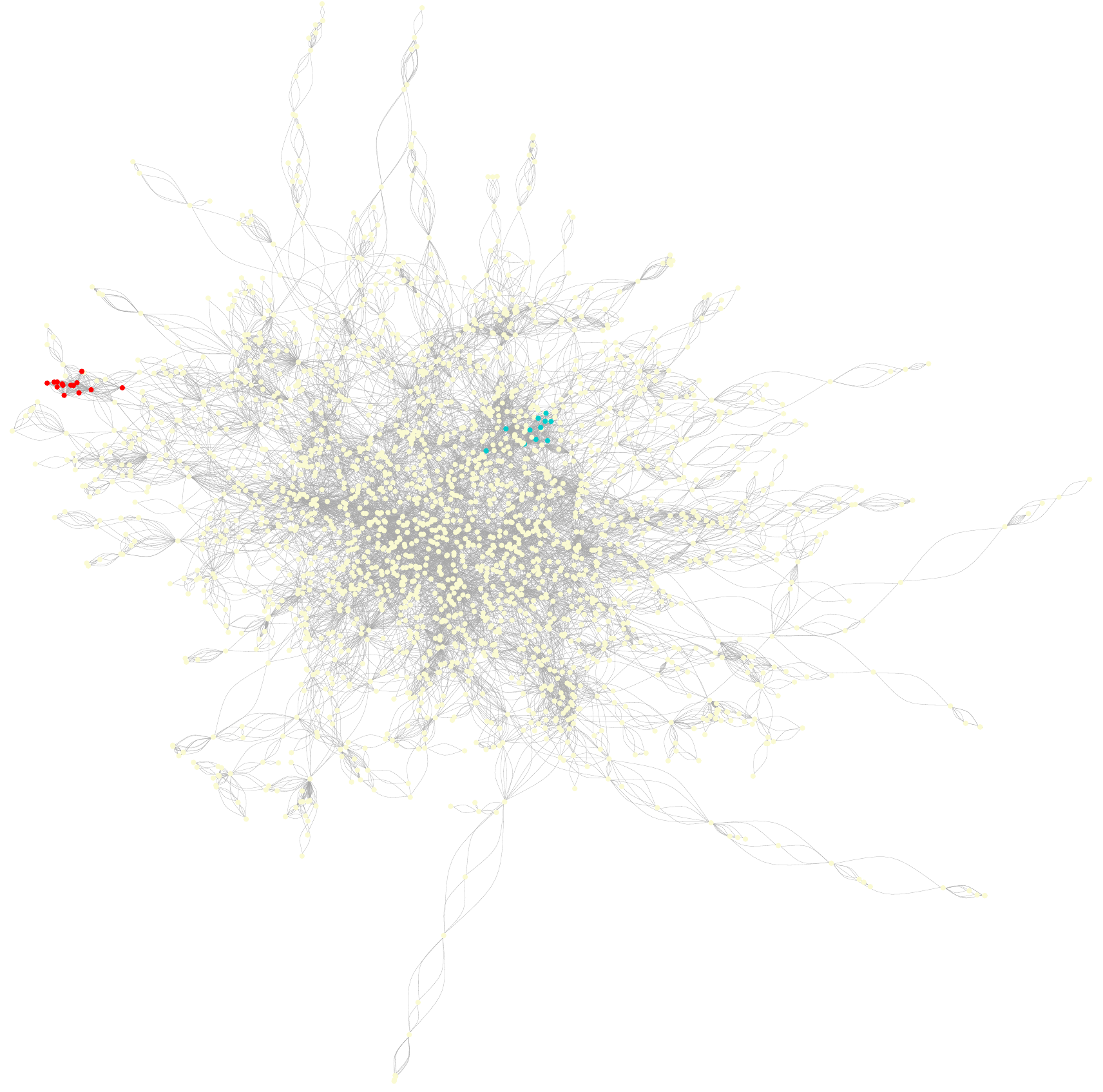}}\\
\subfigure[Biostatistics and medical statistics community]{\includegraphics[height=6cm,width=6cm,angle=0]{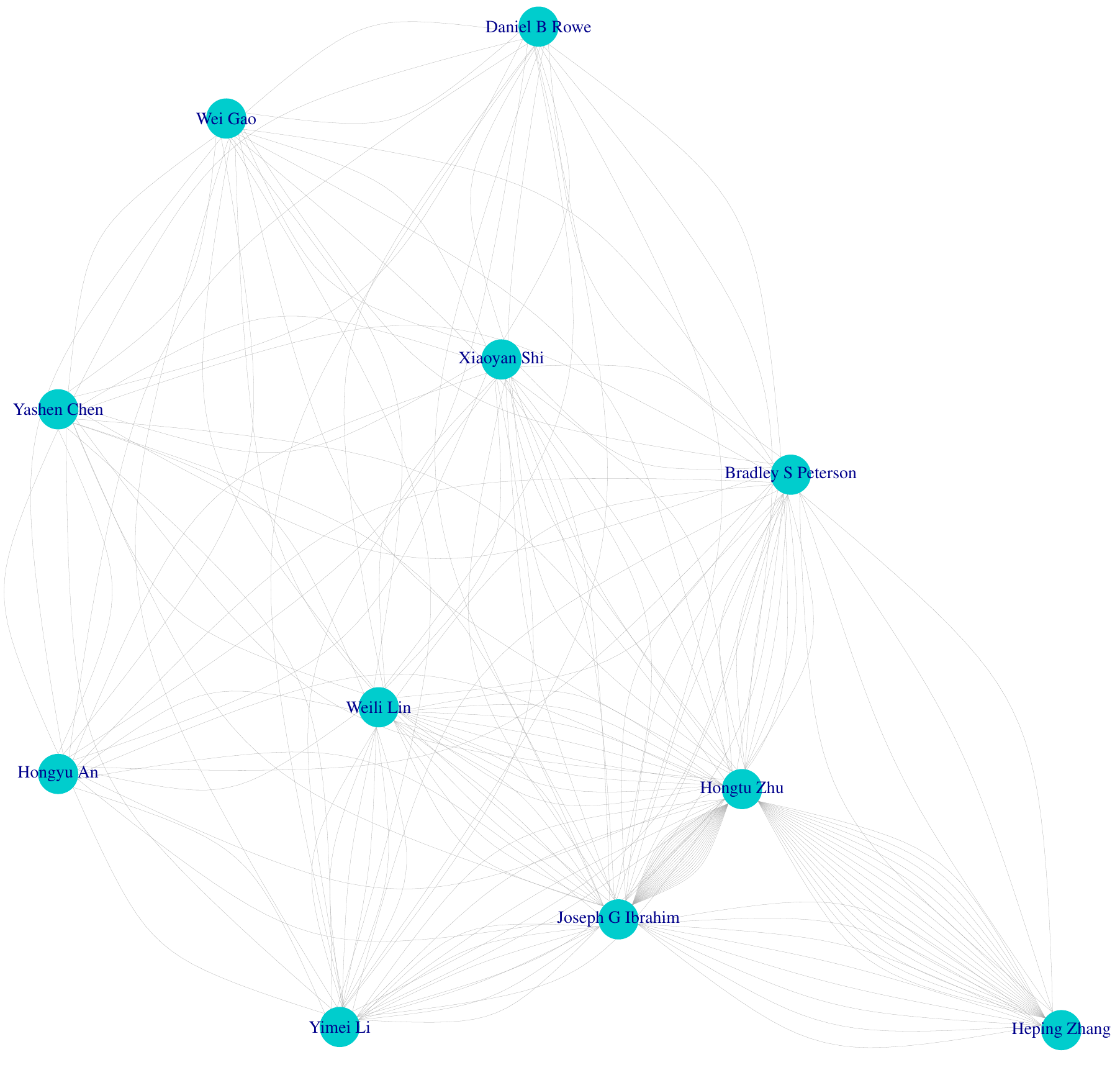}}
\hspace{0.1\textwidth}
\subfigure[Bayesian statistics community]{\includegraphics[height=6cm,width=6cm,angle=0]{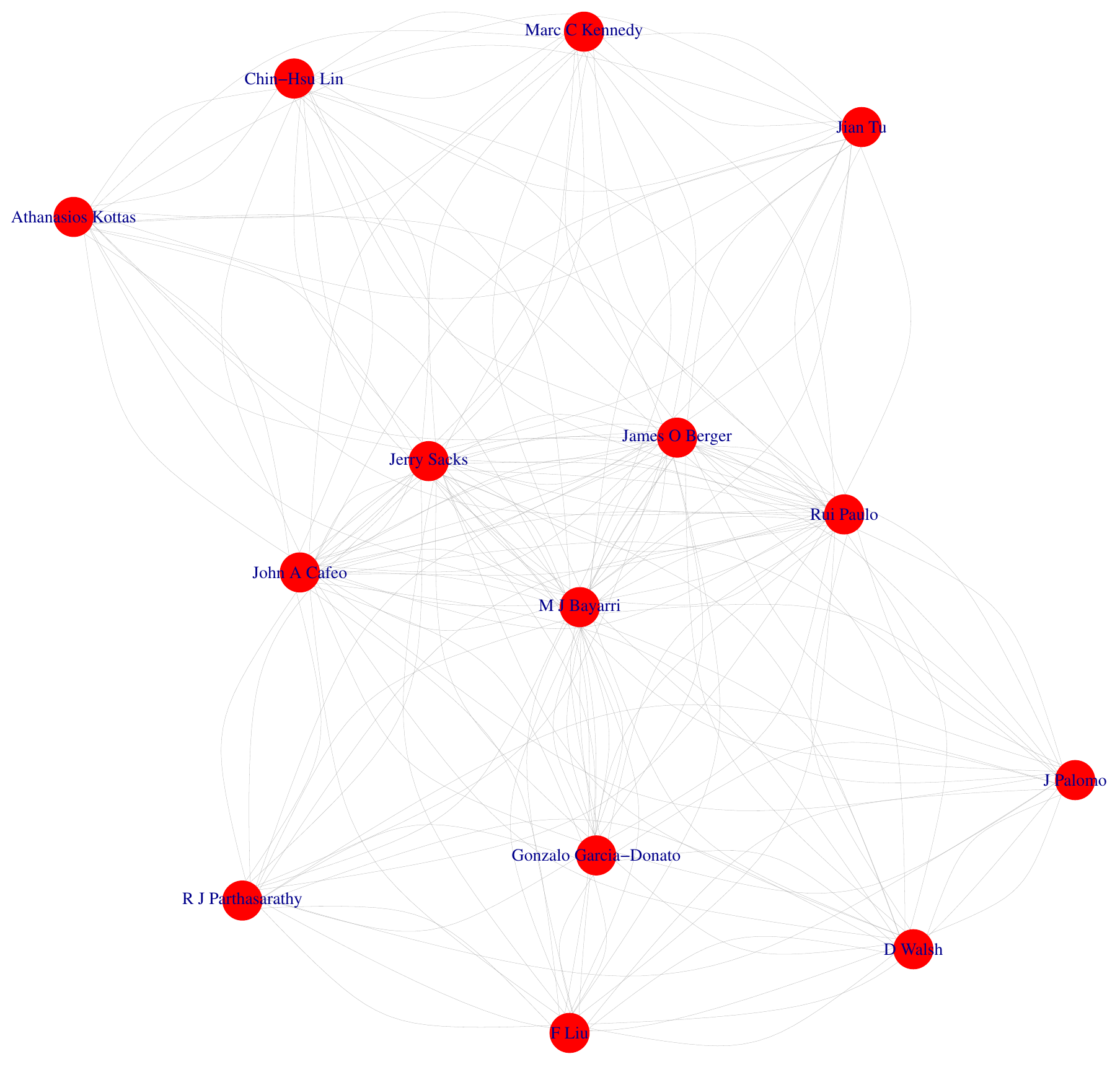}}
\caption{The clustering results for the statisticians' coauthor network by the motif-based spectral clustering. (a) is the general clustering results, and (b) and (c) are specific communities. }\label{coauthormotif}
\end{figure*}

\begin{figure*}[!htbp]{}
\centering
\subfigure[Detected communities]{\includegraphics[height=10cm,width=10cm,angle=0]{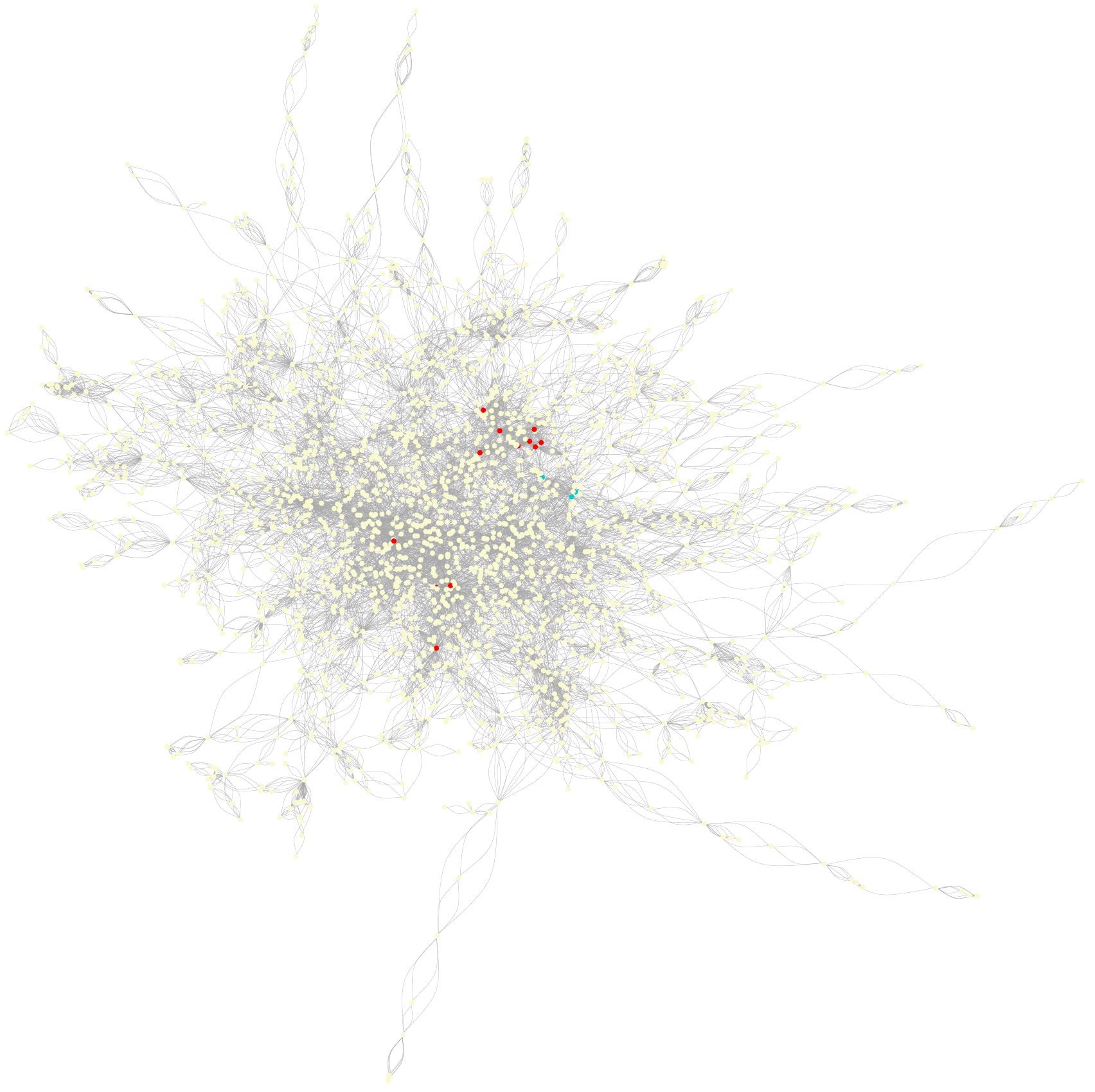}}\\
\subfigure[Biostatistics and medical statistics (at Harvard) community]{\includegraphics[height=5cm,width=5cm,angle=0]{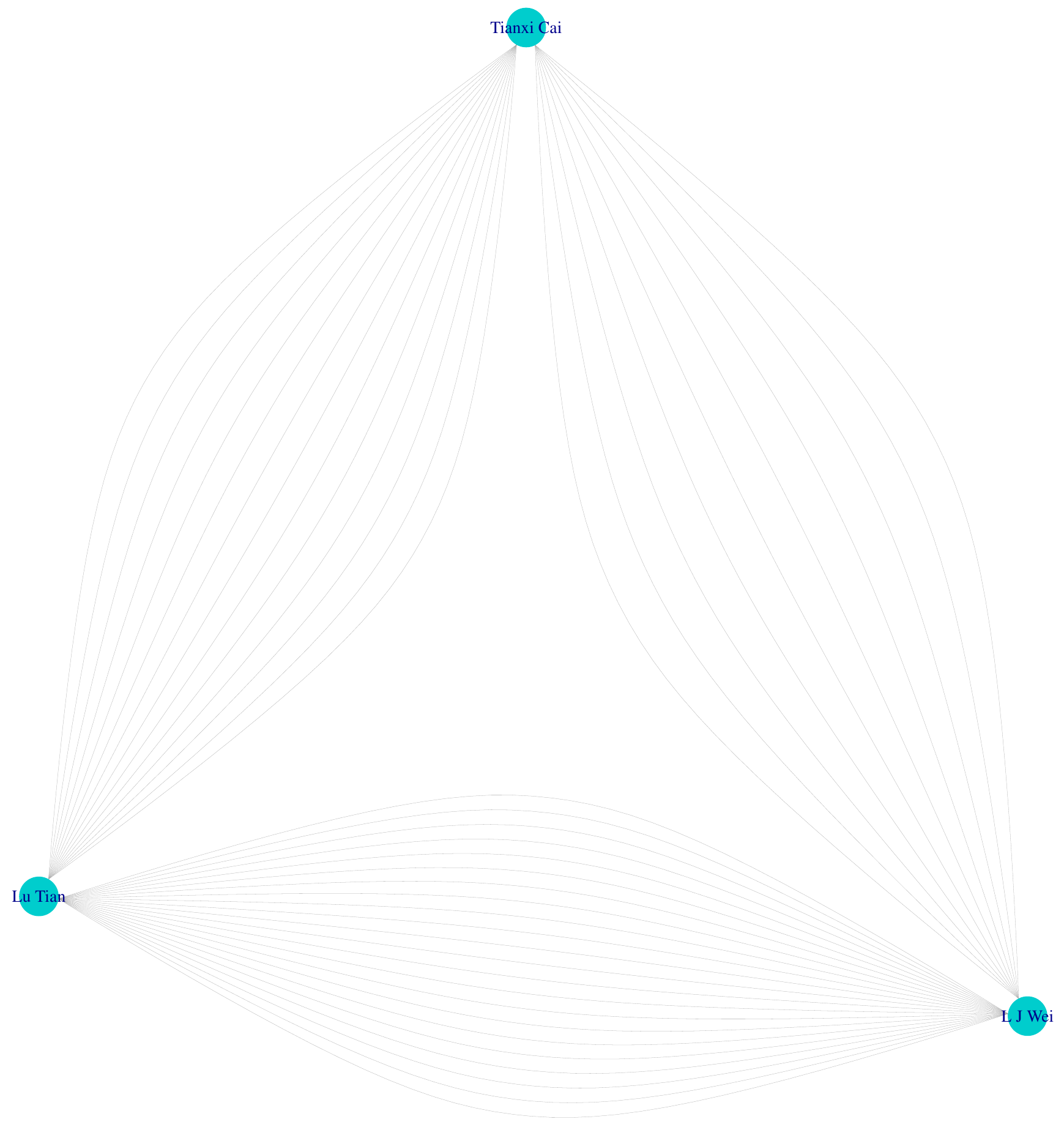}}
\hspace{0.1\textwidth}
\subfigure[Biostatistics/medical statistics and non-parametric statistics \emph{mixed} community ]{\includegraphics[height=6cm,width=6cm,angle=0]{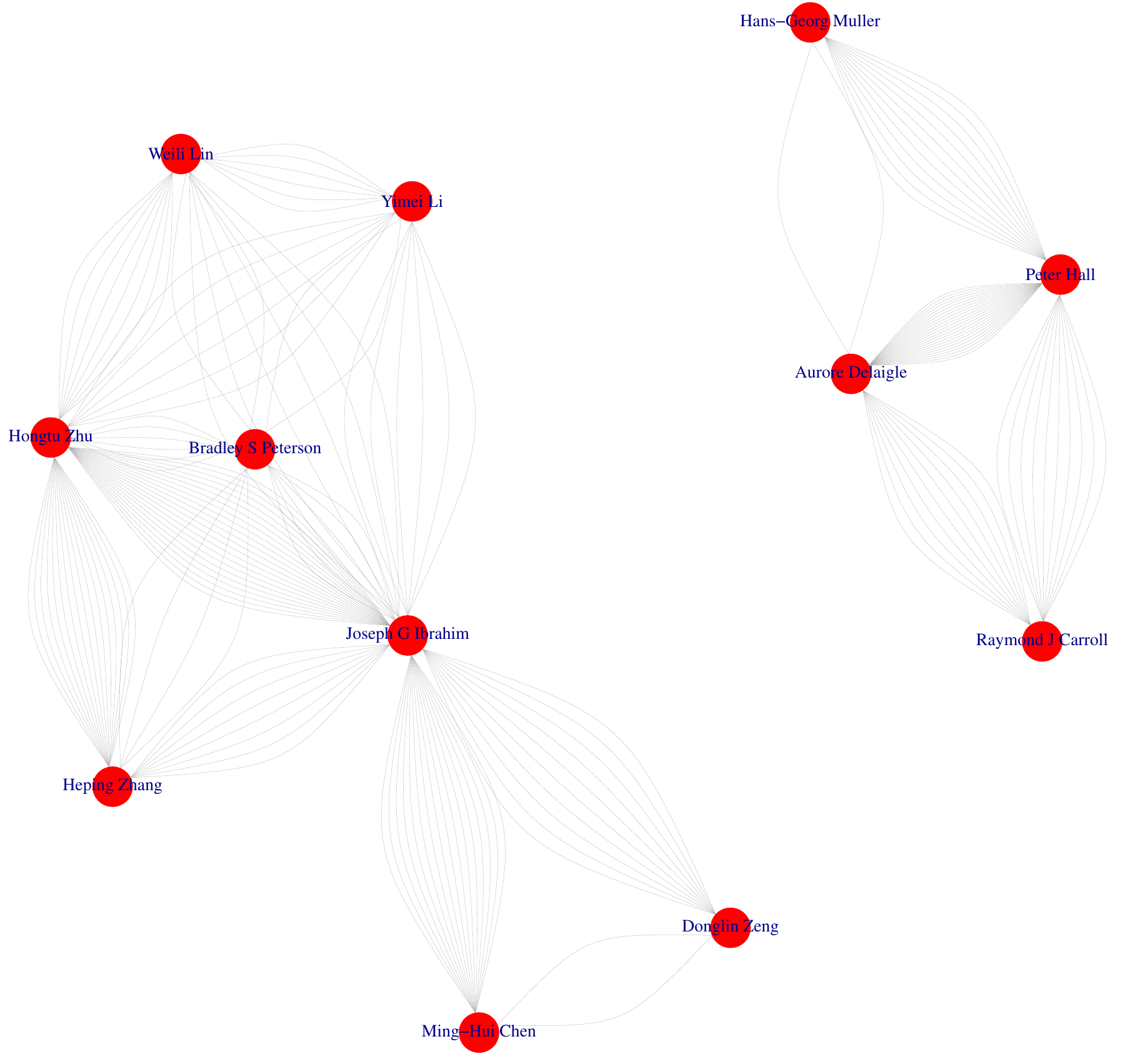}}
\caption{The clustering results for the statisticians' coauthor network by edge-based spectral clustering. (a) is the general clustering results, and (b) and (c) are specific communities.
}\label{coauthoredge}
\end{figure*}

\textbf{Statisticians coauthor network.} This network was also generated based on the aforementioned dataset \citep{ji2016coauthorship}. In particular, each node represents an author and the number of edges between any pair of nodes equals to the number of papers they coauthored. We consider the largest connected components of the network which result in 2263 nodes. We also set $K=3$ as in \citet{ji2016coauthorship}. Figure \ref{coauthormotif} and \ref{coauthoredge} display the clustering results for the motif-based and edge-based spectral clustering, respectively. We can see clearly that both methods detect two main communities, and the remaining communities are so large that we regard them as the background. In particular, the motif-based spectral clustering detected two communities (see Figure \ref{coauthormotif}(b) and \ref{coauthormotif}(c)), which include statisticians engaged in biostatistics/medical statistics and Bayesian statistics, respectively. By contrast, the edge-based spectral clustering detects one community (see Figure \ref{coauthoredge}(b)) with 3 biostatisticians who have worked or studied at Harvard. And the other community (see Figure \ref{coauthoredge}(c)) includes two disconnected components, where one consists of statisticians in biostatistics/medical statistics, and the other includes statisticians in non-parametric statistics. Hence, motif-based spectral clustering detects more reasonable communities than edge-based ones.

\begin{figure*}[!htbp]{}
\centering
\subfigure[Detected communities]{\includegraphics[height=10cm,width=12.3cm,angle=0]{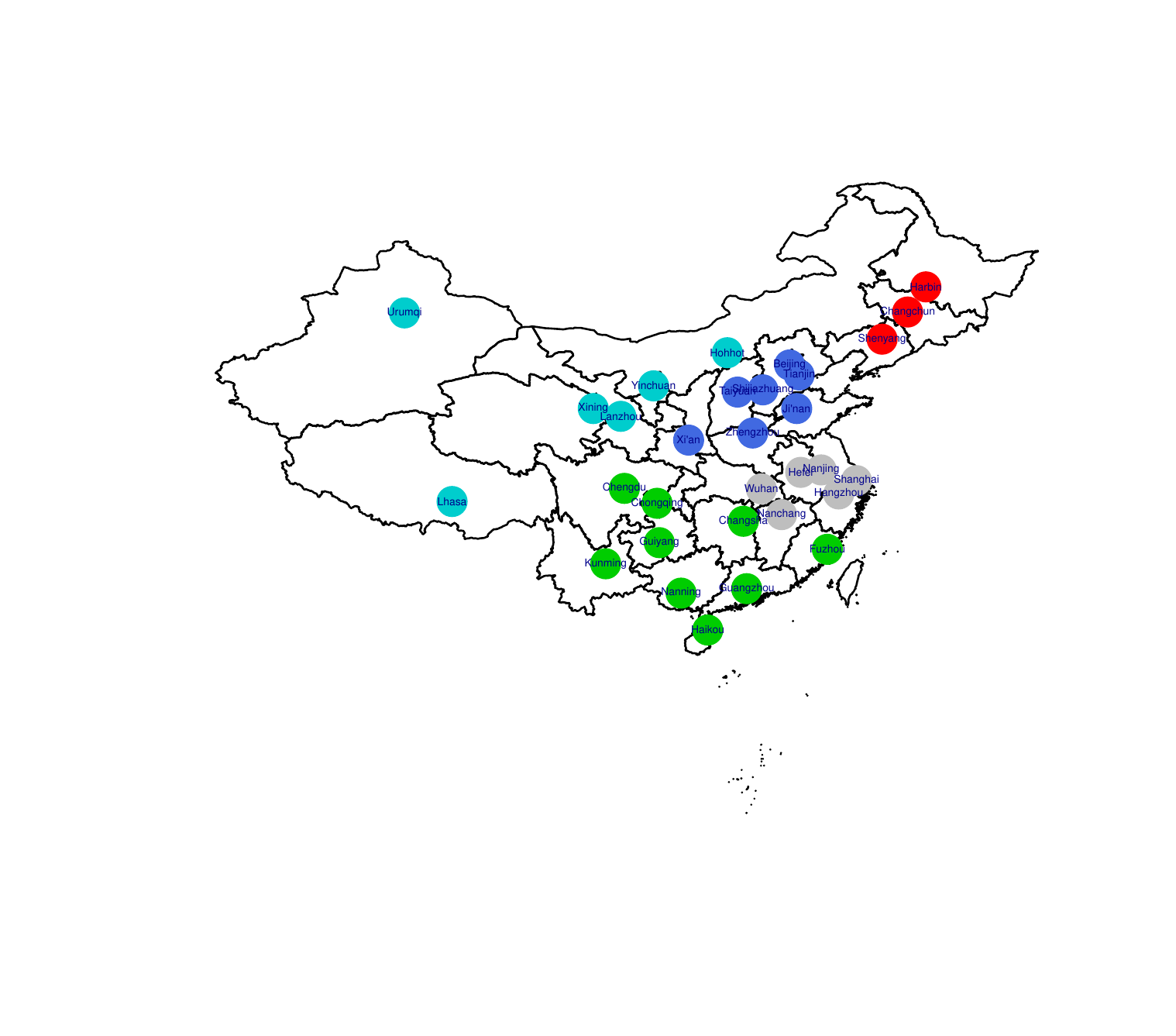}}\\
\vspace{0.5cm}
\subfigure[0.6486]{\includegraphics[height=2.6cm,width=2.8cm,angle=0]{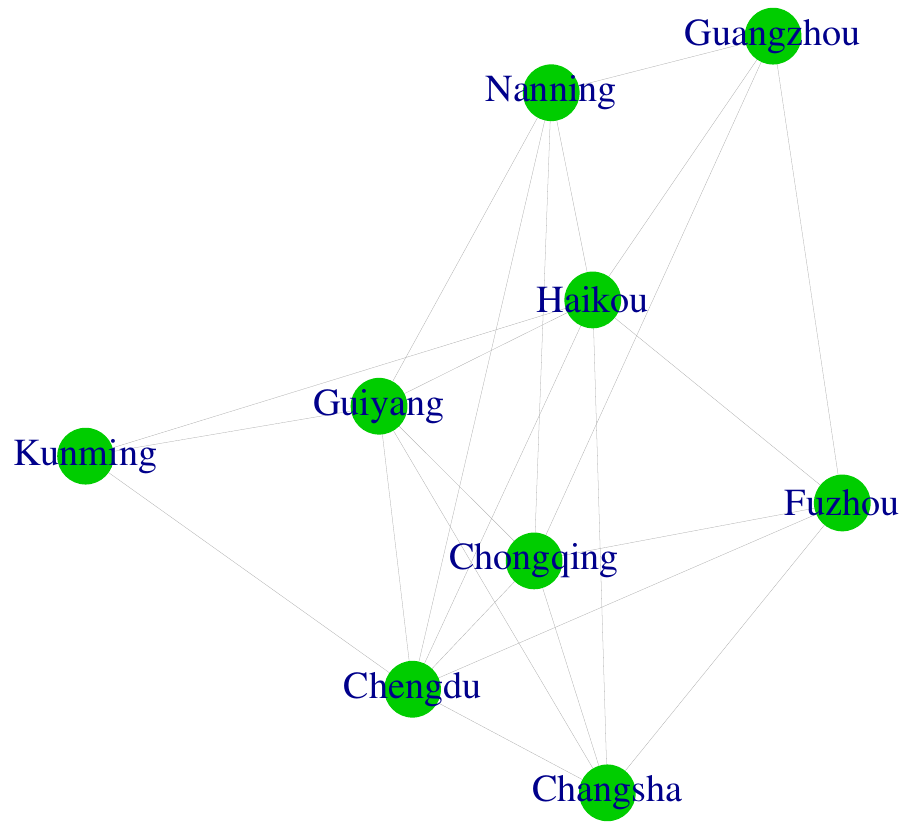}}
\hspace{0.01\textwidth}
\subfigure[0.7895]{\includegraphics[height=2.6cm,width=2.8cm,angle=0]{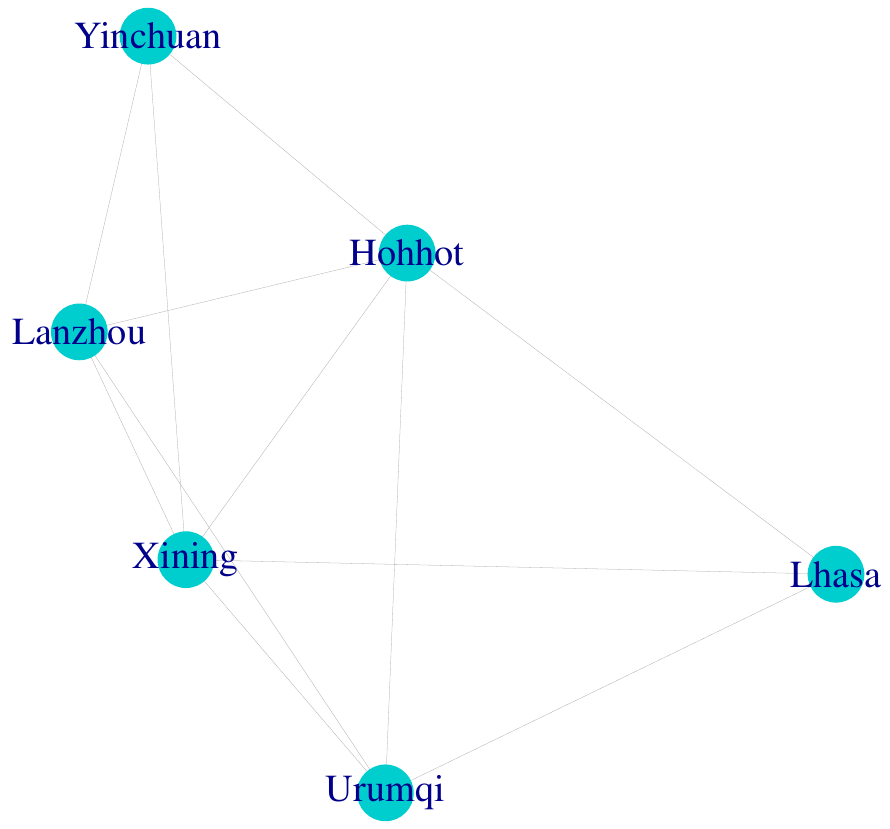}}
\hspace{0.01\textwidth}
\subfigure[0.8667]{\includegraphics[height=2.6cm,width=2.8cm,angle=0]{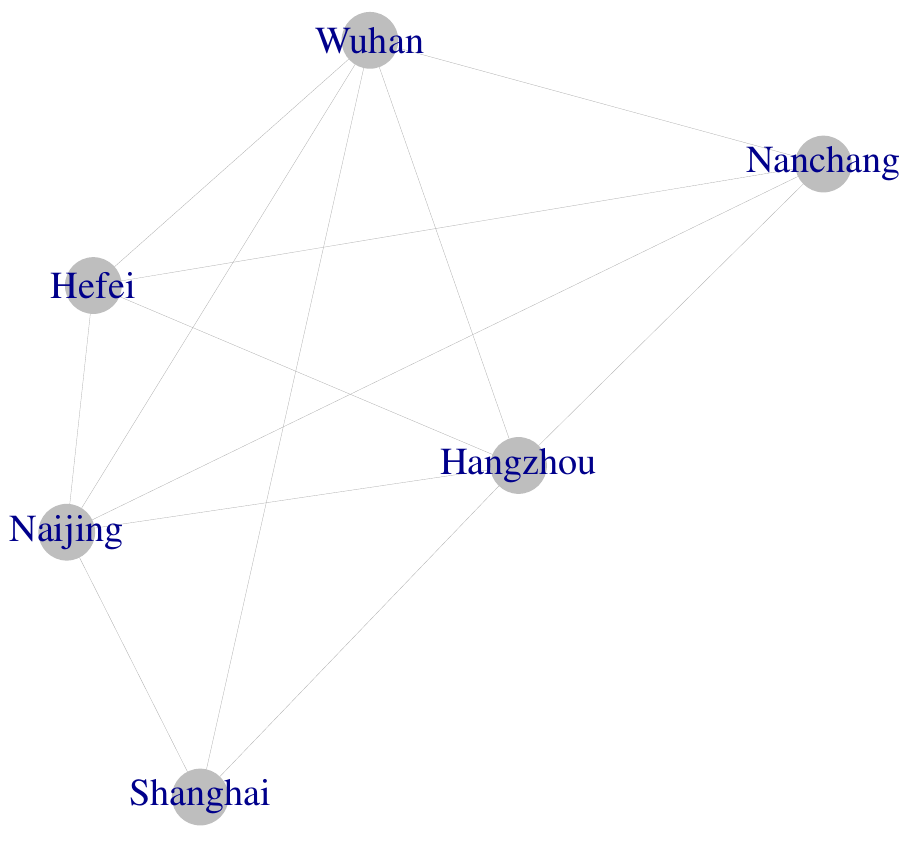}}
\hspace{0.01\textwidth}
\subfigure[1]{\includegraphics[height=2.6cm,width=2.8cm,angle=0]{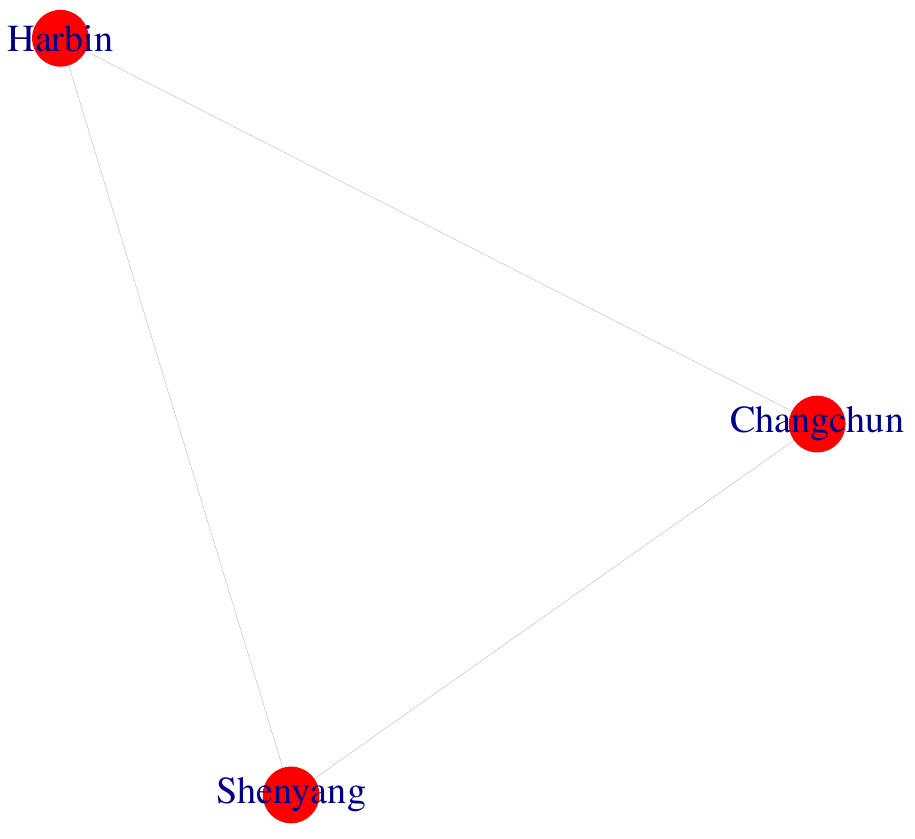}}
\hspace{0.01\textwidth}
\subfigure[0.7612]{\includegraphics[height=2.6cm,width=2.8cm,angle=0]{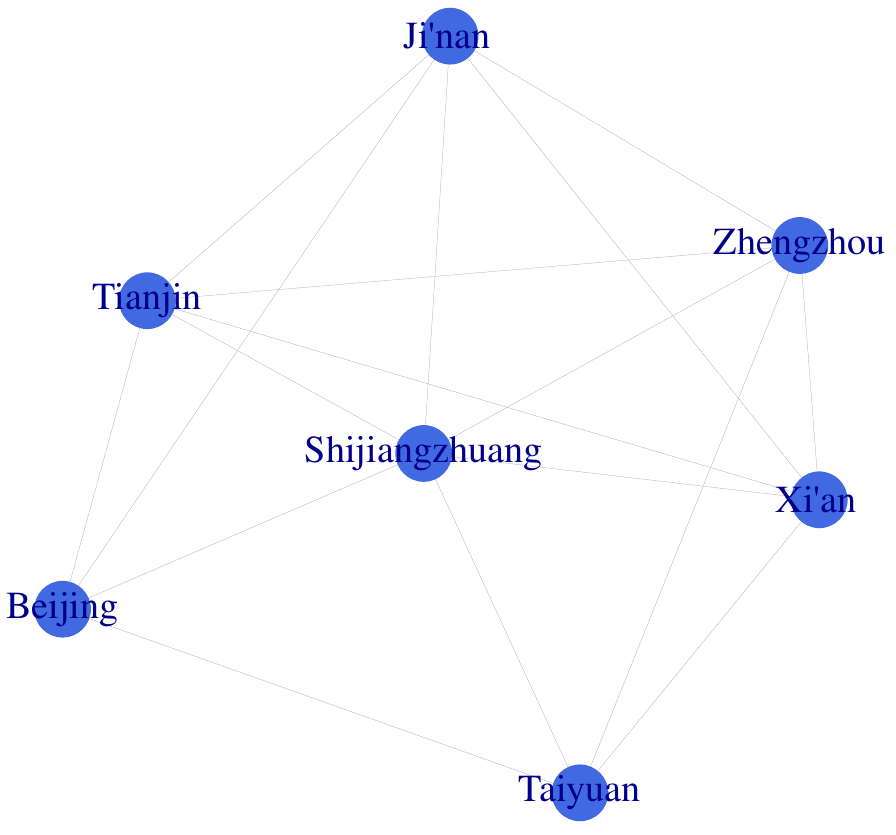}}
\caption{The clustering results for the ${\rm PM}_{2.5}$ data by the motif-based spectral clustering. (a) is the general clustering results, and (b)-(f) are specific communities where their clustering coefficients are also shown.
}\label{pmmotif}
\end{figure*}

\begin{figure*}[!htbp]{}
\centering
\subfigure[Detected communities]{\includegraphics[height=10cm,width=12.3cm,angle=0]{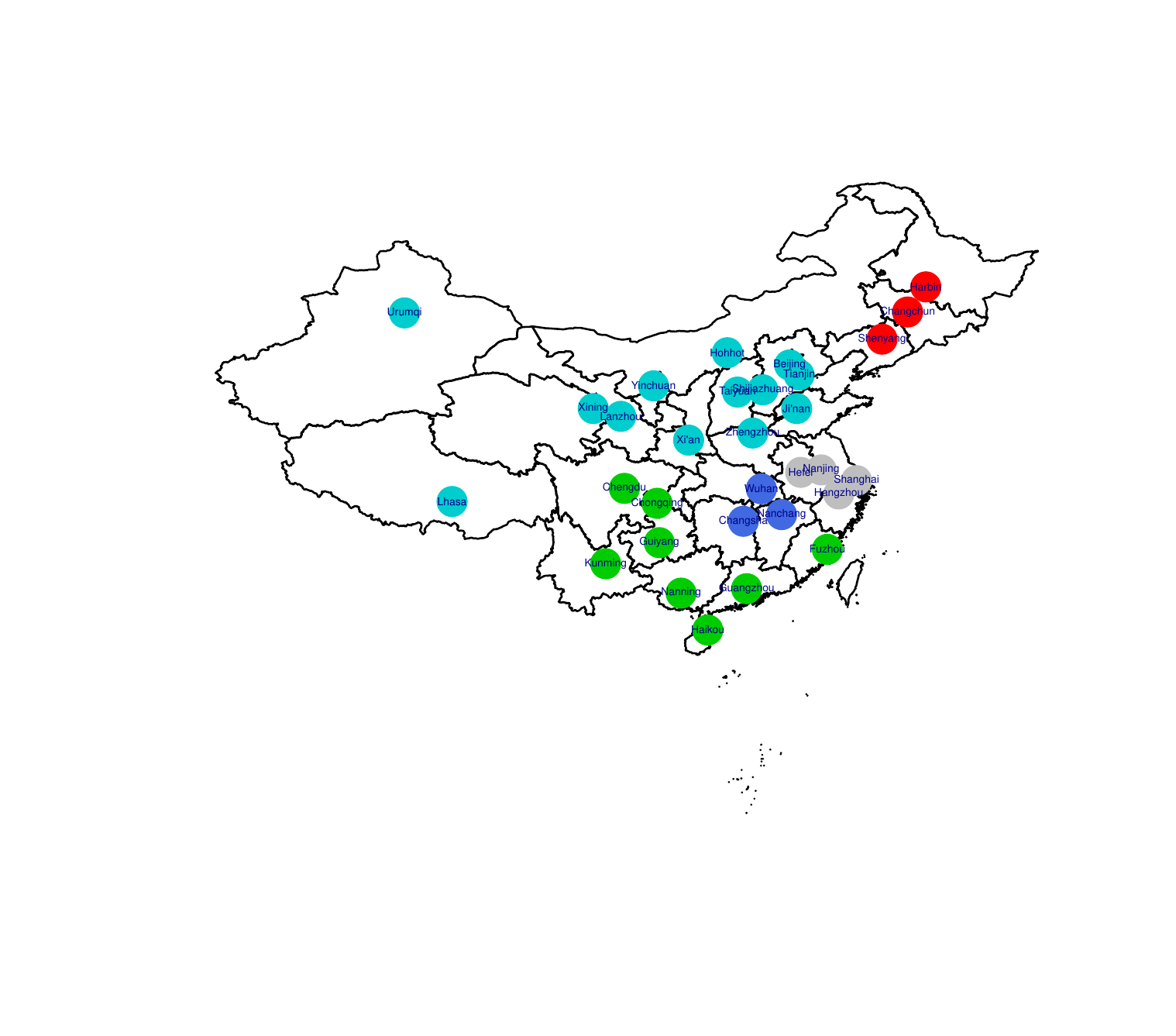}}\\
\vspace{0.5cm}
\subfigure[1]{\includegraphics[height=2.6cm,width=2.8cm,angle=0]{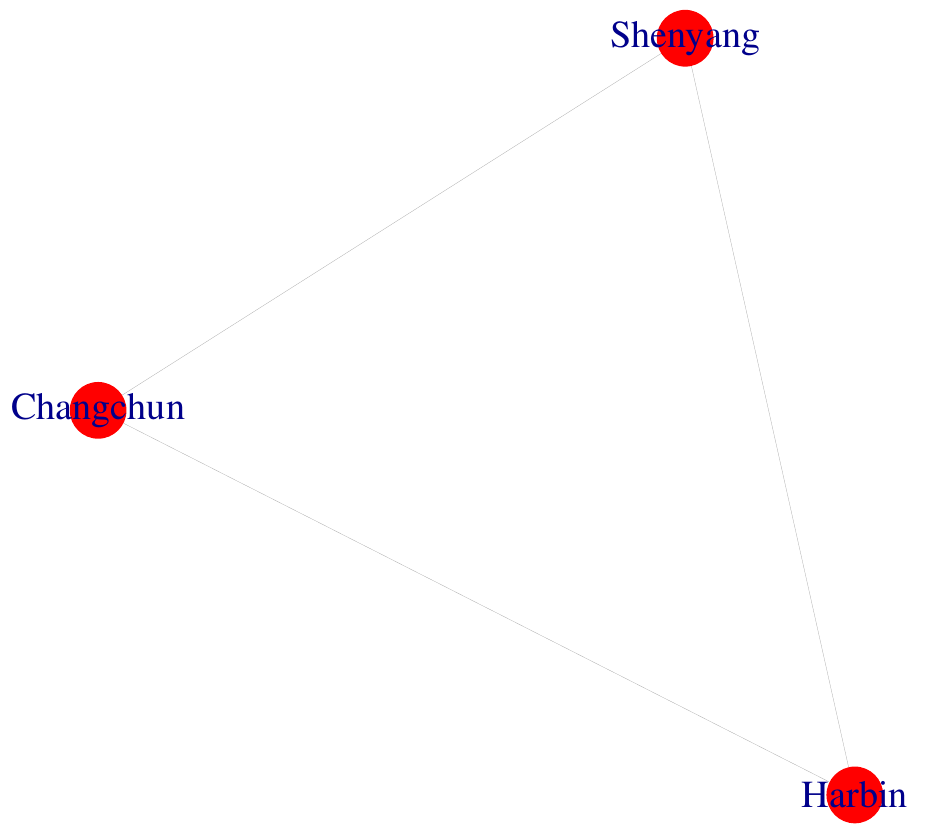}}
\hspace{0.01\textwidth}
\subfigure[0.6400]{\includegraphics[height=2.6cm,width=2.8cm,angle=0]{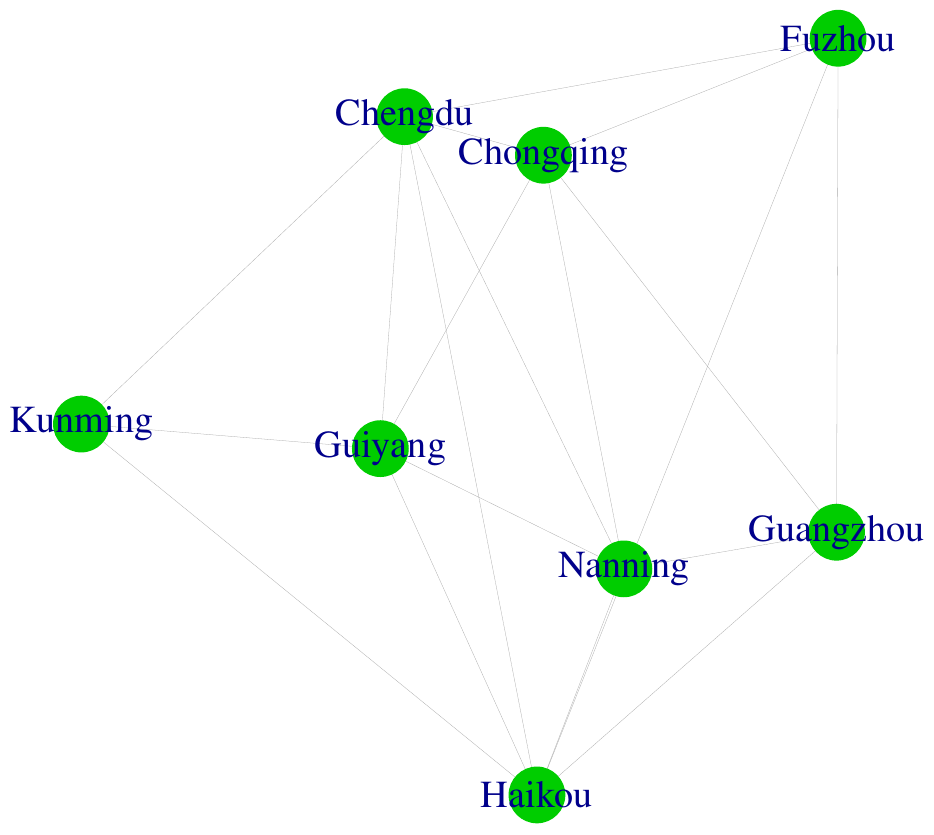}}
\hspace{0.01\textwidth}
\subfigure[1]{\includegraphics[height=2.6cm,width=2.8cm,angle=0]{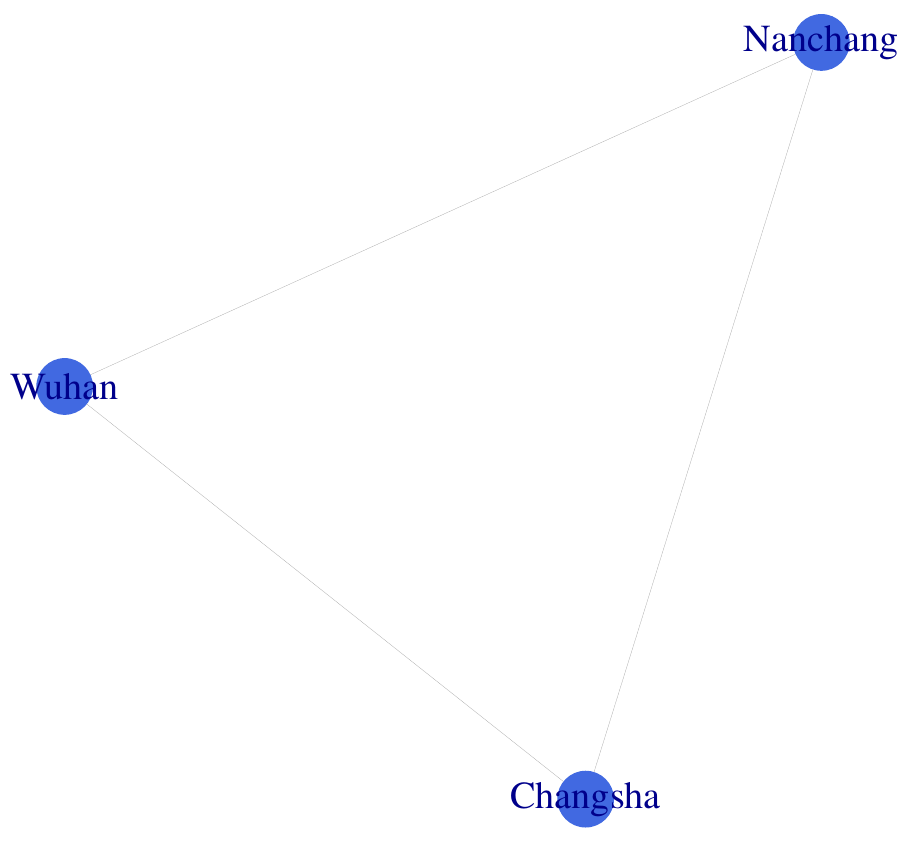}}
\hspace{0.01\textwidth}
\subfigure[0.6279]{\includegraphics[height=2.6cm,width=2.8cm,angle=0]{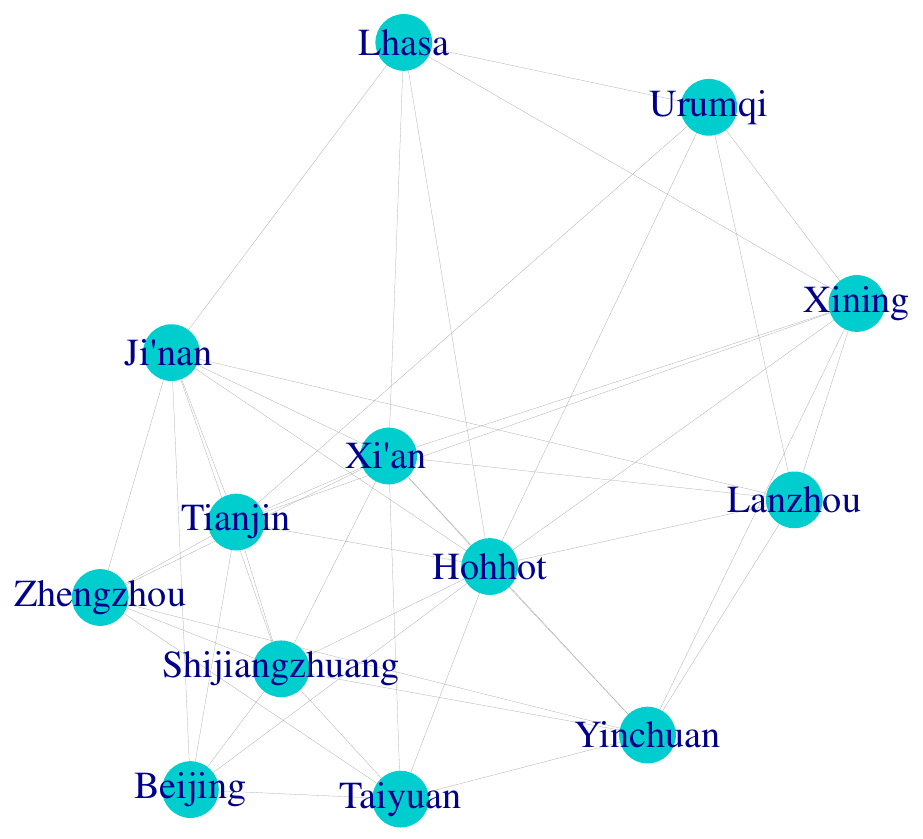}}
\hspace{0.01\textwidth}
\subfigure[0.75]{\includegraphics[height=2.6cm,width=2.8cm,angle=0]{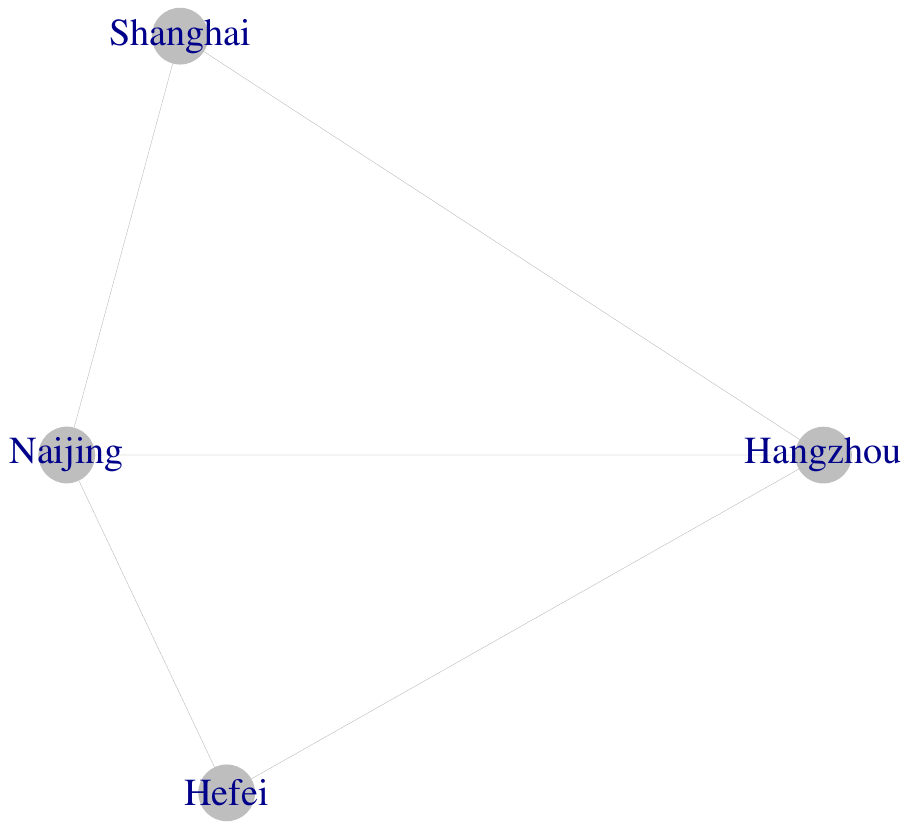}}
\caption{The clustering results for the ${\rm PM}_{2.5}$ data by the edge-based spectral clustering. (a) is the general clustering results, and (b)-(f) are specific communities where their clustering coefficients are also shown.
}\label{pmedge}
\end{figure*}

\textbf{${{\rm \bf PM}_{\bf 2.5}}$ pollution data.} {We collected the ${\rm PM}_{2.5}$ data which consists of daily averaged ${\rm PM}_{2.5}$ concentration in the year 2015 for each of the 31 Chinese capital cities. In more detail, for each city, we have its daily averaged ${\rm PM}_{2.5}$ concentrations for 354 days except 11 days containing missing data}. We aim to study the communities of cities in terms of the ${\rm PM}_{2.5}$ pollution. For this purpose, we should construct the ${\rm PM}_{2.5}$ pollution network first. Graphical models provide a useful tool for constructing the network from such data. Specifically, we treat the ${\rm PM}_{2.5}$ concentration of each city as a random variable (i.e., the node in the resulting network) and employ the graphical lasso \citep{yuan2007model,friedman2008sparse} with the tuning parameter selected by eBIC \citep{foygel2010extended} to obtain a weighted network, whereby the rationality of graphical models, the absolute weight between any pair of nodes (cities) is proportional to the conditional correlation of the two corresponding random variables, given the remaining variables. We set $K=5$, which is consistent with the output by the fast
greedy modularity optimization algorithm \citep{clauset2004finding} that decides the number of communities automatically. Figure \ref{pmmotif} and \ref{pmedge} show the clustering results corresponding to the motif-based and the edge-based spectral clustering, where for ease of interpretation, we show the cities on the Chinese map.
It can be seen from Figure \ref{pmmotif} and \ref{pmedge} that, both methods detect communities of cities such that cities within each community are closely located. It could be explained by the fact that the cities located closely share similar meteorological, economic, and industrial patterns, thus leading to the conditional dependence of their ${\rm PM}_{2.5}$ pollution. However, the communities detected by the two methods are different in some sense. There are two main differences. The first difference is that two communities detected by the motif-based method in Figure \ref{pmmotif}(c) and \ref{pmmotif}(f) are generally merged into one single community (see Figure \ref{pmedge}(e)) by the edge-based method. The second difference is that the community found by the motif-based method in Figure \ref{pmmotif}(d) is generally divided into two communities by the edge-based method (see Figure \ref{pmedge}(d) and \ref{pmedge}(f)). One who has common sense about Chinese geography will feel that the communities detected by the motif-based method are more reasonable. As an explanation, we conduct a two-sample t-test to test whether the average ${\rm PM}_{2.5}$ value of cities within the two communities in Figure \ref{pmmotif}(c) and \ref{pmmotif}(f) (motif-based communities) have a significant difference. The answer is yes, with the p-value smaller than $2.2\times 10^{-16}$. Similarly, we also test the difference between the two communities in Figure \ref{pmedge}(d) and \ref{pmedge}(f) (edge-based communities) in terms of the mean for ${\rm PM}_{2.5}$. Whereas the answer is no, with the p-value being 0.819. The above analysis indicates that the motif-based method yields more reasonable communities than the edge-based method does.

\begin{figure*}[!htbp]{}
\centering
\scriptsize
\subfigure[Citation network]{\includegraphics[height=4.2cm,width=5cm,angle=0]{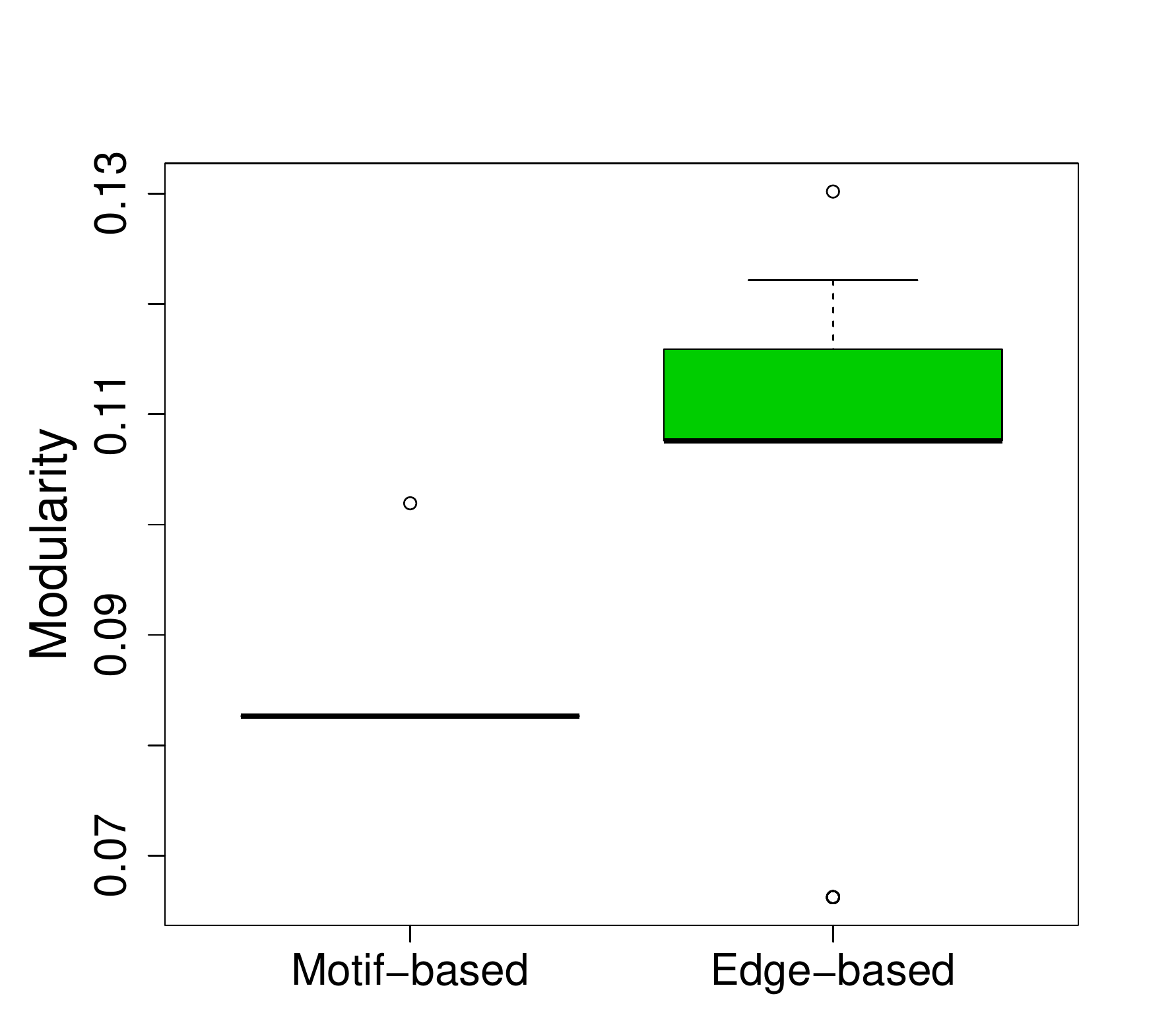}}
\subfigure[Coauthor network]{\includegraphics[height=4.2cm,width=5cm,angle=0]{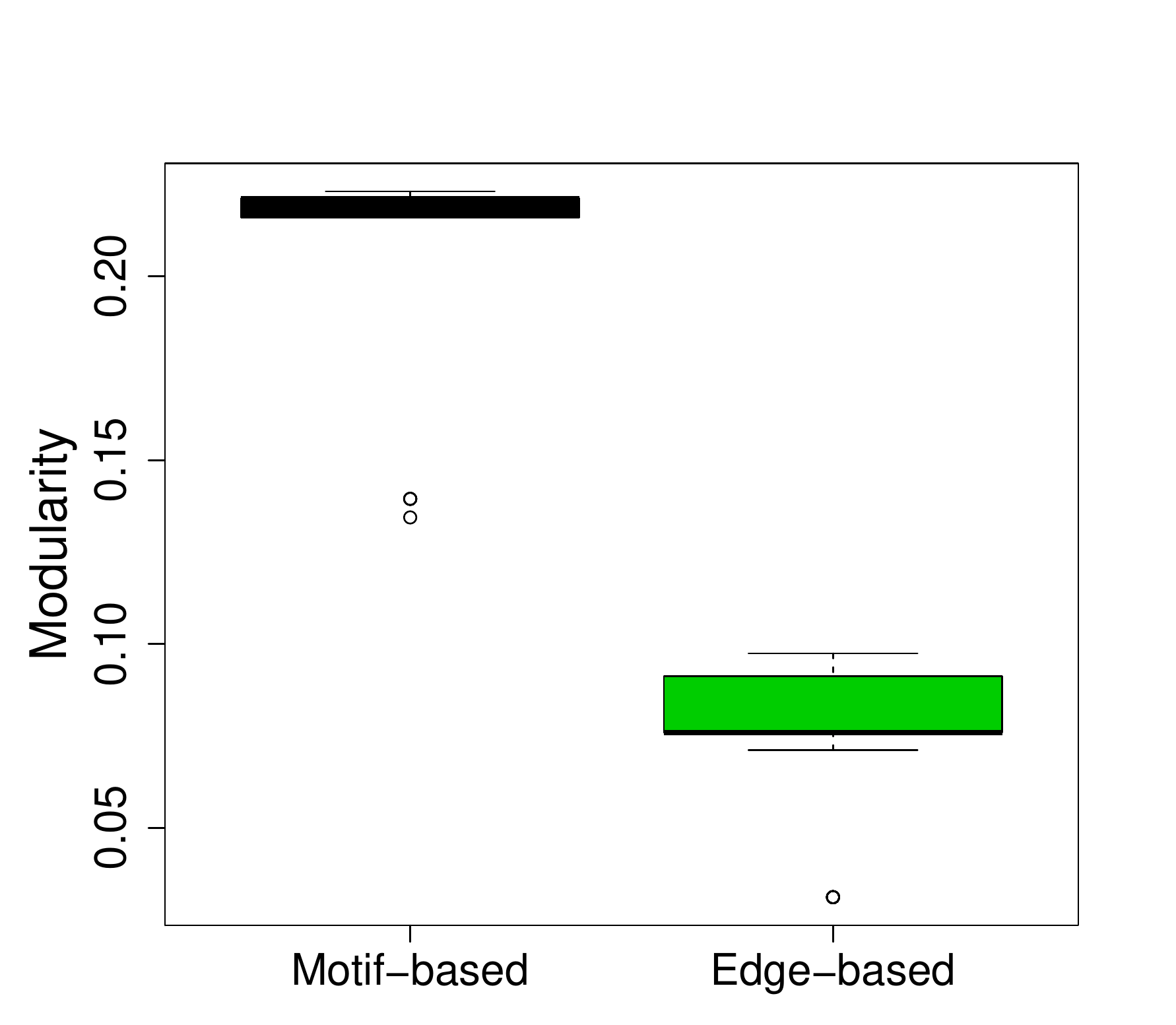}}
\subfigure[Pollution network]{\includegraphics[height=4.2cm,width=5cm,angle=0]{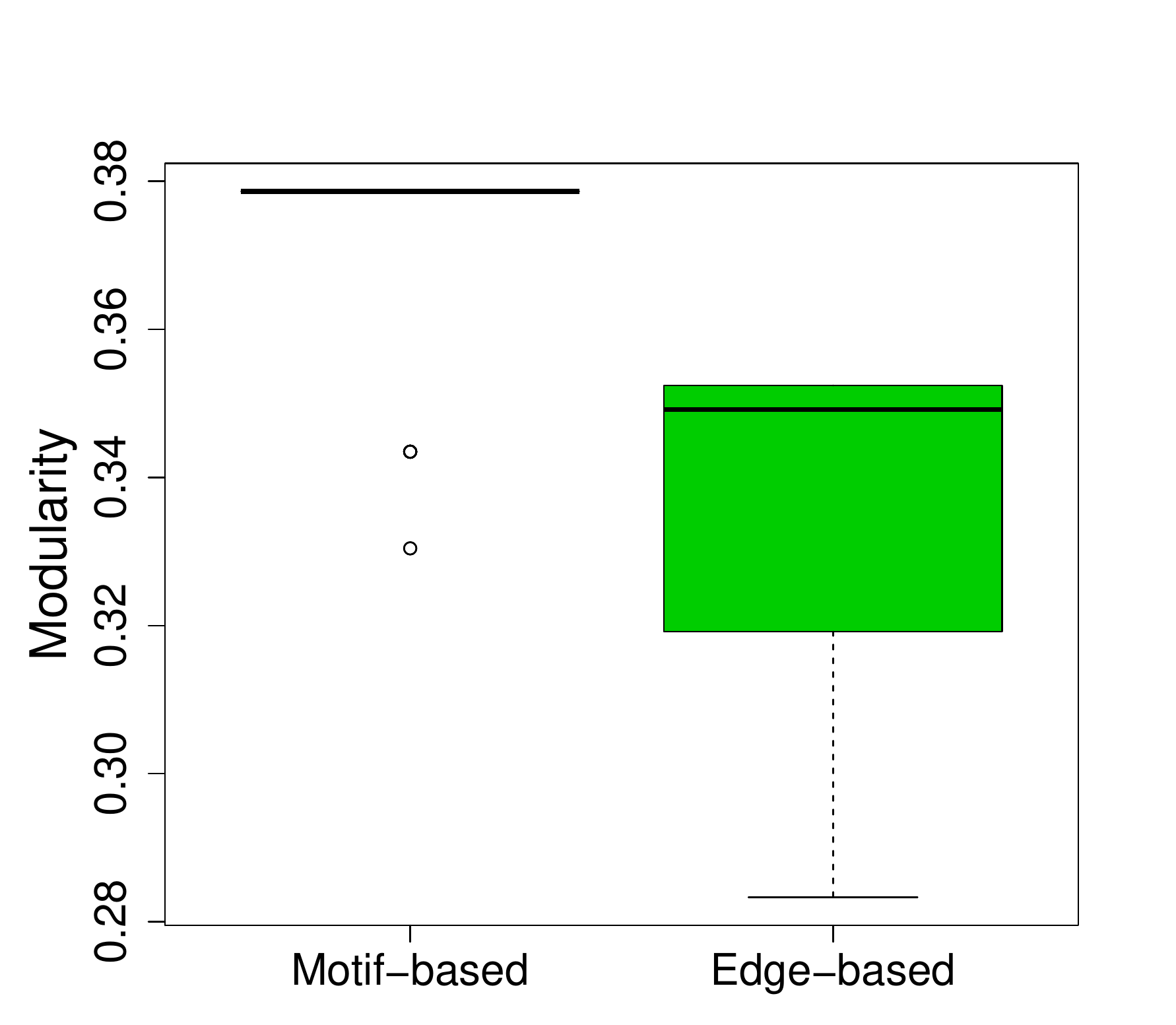}}\\
\caption{The box plot of modularity of two methods on three datasets over 100 replications. The modularity is computed based on the weighted motif adjacency matrix. }\label{modularityreal}
\end{figure*}

\textcolor{black}{\paragraph{Comparison of modularity.} The aforementioned analysis shows that the motif-based spectral clustering can yield more interpretable communities than the edge-based counterpart, which is consistent with the results in \citet{benson2016higher}. As a complement, we also compute the modularity of the communities obtained by both methods; see Section \ref{simulation} for details of modularity. Figure \ref{modularityreal} displays the box plot of the modularity of two methods on three datasets over 100 replications. It turns out that the motif-based method has slightly larger modularity than the edge-based counterpart on the latter two datasets. While it is slightly worse than the edge-based one on the first dataset. In addition, it is obvious that the motif-based method is far more stable than the edge-based method, which makes the motif-based method more desirable in real applications.}

\section{Conclusion}
\label{conclusion}
{
The higher-order structure of networks and the corresponding higher-order spectral clustering have been proven to be insightful and helpful in real network clustering problems. However, there are rare works disclosing the merits of higher-order spectral clustering over the edge-based one systematically and theoretically. Motivated by this, \textcolor{black}{we theoretically studied the clustering performance of the higher-order spectral clustering from a statistical perspective}, where we typically assumed the underlying networks are generated from a weighted stochastic block model. The results showed that when the network is dense and the signal of weights is weak, the higher-order spectral clustering could produce better \textcolor{black}{misclustering upper bound} than the edge-based counterpart.} The methodologies and theories in this paper can be generally used in other problems. For example, the weighted motif adjacency matrix can be thought of as a network denoising trick and can be used for any downstream network analysis. In addition, the theoretical tools in this paper can be used to analyze networks with summary statistics but without edge information, which is of independent interest in the context of privacy-preserving network analysis.

There are many ways that the content of this work could be extended. First, we only specify the expectation of weights in the WSBM. It would be beneficial to derive more delicate results if the distribution of weights is incorporated. Second, we found by experiments that higher-order spectral clustering also works better than edge-based under the degree-corrected stochastic block models. It would be interesting to understand this phenomenon theoretically, which is currently under our consideration. Third, we considered triangles in undirected networks. More types of higher-order motifs and the clustering in directed networks deserve further research. \textcolor{black}{Besides, it is of interest to see whether the motif-based method can be generalized to the clustering of multi-layer networks \citep{paul2020spectral}.} Finally, we take a step forward in understanding how higher-order spectral clustering works, and there is still an urgent need to understand it thoroughly. In addition, it is crucial to develop a network-generating model that really captures the higher-order connectivity patterns in real networks.

\section*{Acknowledgement}
This work is partially done when Xiao Guo was a visiting student at Columbia University. The authors would like to thank Professor Ming Yuan for enlightening discussions in the early stage of this project. Our research is partially supported by National Natural Science Foundation for Outstanding Young Scholars (No.72122018), National Natural Science Foundation of China (No.U1811461), and Natural Science Foundation of Shaanxi Province (No. 2021JQ-429 and No.2021JC-01).


\newpage
\begin{center}
{\large\bf SUPPLEMENTARY MATERIAL}
\end{center}
\begin{description}
\item This file includes the proofs for theorems and the figures on clustering results of real data analysis.
\end{description}
\section*{A. Proof of main results}
\subsection*{\textbf{Proof of Lemma \ref{lemma}}}
Let $\Delta={\rm diag}(\sqrt{n_1},...,\sqrt{n_K})$ and recall (\ref{2.6}), then we have
\begin{align*}
\mathcal W^M &=\Theta((h_1-h_2)I_K+h_21_K1_K^\intercal)\Theta^\intercal\\
&=\Theta\Delta^{-1}\Delta((h_1-h_2)I_K+h_21_K1_K^\intercal)\Delta(\Theta\Delta^{-1})^\intercal,
\end{align*}
where $h_1>h_2>0$. Recalling the structure of $\Theta$, one can easily show $\Theta\Delta^{-1}$ is orthonormal. Denote the eigen-decomposition of $\Delta(h_1-h_2)I_K+h_21_K1_K^\intercal\Delta$ as
$$Z\Sigma Z^\intercal =\Delta((h_1-h_2)I_K+h_21_K1_K^\intercal)\Delta.$$
Then we have
$$\mathcal W^M=\Theta\Delta^{-1}Z\Sigma (\Theta\Delta^{-1}Z)^\intercal= U\Sigma U^\intercal.$$
Thus $U=\Theta\Delta^{-1}Z=\Theta X$ with $X:=\Delta^{-1}Z$ as $\Theta\Delta^{-1}$ and $Z$ are orthonormal, respectively. Moreover, as $Z$ is orthonormal and $\Delta^{-1}Z$ is a square matrix, we can verify that the rows of $\Delta^{-1}Z$ are perpendicular to each other and the $k$th row has length $\sqrt{1/n_k}$, which implies $\|X_{k\ast}-X_{l\ast}\|_2=\sqrt{n_k^{-1}+n_l^{-1}}$.\QEDA

\subsection*{\textbf{Proof of Theorem \ref{spbound}}}
The general framework of the following proof is adapted from \cite{feige2005spectral}, whose arguments are also used in \cite{lei2015consistency,paul2018higher,chin2015stochastic,gao2017achieving}, among others. The proof includes three major steps. \\
\textbf{Step 1:} \textbf{Discretization.} We first reduce controlling $\|W^M-\mathbb E(W^M)\|_2$ to the problem of bounding the supremum of $|x^\intercal (W^M-\mathbb E(W^M))y|$ over all pairs of vectors $x,y$ in a finite set of grid points. For any given pair $(x,y)$ in the grid, the quantity $x^\intercal (W^M-\mathbb E(W^M))y$ is decomposed into the sum of two parts. The first part corresponds to the small entries of both $x$ and $y$, called the \textit{light pairs}, the other part corresponds to large entries of $x$ or $y$, called the \textit{heavy pairs}.  \\
\textbf{Step 2: Bounding the light pairs.} Since the elements of $W^M$ are dependent, we can not use the Bernstein's inequality to control the contribution of the light pairs. Instead, we make use of the typical bounded differences inequality established in \cite{warnke2016method} to bound the light pairs. Note that \cite{paul2018higher}  also used similar arguments to do this in the unweighted SBMs, but the details are different from those we do in the weighted case.\\
\textbf{Step 3: Bounding the heavy pairs.} The contribution from heavy pairs will be bounded using a combinatorial argument on the event that the edge weights (w.r.t. the  weighted motif adjacency matrix $W^M$) in a collection of subgraphs do not deviate much from their expectation. To that end, the concentration inequality in \cite{warnke2017upper} will be used to bound the edge weighted degree.

It should be noted that throughout the proof, we will use $c$, $c'$, $c''$ likewise to denote constants which may be different from line to line.   Now we proceed to prove Theorem \ref{spbound}. Let $S$ denote the unit ball in the $n$-dimensional Euclidean space. Fix $\epsilon\in (0,1)$, for example $\epsilon=\frac{1}{2}$, define an $\epsilon$-net of the ball as follows:
\begin{equation}
\label{A.1}
\mathcal N=\{x=(x_1,...,x_n)\in S: \forall i, \sqrt{n}x_i/\epsilon\in \mathbb Z\},
\tag{A.1}
\end{equation}
where $\mathbb Z$ denotes the set of all integers. Thus $\mathcal N$ consists of all grid points of size $\epsilon\sqrt{n}$ within the unit ball. The Lemma 2.1 in \cite{lei2015consistency} shows that for any $A\in \mathbb R^{n\times n}$,
\begin{equation}
\label{A.2}
\|A\|_2\leq (1-\epsilon)^{-2}\underset{x,y\in \mathcal N}{{\rm sup}}|x^\intercal Ay|.\tag{A.2}
\end{equation}
Therefore, to bound $\|W^M-\mathbb E(W^M)\|_{2}$, we only need to bound $|x^\intercal (W^M-\mathbb E(W^M))y|$ over all possible pairs $(x,y)\in \mathcal N$. For any pair of vector $(x,y)$, we have
\begin{equation}
\label{A.3}
x^\intercal (W^M-\mathbb E(W^M))y=\sum_{i,j}x_iy_j(W^M-\mathbb E(W^M))_{ij}.\tag{A.3}
\end{equation}
We split the pairs $(x_i,y_j)$ into \emph{light pairs }
$${\mathcal L=\mathcal L(x,y)=\{(i,j):|x_iy_j|\leq \frac{\sqrt{D}}{n\tau_{{\rm max}}}\},}$$
and into \emph{heavy pairs }
$${\mathcal H=\mathcal H(x,y)=\{(i,j):|x_iy_j|> \frac{\sqrt{D}}{n\tau_{{\rm max}}}\},}$$
where $D$ and $\tau_{{\rm max}}$ are defined in Theorem \ref{spbound}. For the light pairs, we have the following
results.

\begin{lemma}[\emph{Light pairs}]\label{light}
For some constant $r_1>0$, there exists a constant $c_1(r_1)>0$ such that with probability at least $1-n^{-r_1}$,
\begin{equation}
\label{A.4} \underset{x,y\in \mathcal N}{{\rm sup}}\big |\underset{(i,j)\in \mathcal L}{\sum}x_iy_j(W^M-\mathbb E(W^M))_{ij}\big|\leq c_1\sqrt{D}.\tag{A.4}
\end{equation}
\end{lemma}

To bound the contribution of heavy pairs, we first have the following observation,
\begin{align}
\label{A.5} &\underset{x,y\in \mathcal N}{{\rm sup}}\big |\underset{(i,j)\in \mathcal H}{\sum}x_iy_j(W^M-\mathbb E(W^M))_{ij}\big| \leq \underset{x,y\in \mathcal N}{{\rm sup}}\big |\underset{(i,j)\in \mathcal H}{\sum}x_iy_jW^M_{ij}\big|+\underset{x,y\in \mathcal N}{{\rm sup}}\big |\underset{(i,j)\in \mathcal H}{\sum}x_iy_j\mathbb E(W^M_{ij})\big|.
\tag{A.5}
\end{align}
Recall that $\mathbb E(W_{ij})\leq \alpha_np_{n}$, then
\begin{align}
\label{A.6} \mathbb E(W^{M}_{ij})&=\mathbb E \sum_{k}\mathbf 1(W_{ij}\cdot W_{ik}\cdot W_{jk}>0)(W_{ij}+W_{ik}+W_{jk})\leq 3n\alpha_np_{n}^3.
\tag{A.6}
\end{align}
Then the second term in (\ref{A.5}) can be bounded as follows,
\begin{align}
\label{A.7}
\big |\underset{(i,j)\in \mathcal H}{\sum}x_iy_j\mathbb E(W^M_{ij})\big| &\leq \underset{(i,j)\in \mathcal H}{\sum}\frac{x_i^2y_j^2}{|x_iy_j|}\mathbb E(W^M_{ij})  \nonumber\\
&\leq \frac{n\tau_{{\rm max}}}{\sqrt{D}} {\rm max}_{ij}\mathbb E(W^M_{ij})\sum_{(i,j)}x_i^2y_j^2\nonumber\\
&\leq {\frac{n\tau_{{\rm max}}}{\sqrt{D}}3n\alpha_np_{n}^3\leq \frac{cD}{\sqrt{D}}\leq c\sqrt{D}},
\tag{A.7}
\end{align}
where the second inequality follows from the definition of heavy pairs $\mathcal H$, the third inequality follows from (\ref{A.6}) and the vectors $x$ and $y$ are within the unit ball, and the penultimate inequality follows since
\begin{equation}
\label{A.8}
n\tau_{{\rm max}}\cdot 3n\alpha_np_{n}^3\leq cn^{3}\alpha_np_{n}^5\leq c D,
\tag{A.8}
\end{equation}
by recalling $\tau_{\rm max}=np_{n}^2$.
For the first term in (\ref{A.5}), we have the following result.
\begin{lemma}[\emph{Heavy pairs}]\label{heavy}
For some constants $r_2>0$, $r_3>0$, and any $0<\beta<1$, there exists a constant $c_2(r_2,r_3,\beta)>0$ such that with probability at least $1-n^{-r_2}-{\rm exp}(-r_3({\rm log} n)^\beta)$,
\begin{equation}
\label{A.9} \underset{x,y\in \mathcal N}{{\rm sup}}\big |\underset{(i,j)\in \mathcal H}{\sum}x_iy_jW^M_{ij}\big|\leq c_2\sqrt{D}.\tag{A.9}
\end{equation}
\end{lemma}
Combining the results for the light and heavy pairs and recalling (\ref{A.2}), we obtain
\begin{align}
\label{A.10} \|W^M-\mathbb E(W^M)\|_{2}&\leq 4\underset{x,y\in \mathcal N}{{\rm sup}}\big |{\sum}x_iy_j(W^M-\mathbb E(W^M))_{ij}\big|\leq c\sqrt{D}.\tag{A.10}
\end{align}
\QEDA

\subsection*{\textbf{Proof of Theorem \ref{mis}}}
We make use of the framework in \cite{lei2015consistency} to bound the misclustering error rate. To fix ideas, we give some notation now. $U$ and $\tilde{U}$ denote the $K$ leading eigenvectors of $\mathcal W^M$ and $W^M$, respectively. $\hat{{U}}:=\hat{{\Theta}}\hat{{Y}}$ corresponds to the optimal solution of the higher-order spectral clustering algorithm in that,
\begin{equation}
(\hat{{\Theta}}, \hat{{Y}})=\underset{{\Theta\in  \mathbb M_{n,K},Y\in \mathbb R^{K\times K}}}{{\rm arg\;min}}\;\|\Theta Y-\tilde{U}\|_{\rm \tiny F}^2.\nonumber
\end{equation}
By Lemma \ref{lemma}, in the WSBMs, two nodes are in the same community if and only if the corresponding rows of the population eigenvector matrix $U$ are the same. Building on this result, in what follows, we first bound the deviation of $\hat{U}$ from $U$. Then, within each true cluster, we bound the size of nodes that correspond to a large deviation of $\hat{U}$ from $U$, we bound their size. After that, we show for the remaining nodes that the estimated and true communities are consistent.

First, we bound the deviation of $\hat{U}$ from $U$. Davis-Kahan $\rm sin \Theta$ theorem (Theorem VII.3.1 of \cite{bhatia1997graduate}) provides a useful technical tool for bounding the perturbation of eigenvectors from the perturbation of matrices. In particular, by Theorem 3.1 of \cite{lei2015consistency}, there exists a $K\times K$ orthogonal matrix $O$ such that,
\begin{equation}
\label{A.11}
\|\tilde{U}- UO\|_{\tiny \rm F}\leq \frac{2\sqrt{2K}}{\lambda_K(\mathcal W^M)}\|W^M-\mathcal W^M\|_2,\tag{A.11}
\end{equation}
where we recall that $\lambda_K(\mathcal W^M)$ denotes the minimum non-zero eigenvalue of the population matrix $\mathcal W^M$. Now we proceed to bound the deviation of $\hat{{U}}$ from $U$. Note that
\begin{align}
\label{A.12}
\|\hat{U}- UO\|_{\tiny \rm F}^2&=\|\hat{{U}}-\tilde{U}+\tilde{U}-UO\|_{\tiny \rm F}^2\nonumber \\
& \leq \|UO-\tilde{U}\|_{\tiny \rm F}^2+\|\tilde{U}-UO\|_{\tiny \rm F}^2 \nonumber \\
&=2\|\tilde{U}-UO\|_{\tiny \rm F}^2,\tag{A.12}
\end{align}
where the first inequality follows from our assumption that $\hat{{U}}$ is the global solution of the higher-order spectral clustering algorithm and $UO$ is a feasible solution. Then combine (\ref{A.12}) with (\ref{A.11}), we have
\begin{align}
\label{A.13}
\|\hat{{U}}- UO\|_{\tiny \rm F}^2\leq \frac{16K}{\lambda_K^2(\mathcal W^M)}\|W^M-\mathcal W^M\|_2^2.
\tag{A.13}
\end{align}
Now we calculate the terms on the RHS of (\ref{A.13}). Recall the definition of $\mathcal W^M$ in (\ref{2.6}) and the definitions of $h_1$ and $h_2$ in (\ref{2.4}) and (\ref{2.5}), we can easily obtain
\begin{equation}
\label{A.14}
\lambda_K(\mathcal W^M)=\frac{n}{K}(h_1-h_2)\asymp \frac{n^2\alpha_np_n^3}{K},
\tag{A.14}
\end{equation}
and
\begin{align}
\label{A.15}
\|W^M-\mathcal W^M\|_2&=\|W^M-\mathbb E(W^M)-{\rm diag}(\mathcal W^M)\|_2\asymp \|W^M-\mathbb E(W^M)\|_2.
\tag{A.15}
\end{align}
Combining (\ref{A.14}), (\ref{A.15}), and the bound in Theorem \ref{spbound} with (\ref{A.13}), we have
\begin{align}
\label{A.16}
\|\hat{{U}}- UO\|_{\tiny \rm F}^2\leq \frac{c16K^3}{n\alpha_np_n}:={\rm err}(K,n,c,\alpha_n,p_n),
\tag{A.16}
\end{align}
where for notational convenience, we use ${\rm err}(K,n,c,\alpha_n,p_n)$ to denote ${c16K^3}/{n\alpha_np_n}$ in what follows.

In the sequel, we proceed to bound the fraction of misclustered nodes within each true cluster. By Lemma \ref{lemma}, we can write $U=\Theta X$, where $\|X_{k\ast}-X_{l\ast}\|_2=\sqrt{n_{k}^{-1}+n_{l}^{-1}}$ for all $1\leq k<l\leq K$. Hence $UO=\Theta XO=\Theta X'$ with $X'=XO$, and $\|X'_{k\ast}-X'_{l\ast}\|_2=\sqrt{n_{k}^{-1}+n_{l}^{-1}}$ by the orthogonality of $O$. Define
\begin{equation}
\label{A.17}
\delta_k={\rm min}_{l\neq k}\;\|X'_{k\ast}-X'_{l\ast}\|_2=\sqrt{\frac{1}{n_k}+\frac{1}{{\rm max}\{n_l:l\neq k\}}},
\tag{A.17}
\end{equation}
and
\begin{equation}
\label{A.18}
S_k=\{i\in G_k(\Theta):\; \|(\hat{U})- (UO)_{i\ast}\|_{\tiny \rm F}>\frac{\delta_k}{2}\},
\tag{A.18}
\end{equation}
where $S_k$ is essentially the number of misclustered nodes in the true cluster $k$ (after some permutation) as we will see soon. By the definition of $S_k$, it is obvious
to see
\begin{equation}
\label{A.19}
\sum_{k=1}^K|S_k|\delta_k^2/4\leq \|\hat{{U}}- UO\|_{\tiny \rm F}^2={\rm err}(K,n,c,\alpha_n,p_n).
\tag{A.19}
\end{equation}
Recall that $\delta_k={\rm min}_{l\neq k}\;\|X'_{k\ast}-X'_{l\ast}\|_2=\sqrt{\frac{1}{n_k}+\frac{1}{{\rm max}\{n_l:l\neq k\}}}$, so
$n_k\delta_k^2\geq 1$. Therefore, (\ref{A.19}) entails that
\begin{equation}
\label{A.20}
\sum_{k=1}^K\frac{|S_k|}{n_k}\leq 4\cdot\|\hat{{U}}- UO\|_{\tiny \rm F}^2=4\cdot{\rm err}(K,n,c,\alpha_n,p_n).
\tag{A.20}
\end{equation}

At last, we show that the nodes within in true community $k$ (after some permutation) but outside $S_k$ are correctly clustered. Before moving on, we first prove $|S_k|<n_k$. We have by (\ref{A.19}) that
\begin{equation}
\label{A.21}
|S_k|\leq \frac{4}{\delta_k^2}{\rm err}(K,n,c,\alpha_n,p_n).
\tag{A.21}
\end{equation}
As $n_k\delta_k^2\geq 1$, it suffices to prove
\begin{equation}
\label{A.22}
4\cdot{\rm err}(K,n,c,\alpha_n,p_n)<1.
\tag{A.22}
\end{equation}
which actually follows from the assumption (\ref{3.3}) after some modification of $c$ ($c$ can be different from line to line). As a result, we have
$|S_k|<n_k$ for every $1\leq k\leq K$. And thus, $T_k\equiv G_k\backslash S_k\neq \emptyset$, where we recall that $G_k$ denotes the nodes in the true community $k$. Let $T=\cup _{k=1}^KT_k$, we now show that the rows in $(UO)_{T\ast}$ has a one to one correspondence with those in $\hat{{U}}_{T\ast}$. On the one hand, for $i\in T_k$ and $j\in T_l$ with $l\neq k$,
$\hat{{U}}_{i\ast}\neq \hat{{U}}_{j\ast}$, otherwise one can have the following contradiction
\begin{align}
\label{A.23}
{\rm max} \{\delta_k,\delta_l\}&\leq \|(UO)_{i\ast}-(UO)_{j\ast}\|_2\nonumber\\
&\leq \|(UO)_{i\ast}-\hat{{U}}_{i\ast}\|_2+\|(UO)_{j\ast}-\hat{{U}}_{j\ast}\|_2 \nonumber\\
&<\frac{\delta_k}{2}+\frac{\delta_l}{2},
\tag{A.23}
\end{align}
where the first and last inequality follows from (\ref{A.17}) and (\ref{A.18}), respectively. On the other hand, for $i,j\in T_k$, we have
$\hat{{U}}_{i\ast}= \hat{{U}}_{j\ast}$, because otherwise $U_{T\ast}$ has more than $K$ distinct rows contradicting the fact that the output community size is $K$.

As a result, we have proved the membership is correctly recovered outside of $\cup _{k=1}^K S_k$ and the rate of misclustered nodes in $S_k$ is bounded as in (\ref{A.20}), and therefore the conclusion of Theorem \ref{mis} is arrived.\QEDA

\section*{B. Proof of auxiliary Lemmas}
\subsection*{\textbf{Proof of Lemma \ref{light}}}
We will make use of the typical bounded differences inequality established in \cite{warnke2016method} to control the contribution of the light pairs. To fix ideas, we reproduce the result that we will use in the following proposition.
\begin{proposition}[Theorem 1.2 of \cite{warnke2016method}]\label{typicalbound}
Let $X=(X_1,...,X_N)$ be a family of independent random variables with $X_k$ taking values in a set $\Lambda_k$.
Let $\Gamma\subseteq \prod_{j\in [N]}\Lambda_j$ be an event and assume that function $f:\prod_{j\in [N]}\Lambda_j \rightarrow \mathbb R$ satisfies the following {\rm typical Lipschitz condition}. \\
{\rm(TL)} There are numbers $(c_k)_{k\in [N]}$ and $(d_k)_{k\in [N]}$ with $c_k\leq d_k$ such that, whenever $x,\tilde{x}\in \prod_{j\in[N]}\Lambda_j$ differ only in the $k$th coordinate, we have
\begin{equation}
\label{B.1}
\big|f(x)-f(\tilde{x}) \big|\leq \begin{cases}
c_k, & \mbox{if }\; \Gamma \;\mbox{holds}, \\
d_k, & otherwise.
\end{cases}
\tag{B.1}
\end{equation}
Then for all numbers $(\gamma_k)_{k\in[N]}$ with $\gamma_k\in (0,1]$ there is an event $\mathcal B=\mathcal B(\Gamma, (\gamma_k)_{k\in[N]})$ satisfying
\begin{equation}
\label{B.2}
\mathbb P(\mathcal B)\leq \mathbb P(X\notin \Gamma)\underset{k\in[N]}{\sum}{\gamma_k^{-1}}\quad { and}\quad \urcorner\mathcal B\subseteq \Gamma,
\tag{B.2}
\end{equation}
such that for all $t\geq 0$ we have
\begin{equation}
\label{B.3}
\mathbb P(f(X)\geq \mathbb Ef(X)+t\; and\;\urcorner\mathcal B)\leq {\rm exp}\big( -\frac{t^2}{2\sum_{k\in[N]}(c_k+e_k)^2} \big),
\tag{B.3}
\end{equation}
where $e_k=\gamma_k(d_k-c_k).$
In many applications, the following simple consequence of {(\ref{B.2})} and {(\ref{B.3})} suffices:
\begin{align}
\label{B.4}
&\mathbb P(f(X)\geq \mathbb Ef(X)+t)\leq {\rm exp}\big(-\frac{t^2}{2\sum_{k\in[N]}(c_k+\gamma_k(d_k-c_k))^2}\big)+\mathbb P(X\notin\Gamma)\underset{k\in[N]}{\sum}\gamma_k^{-1}.
\tag{B.4}
\end{align}

\end{proposition}

Now we proceed to bound the light pairs. Define $$u_{ij}=x_iy_j\mathbf 1((i,j)\in \mathcal L)+x_jy_i\mathbf 1((j,i)\in \mathcal L)$$ for all $i,j=1,...,n$. Then
\begin{align}
\label{B.5}
\underset{(i,j)\in \mathcal L}{\sum}x_iy_j(W^M-\mathbb E(W^M))_{ij}&=\underset{i<j}{\sum}\underset{k\neq i,j}{\sum}\mathbf 1(W_{ij} W_{ik} W_{jk}>0)(W_{ij}+W_{ik}+W_{jk})u_{ij}\nonumber\\
&-\underset{i<j}{\sum}\underset{k\neq i,j}{\sum}\mathbb E\big(\mathbf 1(W_{ij} W_{ik} W_{jk}>0)(W_{ij}+W_{ik}+W_{jk})\big)u_{ij}.
\tag{B.5}
\end{align}
To use the result in Proposition \ref{typicalbound}, we define
\begin{equation}
\label{B.6}
{f(W)=\underset{i<j}{\sum}\underset{k\neq i,j}{\sum}\mathbf 1(W_{ij} W_{ik} W_{jk}>0)(W_{ij}+W_{ik}+W_{jk})u_{ij}.}
\tag{B.6}
\end{equation}
It is obvious that $f$ is a function of independent variables $W_{ij}(i,j=1,...,n)$. Now we proceed to bound the effect on $f$ when only one element of $W_{ij}$ is changed, namely, we specify (\ref{B.1}) in our setting. Suppose $W_{ij}$ changes, then the effect on $f$ may be ``large'' on the term involving $u_{ij}$. In particular, the effect can be bounded as
\begin{equation}
\label{B.7}
\underset{k\neq i,j}{\sum}\mathbf 1( W_{ik}\cdot W_{jk}>0)(1+W_{ik}+W_{jk})u_{ij}.
\tag{B.7}
\end{equation}
On the other hand, the effect on $f$ may be ``small'' on the term involving $u_{ik}$ or $u_{jk}$. In particular, the effect can be bounded as $3u_{ik}$ or $3u_{jk}$, respectively, provided that $k$ is the common neighborhood of $i$ and $j$. To further bound the ``large'' effect and ``small'' effect, we define
\begin{equation}
\label{B.8}
{\tau_{ij}=\underset{k\neq i,j}{\sum}\mathbf 1( W_{ik}\cdot W_{jk}>0)(1+W_{ik}+W_{jk})},
\tag{B.8}
\end{equation}
by recalling the formula in (\ref{B.7})
and define the ``good set'' $\Gamma$
\begin{equation}
\label{B.9}
{\Gamma=\{(W_{ij}): \underset{ij}{\rm max}\,\tau_{ij}\leq c\tau_{\rm max}\},}
\tag{B.9}
\end{equation}
where $c$ is large enough.
We want to prove that the probability of $W$ outside the good set vanishes when $n$ goes to infinity. To this end, we first bound the expectation of
$$\delta_{ij}^k:=\mathbf 1( W_{ik}\cdot W_{jk}>0)(1+W_{ik}+W_{jk}).$$
We have
\begin{align}
\label{B.10}
\mathbb E(\delta_{ij}^k)&{=\mathbb E\big(\mathbf 1( W_{ik}\cdot W_{jk}>0)(1+W_{ik}+W_{jk})\big)}\leq c'p_{n}^2.
\tag{B.10}
\end{align}
Now we use the Bernstein's inequality to bound the the probability of $W$ outside the good set $\Gamma$.
{
\begin{align}
\label{B.11}
\mathbb P(W\notin \Gamma)&\leq n^2\mathbb P(\tau_{ij}>c\tau_{\rm max}) \nonumber\\
&\leq n^2\mathbb P\big(\underset{k\neq i,j}{\sum}\mathbf 1( W_{ik}\cdot W_{jk}>0)(1+W_{ik}+W_{jk})>c\tau_{\rm max}\big)  \nonumber\\
&\leq n^2\mathbb P\big(\underset{k\neq i,j}{\sum}\big[\mathbf 1( W_{ik}\cdot W_{jk}>0)(1+W_{ik}+W_{jk})-\mathbb E(\delta_{ij}^k)\big]>c\tau_{\rm max}-c'np_{\rm max}\big) \nonumber\\
&\leq n^2\mathbb P\big(\underset{k\neq i,j}{\sum}\big[\mathbf 1( W_{ik}\cdot W_{jk}>0)(1+W_{ik}+W_{jk})-\mathbb E(\delta_{ij}^k)\big]>c''\tau_{\rm max}\big)\nonumber\\
&\leq n^2 {\rm exp}\big(-\frac{c\tau_{\rm max}^2}{c'np_{n}^2+c''\tau_{\rm max}}   \big)\leq n^2 {\rm exp}\big(-c'''\tau_{\rm max}   \big)\nonumber\\
&\leq n^2{\rm exp}(-{\rm log} n^c) \leq n^{-c+2}\leq n^{-c'},
\tag{B.11}
\end{align}
}where we used the following facts, $\tau_{\rm max}\geq c{\rm log}n$ which is implied by $\tau_{\rm max}=np_{n}^2$ and $p_{n}^2\geq c {\rm log} n/n$, $\mathbb E((\delta_{ij}^k)^2)\leq c\mathbb E(\delta_{ij}^k)$ and (\ref{B.10}).

Next, we continue to bound the effect of changing one element $W_{ij}$ of $W$ on $f(W)$. Before that, we first have the following observation. Recalling (\ref{B.8}), under the good event $\Gamma$, we have
\begin{align}
\label{B.12}
&\underset{k\neq i,j}{\sum}\mathbf 1( W_{ik}\cdot W_{jk}>0)\leq \tau_{ij}=\underset{k\neq i,j}{\sum}\mathbf 1( W_{ik}\cdot W_{jk}>0)(1+W_{ik}+W_{jk})\leq c\tau_{\rm max},
\tag{B.12}
\end{align}
which means that the number of $k$ being the common neighborhood of $i$ and $j$ is not larger than $c\tau_{\rm max}$ under the good event $\Gamma$. Recall that,
the ``large'' effect can be bounded as $$\underset{k\neq i,j}{\sum}\mathbf 1( W_{ik}\cdot W_{jk}>0)(1+W_{ik}+W_{jk})u_{ij},$$ and the ``small'' effect can be bounded as $3u_{ik}$ or $3u_{jk}$, respectively, provided that $k$ is the common neighborhood of $i$ and $j$. Therefore, combining the ``large'' and ``small'' effects, under the good event $\Gamma$, the total effect of changing one element $W_{ij}$ in $f(W)$ can be upper bounded as
\begin{equation}
\label{B.13}
c_{ij}\leq c\tau_{\rm max}u_{ij}+3\sum_{k:W_{ik}W_{jk}>0}u_{ik}\leq c\tau_{\rm max}u_{ij}+c'\tau_{\rm max}u_{ik}.
\tag{B.13}
\end{equation}
On the other hand, when the bad event $\urcorner \Gamma$ occurs, the total effect of changing one element $W_{ij}$ in $f(W)$ becomes
\begin{equation}
\label{B.14}
d_{ij}\leq 3nu_{ij}+3nu_{ik}.
\tag{B.14}
\end{equation}
We are now moving towards (\ref{B.3}) in our setting. Define $\gamma_{ij}=\frac{1}{n}$ for all $i,j$. Then $e_{ij}=\gamma_{ij}(d_{ij}-c_{ij})=o(c_{ij})$ by recalling the definition of $\tau_{\rm max}$. And we have
\begin{align}
\label{B.15}
\sum_{ij}c_{ij}^2\leq c\tau_{\rm max}^2\sum_{ij}u_{ij}^2,
\tag{B.15}
\end{align}
and
\begin{align}
\label{B.16}
\mathbb P(\mathcal B)&\leq\mathbb P(W\notin\Gamma)\underset{i,j}{\sum}\gamma_{ij}^{-1}=n^3\mathbb P(W\notin\Gamma)\leq {\rm exp}\big(-(c'-3){\rm log}n\big).\tag{B.16}
\end{align}
Therefore, combining (\ref{B.15}) with (\ref{B.3}), we have for large enough $c$ that
\begin{align}
\label{B.17}
{\mathbb P(f(W)-\mathbb Ef(W)>c\sqrt{D} \cap \mathcal \urcorner \mathcal B)}&\leq {\rm exp} \big(-\frac{c^2D}{c'\tau_{\rm max}^2\sum_{ij}u_{ij}^2}\big)\nonumber\\
&\leq {\rm exp} \big(-\frac{c^2\tau_{\rm max}^2n\alpha_np_{n}}{c'\tau_{\rm max}^2\sum_{ij}u_{ij}^2}\big)\nonumber\\
&\leq {\rm exp}(-cn\sqrt{n\alpha_np_{n}})\leq {\rm exp}(-c''n)\nonumber,
\tag{B.17}
\end{align}
where the second inequality follows from the fact that
\begin{align*}
\sum_{ij}u_{ij}^2&\leq c\,{\rm max}_{ij}|u_{ij}|\sum_{i,j}|x_iy_j|\leq c\frac{\sqrt{D}}{n\tau_{\rm max}}\|x\|_2\|y\|_2\leq c'\frac{\sqrt{D}}{n\tau_{\rm max}},
\end{align*}
and the last inequality is implied by {$\alpha_np_{n}>c''({\rm log}n)^\beta/n$} for some small constant $c''$. Finally, it is obvious that the event $\{W\notin \Gamma\}$ does not depend on the choice of vectors $x$ and $y$, thus taking the supremum over all $x$ and $y$, and combining (\ref{B.16}) and (\ref{B.17}) with (\ref{B.4}), we obtain
\begin{align}
\label{B.18} \mathbb P(\underset{x,y\in \mathcal N}{{\rm sup}}\big |\underset{(i,j)\in \mathcal L}{\sum}x_iy_j(W^M-\mathbb E(W^M))_{ij}\big|\geq c_1\sqrt{D})\leq {\rm exp}\big(-(c''-{\rm log}5)n  \big)+{\rm exp}\big(-(c'-3){\rm log}n\big),
\tag{B.18}
\end{align}
where we used the fact that $|\mathcal N|\leq {\rm exp}(n{\rm log}5)$. Since $c''$ and $c'$ are large enough, the probability in (\ref{B.18}) decays polynomially with $n$. \QEDA

\subsection*{\textbf{Proof of Lemma \ref{heavy}}}
The proof follows a similar strategy as in \cite{feige2005spectral,paul2018higher,lei2015consistency}, among others. We will focus on the heavy pairs $(i,j)$ such that $x_i>0,y_j>0$ and the set $$\mathcal H_1=\{(i,j)\in\mathcal H: x_i>0,y_j>0\}.$$
The other three cases can be treated similarly. We will need the following notation:
\begin{itemize}
\item [$\bullet$] $I_1=\{i: \frac{1}{2n^{1/2}}\leq x_i\leq \frac{1}{n^{1/2}}\}$, $I_s=\{i: \frac{2^{s-1}}{2n^{1/2}}\leq x_i\leq \frac{2^{s}}{2n^{1/2}}\}$ for $s=2,3,...,\lceil{\rm log}_22n^{1/2}\rceil$.\\
 $J_1=\{j: \frac{1}{2n^{1/2}}\leq y_j\leq \frac{1}{n^{1/2}}\}$, $J_t=\{j: \frac{2^{t-1}}{2n^{1/2}}\leq y_j\leq \frac{2^{t}}{2n^{1/2}}\}$ for $t=2,3,...,\lceil{\rm log}_22n^{1/2}\rceil$.
\item [$\bullet$] $e(I,J)$ denotes the weights of distinct edges in the motif-based network $W^M$ between nodes sets $I$ and $J$.
$e(I,J)=\sum_{i\in I}\sum_{j\in J}W^M_{ij}, \,\mbox{if }\, I\cap J=\emptyset$; $\sum_{(i,j)\in I\times J\backslash (I\cap J)^2} W^M_{ij}+\sum_{(i,j)\in (I\cap J)^2,i<j}W^M_{ij},\,\mbox{if}\; I\cap J\neq\emptyset$.
\item [$\bullet$] $\mu(I,J)=\mathbb E (e(I,J))$, and $\bar\mu=|I||J|\frac{\Delta}{n}$, where we define $\Delta=n^2\alpha_np_{n}^3.$
\item [$\bullet$] $\lambda_{st}=e(I_s,J_t)/\bar\mu_{st}$, where $\bar\mu_{st}=\bar\mu(I_s,J_t)$.
\item [$\bullet$] $\alpha_s=|I_s|2^{2s}/n,\beta_t=|J_t|2^{2t}/n$, and $\sigma_{st}=\lambda_{st}\frac{\sqrt{D}}{\tau_{\rm max}}2^{-(s+t)}$.
\end{itemize}
We begin by providing the following two results which play a key role in the proof.
\begin{lemma}[{Bounded degree}]\label{boundeddegree}
Let $$d_{i}^M=\sum_{j}\sum_{k\neq i,j}(\mathbf 1(W_{ij}\cdot W_{ik}\cdot W_{jk}>0)(W_{ij}+W_{ik}+W_{jk}))$$ denote the {\rm weighted triangle degree} of node $i$. If $\alpha_np_{n}^2>c''({\rm log}n)^\beta/n$ for any $0<\beta<1$, then for
all $i$ and some constant $r_4,r_5>0$, there exists a constant $c_3(r_4,r_5)>0$ such that $$d_{i}^M\leq c_3\Delta,$$
with probability larger than $1-n^{-r_4}-{\rm exp}(-r_5({\rm log} n)^\beta)$.
\end{lemma}
\begin{lemma}[{Bounded discrepancy}]\label{boundeddiscrepancy}
For a constant $r_6>0$, there exists constants $c_4=c_4(r_6)>0$, $c_5=c_5(r_6)>0$ such that with probability larger than $1-2n^{-r_6}$, for any $I,J\subseteq [n]$ with $[I]\leq [J]$, at least one of the following statements holds:

$(a)\;\frac{e(I,J)}{\bar{\mu}(I,J)}\leq e c_4$,

$(b)\;e(I,J){\rm log}\frac{e(I,J)}{\bar\mu(I,J)}\leq  c_5\tau_{\rm max}|J|{\rm log}\frac{n}{|J|}$.
\end{lemma}
Define $\mathcal E$ be the event that the results of Lemma \ref{boundeddegree} and \ref{boundeddiscrepancy} hold. In the sequel, we proceed to bound the heavy pairs under the event $\mathcal E$. We first have the following facts.
\begin{align}
&\underset{(i,j)\in\mathcal H_1}{\sum}x_iy_jW^M_{ij}\nonumber\\
&\leq 2 \underset{2^{(s+t)}\geq \frac{\sqrt{D}}{\tau_{\rm max}}}{\sum}e(I_s,J_t)\frac{2^s}{2n^{1/2}}\frac{2^t}{2n^{1/2}} \nonumber\\
&\leq 2\frac{1}{4}\underset{2^{(s+t)}\geq \frac{\sqrt{D}}{\tau_{\rm max}}}{\sum}\frac{e(I_s,J_t)}{|I_s||J_t|n\alpha_np_{\rm max}^3}\frac{D}{\tau_{\rm max}}\frac{2^s|I_s|2^t|J_t|}{n^{2}}\nonumber\\
&=\frac{1}{2}\sqrt{D}\underset{2^{(s+t)}\geq \frac{\sqrt{D}}{\tau_{\rm max}}}{\sum}\frac{e(I_s,J_t)}{\bar{\mu}(I_s,J_t)}\frac{\sqrt{D}}{\tau_{\rm max}}2^{-(s+t)}\frac{2^{2s}|I_s|2^{2t}{J_t}}{n^{2}}\nonumber\\
&= \frac{1}{2}\sqrt{D}\underset{2^{(s+t)}\geq \frac{\sqrt{D}}{\tau_{\rm max}}}{\sum}\alpha_s\beta_t\sigma_{st}.\nonumber
\end{align}
We will prove that ${\sum}_{2^{(s+t)}\geq \frac{\sqrt{D}}{\tau_{\rm max}}}\alpha_s\beta_t\sigma_{st}$ is bounded by some constant. To that end, following \cite{feige2005spectral,lei2015consistency,paul2018higher}, we split the set of pairs $\mathcal C:\{(s,t):2^{(s+t)}\geq \frac{\sqrt{D}}{\tau_{\rm max}},|I_s|\leq |J_t|\}$ into six sets and we will show that the contribution of each part is bounded. We will use the following facts repeatedly,
\begin{equation}
\sum_s\alpha_s\leq 4(1/2)^{-2}=16,\quad\quad \sum_t\beta_t\leq 4(1/2)^{-2}=16.\nonumber
\end{equation}
\begin{itemize}
\item [$\bullet$] $ \mathcal C_1: \{(s,t)\in\mathcal C,\sigma_{st}\leq 1\}$.
$$\sum_{s,t}\alpha_s\beta_t\sigma_{st}\mathbf 1\{(s,t)\in\mathcal C_1\}\leq \sum_{s,t}\alpha_s\beta_t\leq 256.$$
\item [$\bullet$] $\mathcal C_2: \{(s,t)\in\mathcal C\backslash \mathcal C_1,\lambda_{st}\leq c e\}$.\\
Note that $$\sigma_{st}=\lambda_{st}\frac{\sqrt{D}}{\tau_{\rm max}}2^{-(s+t)}\leq \lambda_{st}\leq ce,$$ so we have
$$\sum_{s,t}\alpha_s\beta_t\sigma_{st}\mathbf 1\{(s,t)\in\mathcal C_2\}\leq ce.$$
\item [$\bullet$] $ \mathcal C_3: \{(s,t)\in\mathcal C\backslash ( \mathcal C_1\cup \mathcal C_2),2^{s-t}\geq \frac{\sqrt{D}}{\tau_{\rm max}} \}$.\\
Note that by Lemma \ref{boundeddegree}, $$\lambda_{st}=\frac{e(I_s,J_t)}{\bar{\mu}_{st}}\leq c\frac{|I_s|\Delta}{|I_s||J_t|\Delta/n}\leq c\frac{n}{|J_t|}.$$ Hence
$$\sigma_{st}=\lambda_{st}\frac{\sqrt{D}}{\tau_{\rm max}}2^{-(s+t)}\leq c\frac{n}{|J_t|}\frac{\sqrt{D}}{\tau_{\rm max}}2^{-(s+t)}.$$
And consequently \begin{align*}
\sum_t\beta_t\sigma_{st}&\leq c\sum_t2^{2t}\frac{|J_t|}{n}\cdot \frac{n}{|J_t|}\frac{\sqrt{D}}{\tau_{\rm max}}2^{-(s+t)}=c\sum_t\frac{\sqrt{D}}{\tau_{\rm max}2^{s-t}}\leq c,
\end{align*}
where the last inequality follows from the fact that the non-zero summands over $t$ are bounded by 1 for a geometric sequence. Finally, we have $$\sum_{s,t}\alpha_s\beta_t\sigma_{st}\mathbf 1\{(s,t)\in\mathcal C_3\}\leq c\sum_{s}\alpha_s\leq c'.$$
\item [$\bullet$] $ \mathcal C_4: \{(s,t)\in\mathcal C\backslash ( \mathcal C_1\cup \mathcal C_2\cup \mathcal C_3), {\rm log}\lambda_{st}>\frac{1}{4}[2t\,{\rm log}2+{\rm log}(1/\beta_t)]  \}$.\\
By Lemma \ref{boundeddiscrepancy}, we have
\begin{align*}
\lambda_{st}{\rm log}\lambda_{st}|I_s||J_t|\frac{\Delta}{n}&=\frac{e(I_s,J_t)}{\bar{\mu}(I_s,J_t)}{\rm log}\frac{e(I_s,J_t)}{\bar{\mu}(I_s,J_t)}\bar{\mu}(I_s,J_t)\leq c_5\tau_{\rm max}|J_t|{\rm log}\frac{n}{|J_t|}.
\end{align*}
Note that $\frac{\tau_{\rm max}}{\Delta}=\frac{\tau_{\rm max}^2}{D}$ by the definition of $D,\Delta$ and $p_n\geq \frac{{\rm log}n}{n}$. Then we have
\begin{align}
\label{B.19}
\sigma_{st}\alpha_s&\leq \lambda_{st}\frac{\sqrt{D}}{\tau_{\rm max}}2^{-(s+t)}\frac{|I_s|2^{2s}}{n} \nonumber\\
&\leq c_5\frac{1}{{\rm log}\lambda_{st}}\frac{\tau_{\rm max}^2}{D}{\rm log}\big(\frac{n}{|J_t|}\big)\frac{\sqrt{D}}{\tau_{\rm max}}2^{s-t}\nonumber\\
&\leq c_5\frac{1}{{\rm log}\lambda_{st}}\frac{2^{s-t}}{\sqrt{D}/\tau_{\rm max}}\big[2t{\rm log}2+{\rm log}(\frac{1}{\beta_t})  \big]\tag{B.19}\\
&\leq 4c_5\frac{2^{s-t}}{\sqrt{D}/\tau_{\rm max}}.\nonumber
\end{align}
And then
\begin{align}
&\underset{s,t}{\sum}\alpha_s\beta_t\sigma_{st}\mathbf 1\{(s,t)\in \mathcal C_4\}=\underset{t}{\sum}\beta_t\underset{s}{\sum}\sigma_{st}\alpha_s\mathbf 1\{(s,t)\in \mathcal C_4\}\nonumber\\
&\leq 4c_5\underset{t}{\sum}\beta_t \underset{s}{\sum}\frac{2^{s-t}}{\sqrt{D}/\tau_{\rm max}}\mathbf 1\{(s,t)\in \mathcal C_4\}\leq 4c_5\underset{t}{\sum}\beta_t\leq c',
\nonumber
\end{align}
where we used the fact that $(s,t)\notin \mathcal C_3$.
\item [$\bullet$] $ \mathcal C_5: \{(s,t)\in\mathcal C\backslash ( \mathcal C_1\cup \mathcal C_2\cup \mathcal C_3\cup \mathcal C_4), 2t{\rm log}2\geq {\rm log}(1/\beta_t)  \}$.\\
Since $(s,t)\notin \mathcal C_4$, we have
$${\rm log}\lambda_{st}\leq \frac{1}{4}[2t{\rm log}2+{\rm log}\beta_t^{-1}]\leq ct{\rm log}2,$$
and thus $\lambda_{st}\leq c 2^t$.
Because $(s,t)\notin \mathcal C_1$, we have
$$1\leq \sigma_{st}=\lambda_{st}\frac{\sqrt{D}}{\tau_{\rm max}}2^{-(s+t)}\leq c\frac{\sqrt{D}}{\tau_{\rm max}}2^{-s}.$$
And because $(s,t)\notin \mathcal C_2$, we have ${\rm log}\lambda_{st}\geq c'$. Thus by (\ref{B.19}) and the definition of $\mathcal C_5$, we have
$$\sigma_{st}\alpha_s\leq c\frac{2^{s-t}}{\sqrt{D}/\tau_{\rm max}}\cdot 4t {\rm log}2.$$
Consequently,
\begin{align}
\underset{s,t}{\sum}\alpha_s\beta_t\sigma_{st}\mathbf 1\{(s,t)\in \mathcal C_5\} &=\underset{t}{\sum}\beta_t\underset{s}{\sum}\sigma_{st}\alpha_s\mathbf 1\{(s,t)\in \mathcal C_5\}\nonumber\\
&\leq c\underset{t}{\sum}\beta_t\underset{s}{\sum}\frac{2^{s-t}}{\sqrt{D}/\tau_{\rm max}}4t\nonumber\\
&\leq c\underset{t}{\sum}\beta_tt2^{-t}\underset{s}{\sum}\frac{2^{s}}{\sqrt{D}/\tau_{\rm max}}\nonumber\\
&\leq c\underset{t}{\sum}\beta_t\leq c',\nonumber\end{align}
where the last inequality follows from the fact that the non-zero summands over $s$ are bounded by 1 for a geometric sequence.
\item [$\bullet$] $ \mathcal C_6: \{(s,t)\in\mathcal C\backslash ( \mathcal C_1\cup \mathcal C_2\cup \mathcal C_3\cup \mathcal C_4\cup \mathcal C_5) \}$.\\
Since $(s,t)\notin \mathcal C_4\cup\mathcal C_2 \cup\mathcal C_5, $ , we have
$$0<{\rm log}\lambda_{st}\leq \frac{1}{4}[2t{\rm log}2+{\rm log}\beta_t^{-1}]\leq \frac{1}{2}{\rm log}\beta_t^{-1}\leq {\rm log}\beta_t^{-1}.$$
Then
\begin{align*}
\underset{s,t}{\sum}\alpha_s\beta_t\sigma_{st}\mathbf 1\{(s,t)\in \mathcal C_6\}
&=\underset{s}{\sum}\alpha_s\underset{t}{\sum}\beta_t\lambda_{st}\frac{\sqrt{D}}{\tau_{\rm max}}2^{-(s+t)}\mathbf 1((s,t)\notin \mathcal C_6)\nonumber\\
&\leq \underset{s}{\sum}\alpha_s\underset{t}{\sum}\frac{\sqrt{D}}{\tau_{\rm max}}2^{-(s+t)}\mathbf 1((s,t)\notin \mathcal C_6)\nonumber\\
&\leq c\underset{s}{\sum}\alpha_s\leq c'.\nonumber
\end{align*}
\end{itemize}
Putting these pieces together, and combing (\ref{A.6}), for any fixed $x,y\in \mathcal H$, we have
$$\underset{(i,j)\in\mathcal H}{\sum}x_iy_jW^M_{ij}\leq c \sqrt{D},$$
under the event $\mathcal E$. Taking the supremum and combining the results of Lemma \ref{boundeddegree} and \ref{boundeddiscrepancy}, we arrive the conclusion of Lemma \ref{heavy}.
\QEDA

\subsection*{\textbf{Proof of Lemma \ref{boundeddegree}}}
We will make use of \cite{warnke2017upper} to bound the degree. For reference, we reproduce the result as the following proposition.
\begin{proposition}[Theorem 9 of \cite{warnke2017upper}]\label{dependencebound}
Let $(Y_i),i\in\mathcal I$ be a collection of non-negative random variables with $\sum_{i\in \mathcal I}\mathbb E(Y_i)\leq \mu$. Assume that $\sim$ is a symmetric relation on $\mathcal I$ such that each $Y_i$ with $i\in \mathcal I$ is independent of $\{Y_j: j\in\mathcal I,j\not\sim i\}$. Let $Z_C={\rm max}\sum_{i\in\mathcal J}Y_i$, where the maximum is taken over all sets $\mathcal J\subset \mathcal I$ such that ${\rm max}_{j\in\mathcal J}\sum_{i\in \mathcal J,i\sim j}Y_i\leq C$. Then for all $C,t>0$, we have
$$\mathbb P(Z_C\geq \mu+t)\leq {\rm min}\Big\{{ \rm exp}\Big(-\frac{t^2}{2C(\mu+t/3)}\Big),\Big(1+\frac{t}{2\mu}\Big)^{-t/2C}\Big\}.$$
\end{proposition}
Now we proceed to bound the degree. First, we have the observation that
\begin{align*}
&\mathbb Ed_{i}^M=\mathbb E[\sum_{j}\sum_{k\neq i,j}(\mathbf 1(W_{ij}W_{ik} W_{jk}>0)(W_{ij}+W_{ik}+W_{jk}))]\\
&\leq 3n^2\alpha_np_{\rm max}^3\leq c\Delta,
\end{align*}
where we recall that $\Delta=n^2\alpha_np_{n}^3.$
Then let $I_i$ be the set of all triangles that includes node $i$ and denote $$M_\theta:=\mathbf 1(W_{ij}\cdot W_{ik}\cdot W_{jk}>0)(W_{ij}+W_{ik}+W_{jk})$$ with the index $\theta=\{i,j,k\}\in I_i$. Recall the good event $\Gamma$ we defined in (\ref{B.9}), then under the good event, every pair of nodes has at most $C=c\tau_{\rm max}$ common neighbors, and the good event holds with probability larger than $1-n^{-c''}$. It is clear that two triangles belonging to the set $I_i$ are independent if they do not share any edges. Let $\sim$ denote a relation such that $\theta_1\sim\theta_2$ holds if $\theta_1$ and $\theta_2$ share an edge. For any $M_{ijk}$, the good event restricts the number of triangles in the set $I_i$ that are dependent with $M_{ijk}$ to $2C$.

Define $J_i\subseteq I_i$ as
$$J_i=\{\theta: \underset{\theta_1\in J_i}{{\rm max}}\;|\theta_2\in J;\theta_1\;{\mbox{and}}\;\theta_2\; \mbox{ share at least one edge} |\leq 2C\}.$$
Then, we immediately have
$$\underset{\theta_2\in J_i}{{\rm max}}\;\underset{\theta_1\in J_i:\theta_1\sim\theta_2}{\sum}M_\theta\leq 3\cdot2C\leq c\tau_{\rm max}, \quad \sum_{\theta\in \mathcal I_i}\mathbb E(M_\theta)\leq \Delta.$$
Take $t=\mu=\Delta$, then the results in Proposition \ref{dependencebound} imply,
\begin{align}
&\mathbb P(\underset{J_i}{{\rm max}}\;\underset{\theta\in J_i}{\sum}{M_\theta}\geq 2\Delta)\nonumber\\
&\leq {\rm min}\Big\{{\rm exp}\Big(-\frac{\Delta^2}{6C(\Delta+\Delta/3)}\Big),\Big(1+\frac{\Delta }{2\Delta}\Big)^{-\Delta/6C}\Big\}\nonumber\\
&= {\rm min}\Big\{{\rm exp}\Big(-\frac{\Delta}{c\tau_{\rm max}}\Big),(\frac{3}{2})^{-\Delta/c\tau_{\rm max}}\Big\}\nonumber\\
&\leq {\rm exp}(-c'({\rm log}n)^\beta),\nonumber
\end{align}
where the last inequality follows from the assumption that $\frac{\Delta}{\tau_{\rm max}}\geq cn\alpha_np_{n}\geq c'{({\rm log}n)^\beta}$ with $0<\beta<1$. Under the good event, we have $J^*_i=I_i$, and thus $\underset{J_i}{{\rm max}}\;\underset{\theta\in J_i}{\sum}{M_\theta}=d_i^M$. Consequently,
\begin{align*}
\mathbb P(d_i^M\geq 2\Delta)&\leq {\rm exp}(-c'({\rm log}n)^\beta)+P(W\notin \Gamma)\\
&\leq n^{-c''}+{\rm exp}(-c'({\rm log}n)^\beta).
\end{align*}
\QEDA
\subsection*{\textbf{Proof of Lemma \ref{boundeddiscrepancy}}}
Recall that $\bar\mu=|I||J|\frac{\Delta}{n}$ with $\Delta=n^2\alpha_np_{n}^3.$ So if $|J|>\frac{n}{e}$, then
$$\frac{e(I,J)}{\bar{\mu}(I,J)}=\frac{\sum_{i\in I}{\rm max}_id^M_i}{|I|\Delta/e}\leq \frac{c|I|\Delta}{|I|\Delta/e}\leq c'e,$$
which is the case $(a)$ in the lemma.

Now we prove the case $(b)$ also holds. Denote $S$ as the set of all 3-tuples such that each tuple has one vertex in each of the sets $I$ and $J$. Then
$$e(I,J)=\sum_{\theta\in S(I,J)}W^M_\theta,$$ and $$\mathbb E(\sum_{\theta\in S(I,J)}W^M_\theta)\leq |I||J|\Delta/n=\bar{\mu}(I,J).$$
Since $e(I,J)$ is the sum of dependent variables, next we use the concentration inequality in Proposition \ref{dependencebound} to bound $e(I,J)$.
Recall the good event $\Gamma$ we defined in (\ref{A.19}), under the good event, every pair of nodes has at most $C=c\tau_{\rm max}$ common neighbors, and the good event holds with probability larger than $1-n^{-c''}$. Define $S^{*}\subset S$ be the set such that any $W_\theta^M$ depends on $2C$ other $W_\theta^M$.
Then under the good event, we have
$$\underset{\theta_1\in S^*}{{\rm max}}\underset{\theta_2\in S^*(I,J),\theta_2\sim\theta_1}{\sum}W^M_\theta\leq c\tau_{\rm max}.$$ Let $t=(l-1)\bar{\mu}(I,J)$, and apply Proposition \ref{dependencebound}, we can obtain
\begin{align}
\mathbb P(e(I,J)\geq l\bar{\mu}(I,J))&\leq {\rm exp}\big(-\frac{t}{4\tau_{\rm max}}{\rm log}(1+\frac{t}{2\bar{\mu}(I,J)})\big)   \nonumber\\
&\leq{\rm exp}\big(-\frac{(\bar{\mu}(I,J)+t){\rm log}(1+\frac{t}{\bar{\mu}(I,J)})-t}{\tau_{\rm max}}\big)  \nonumber\\
&={\rm exp}\big(\frac{l\;{\rm log}\;l\bar{\mu}(I,J)-(l-1)\bar{\mu}(I,J)}{\tau_{\rm max}}\big),\nonumber\\
&\leq{\rm exp}\big(-\frac{1}{2}\frac{l\;{\rm log}\;l\bar{\mu}(I,J)}{\tau_{\rm max}}\big),\nonumber
\end{align}
where the second inequality follows from the fact that
$$\bar{\mu}(I,J){\rm log}\;l-(l-1)\bar{\mu}(I,J)\leq (l-1)\bar{\mu}(I,J)/2{\rm log}(1+(l-1)/2),$$
for large enough $l$, and the last inequality holds when $l\geq 8$. The remaining part of the proof is almost the same as that in \cite{paul2018higher,feige2005spectral,lei2015consistency}, hence we omit it. \QEDA

\section*{C. Clustering results in real data analysis}

\newpage

\bibliographystyle{plainnat}
\bibliography{highord}
\end{document}